%% file: main.tex
\documentclass[epj]{svjour}
\usepackage[utf8]{inputenc}
\usepackage[normalem]{ulem}
\usepackage{amsmath}
\usepackage{amssymb}
\usepackage{MnSymbol}
\usepackage{bbold}
\usepackage{graphicx}
\usepackage{braket}
\usepackage[dvipsnames]{xcolor}
\usepackage{dsfont}
\usepackage{bm}
\usepackage{upgreek}
\usepackage{indentfirst}
\usepackage{cite}

\usepackage[T1]{fontenc}
\usepackage{microtype}
\usepackage{orcidlink}
\hypersetup{
    colorlinks=true,       
    linkcolor=blue,          
    citecolor=blue,        
    filecolor=blue,      
    urlcolor=blue           
}

\newcommand{\M}{\mathcal{M}}
\newcommand{\Moller}{M{\o}ller }

\begin{document}

\vspace{3cm}

\title{\\ \\ \\ Quantum Complexity and New Directions in Nuclear Physics and High-Energy Physics Phenomenology}

\author{Caroline E. P. Robin\inst{1,2} \thanks{crobin@physik.uni-bielefeld.de} \orcidlink{0000-0001-5487-270X} 
\and Martin J. Savage\inst{3}\thanks{mjs5@uw.edu . On leave from the Institute for Nuclear Theory.} \orcidlink{0000-0001-6502-7106}}

\institute{Fakult\"at f\"ur Physik, Universit\"at Bielefeld, D-33615, Bielefeld, Germany. 
\and GSI Helmholtzzentrum f\"ur Schwerionenforschung, Planckstra{\ss}e 1, 64291 Darmstadt, Germany. 
\and InQubator for Quantum Simulation (IQuS), Department of Physics, University of Washington, Seattle, WA 98195, USA.   }

\abstract{
Advances in quantum information science (QIS) 
are providing transformative insights into the complexity of quantum many-body systems,  
potentially defining new frontiers in nuclear and high-energy physics. 
This review explores how QIS-derived techniques are fostering new analytic frameworks and algorithms—both classical and quantum—to tackle (some of the) present barriers 
to discovery in fundamental physics, 
with applicability to other science domains.
We highlight how these techniques
are shedding new light on the structure and dynamics 
of hadrons, nuclei, matter in extreme conditions, and beyond. 
Importantly, they are expected to play an essential role in 
the development of large-scale quantum simulations of such systems, 
particularly in setting the balance among quantum and classical computational resources.
}


%
\maketitle

\tableofcontents

\input{Section_Introduction}

\input{Section_Complexity}

\input{Section_Info_Flow}

\input{Section_Motivations}

\section{Applications in Low-Energy Nuclear Physics }
\label{sec:ApplicationsLENP}

There is a growing effort to understand how aspects of entanglement, magic and quantum complexity in general, are connected with emergent properties of the nuclear and hyper-nuclear forces, 
the structure and reactions of nuclei, and various many-body nuclear phenomena. 
In this section, we provide an overview of recent investigations of quantum complexity 
in low-energy nuclear physics.

\input{Section_Nuclear_Forces}
\input{Section_Nuclei}
\input{Section_NBody_methods}

\section{Applications in High-Energy and Neutrino Physics }
\label{sec:ApplicationsHEP}

High-energy physics processes exhibit quantum complexity in ways that are fundamental to each process, 
and that are also expected to provide foundations for the complexity of nuclear forces and nuclei discussed above.
Important progress has been made in connecting the symmetries and 
structure of the input coefficients in the Standard Model and modest extensions
to quantum complexity.
In the QCD sector, 
the quantum complexity of systems of quarks and gluons is a significant recent focus, from gluon-gluon scattering, 
through string-breaking and hadronization, 
through thermalization in non-Abelian gauge theories, to heavy-ion collisions.
While there has been great progress in making robust connections between QCD and nuclei, 
significant work remains to be accomplished in order to have first principles calculations in terms of quarks and gluons in each of these areas.
Efforts to understand these connections in terms of quantum complexity are in their infancy.
In the neutrino sector, where low-energy dynamics of neutrinos interacting with themselves in exotic environments, 
such as supernovae, cannot be 
directly informed from laboratory experiments, quantum complexity is beginning to  play a role in assessing simulation requirements.
In this section, we review recent advances in each of these areas.

\input{Section_Fundamental_Particles}

\input{Section_Hadron_Structure}

\input{Section_QuantumThermo}

\input{Section_StringBreaking}

\input{Section_HeavyIons}

\input{Section_Neutrinos}

\section{Summary and Thoughts}
\noindent

Quantum information science is reshaping our understanding of key aspects of nuclear and high-energy physics.
As technological progresses, theoretical tools are being developed to enable 
studies of physical systems from a rapidly advancing quantum-information-theoretic perspective.  
This change in viewpoint is providing new analytic frameworks and algorithms,
leading toward an organizational structure that enables optimal use of quantum computers to 
bypass classical computational limits, imposed by large-scale multipartite entanglement and magic.
Ultimately, this paradigm shift is poised to accelerate research into hadronic structure, nuclear structure and reactions, neutrino dynamics, high-energy collisions and more.

Strong-interaction matter, with various emerging structures across scales, ranging from quarks and gluons to atomic nuclei and their collective motion, provides a rich setting to investigate aspects of quantum complexity, their interplay, and connections with emergent physical phenomena.
Understanding how magic and entanglement are generated from few-body scattering processes, how complexity spreads in many-body nuclear environments in a multipartite way, and how information flows across energy scales, is beginning to shed light on the mechanisms underlying the emergence of collective behaviors, new degrees of freedom, and effective forces.

At the lower energy scales, atomic nuclei are particularly rich many-body systems, 
which exhibit significant multipartite entanglement and magic already in their ground states. This complexity is related, in part, to the non-perturbative character of the nuclear force, and to the presence of two types of fermions which appears to favor collective distributions of information. 
With their self-bound mesoscopic character, {\it nuclei would seem to lie near complexity barriers}, 
at the interface between various collective and single-particle regimes where entanglement, magic, and non-Gaussianity come together and interplay.
This places nuclei beyond classical tractability and makes them compelling candidates for realizing potential quantum advantages.

Frameworks for describing complexity phases and their interplay
are being developed, building on past knowledge from nuclear physics and emerging concepts from quantum information and ML/AI, potentially providing 
good initial states for further processing on quantum computers.
Overall, a detailed understanding of the complexity features of nuclei is still in its early stages. 
While multipartite entanglement and magic have been studied within practical approximations, 
full {\it ab-initio} treatments, and understanding the role of three-nucleon forces in generating complexity, are still to be explored.
On the reaction side, studies have so far been largely limited to bipartite entanglement, and understanding these dynamical processes in terms of entanglement, magic, and their mutual response, remains uncharted.
\\

Remarkably, 
all of the quantum complexity of nuclei and nuclear forces are emergent from QCD,
defined by just a handful of input parameters from the Standard Model,
and recent 
studies of the quantum complexity in (high-energy) gluon-gluon scattering are establishing basic elements toward connecting QCD and nuclei.
Making reliable predictions throughout 
the evolution of a quark-gluon plasma, 
formed in heavy-ion collisions,
through fragmentation and hadronization into final state hadrons and nuclei,
encounters one or more quantum complexity barriers.
The considerations and progress discussed above,
including in 
string breaking, 
heavy-ion collisions, 
thermalization and non-equilibrium dynamics,
gluon-gluon scattering, 
the structure of hadrons, 
and nuclear forces,
are all essential ingredients that are brought together in such processes. 
They dictate the need for quantum computing resources 
and the stages of the evolution in which these resources are required.
``Stitching'' a 
quantum complexity driven
path together is at its earliest stages, 
but is already providing new insights that are  
guiding conceptual and simulations developments.
We imagine more substantial 1+1D quantum simulations and beyond exploratory 2+1D simulations, 
including error-quantification,
aimed toward improving predictions for 
high-energy QCD collisions, including quantum correlations,
with increasingly more accurate projections for quantum resource requirement estimates.
These will enable further numerical studies of the evolution of quantum complexity in accessible processes, and possible identifications of new quantum correlations that could enhance 
future new-physics discovery capabilities in experiments.

Most of the focus of studies of quantum complexity in particle physics have centered around two-particle systems, such as the $t\overline{t}$ measurements and analysis, and the structures of the electroweak interactions and Higgs sector.
These further highlight the role between complexity and symmetry.
The $t\overline{t}$ measurements of magic provide a demonstration of what might be possible for other processes, and how quantum complexity, and possibly new quantum correlation based  observables, have the potential to aid in the discovery of new physics.  
\\

To close, 
new insights into key aspects of nuclear physics and high-energy physics,
along with (quantum) algorithms and simulation strategies,
are being made possible by integrating
concepts and techniques of quantum complexity developing in quantum information science. 
Ultimately, it is hoped that these new methods and capabilities will enable predictive modeling of extreme, 
non-equilibrium matter, and accelerate discoveries of new physics and phenomena in fundamental physics.

\section*{Acknowledgments}
\addcontentsline{toc}{section}{Acknowledgments}
\noindent
We would like to thank our colleagues, and the community more generally, 
for creating and discovering much of the work we have reviewed,
and providing a stimulating environment in which we are able to make our contributions to this exciting field of research. 
We are grateful to the organizers and participants
of workshops that brought together many of the ideas and themes in this area,
including the 2024 IQuS workshops on 
{\it Pulses, Qudits and Quantum Simulations}
and {\it Entanglement in Many-Body Systems: From Nuclei to Quantum Computers and Back},
as well as 
the First~\footnote{\url{https://mbqm.tii.ae/}} and 
Second~\footnote{\url{https://iqus.uw.edu/events/iqus-workshop-2025-2/}} 
International Workshops on Many-Body Quantum Magic.
This work was supported, in part, by Universit\"at Bielefeld, by GSI Helmholtzzentrum f\"ur Schwerionenforschung, 
by the Ministerium f\"ur Kultur und Wissenschaft des Landes Nordrhein-Westfalen (MKW NRW) under the funding code NW21-024-A, and by the Deutsche Forschungsgemeinschaft (DFG, German Research Foundation) through the CRC-TR 211 'Strong-interaction matter under extreme conditions'– project number 315477589 – TRR 211 (Caroline). 
This work is also supported 
by U.S. Department of Energy, Office of Science, Office of Nuclear Physics, InQubator for Quantum Simulation (IQuS)\footnote{\url{https://iqus.uw.edu}} under Award Number DOE (NP) Award DE-SC0020970 via the program on Quantum Horizons: QIS Research and Innovation for Nuclear Science\footnote{\url{https://science.osti.gov/np/Research/Quantum-Information-Science}},
and, in part,
through the Department of Physics\footnote{\url{https://phys.washington.edu}}
and the College of Arts and Sciences\footnote{\url{https://www.artsci.washington.edu}} at the University of Washington (Martin).

\appendix

\section*{\hypertarget{appendices}{Appendices}}

\addtocontents{toc}{%
  \protect\par\noindent
  \protect\hyperlink{appendices}{Appendices}%
  \protect\par
}

\section{Two-Qudit Quantum Correlations}
\label{sec:TwoQudits}
\begin{figure}[!ht]
    \centering
    \includegraphics[width=0.9\columnwidth]{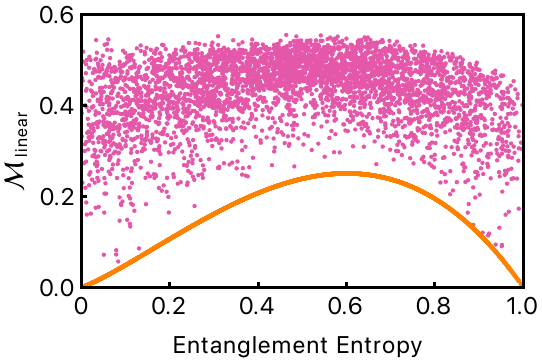}\\
 \vspace{0.5cm}
    \includegraphics[width=0.9\columnwidth]{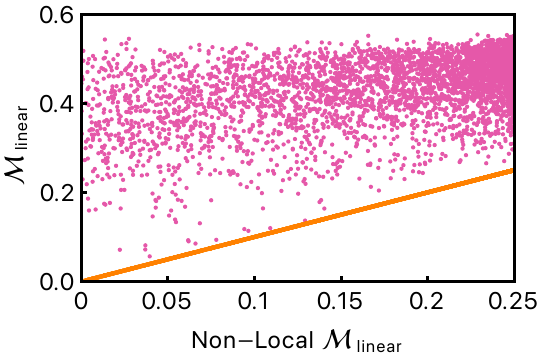}
    \caption{
    The upper panel shows the total linear magic as a function of the 
    entanglement entropy (points) in 2-qubit wavefunctions from
    4K  random samples
    from the SU(4) Haar measure, and the curve corresponds to the non-local linear magic.
    The lower panel shows the linear magic as a function of the non-local linear magic (points), 
along with a straight line with unit slope.
    }
    \label{fig:2qubitsampling}
\end{figure}

\noindent
Quantum correlations in a 2-qubit wavefunction are reviewed here.
The 2-qubit wavefunction has the general form of 
\begin{align}
    |\psi\rangle & =  c_{00} |00\rangle + c_{01} |01\rangle + c_{10} |10\rangle + c_{11} |11\rangle
    = \hat U_{SU(4)}  |00\rangle
    ,
\end{align}
where $\hat U_{SU(4)}$ is an SU(4) unitary transformation defined generally by 15 angles.
To gain intuition about correlations among the various measures of quantum complexity discussed in the main text, it is helpful to consider their values for arbitrary random states.
Using states generated randomly from the SU(4) Haar measure,
the bi-partite entanglement entropy, the 2-tangle, total and non-local linear magic, and anti-flatness are computed. 
The results of this sampling are shown in Fig.~\ref{fig:2qubitsampling}.
The non-local linear magic, which provides a lower bound to the total linear magic, is equal to four times the anti-flatness.
Further, the 2-tangle is directly related to the entanglement entropy, and is also the square of the concurrence, $\tau_2=C^2$.

In terms of eigenvalues of the reduced density matrix for either qubit,
$\rho_A = {\rm diag}(\lambda,1-\lambda)$, the quantities considered here are 
\begin{eqnarray}
    S_E & = & -\lambda\log_2\lambda - (1-\lambda)\log_2 (1-\lambda)
    \ ,\ 
    \tau_2\ =\ 4\lambda (1-\lambda) 
    \ ,
    \nonumber\\
    {\cal F}_A & = &  {1\over 4} {\cal M}_{\rm lin}^{(NL)}\ =\ \lambda(1-\lambda)(1-2\lambda)^2
    \ .
\end{eqnarray}
The latter relation coincides with $1/4$ of
the upper bound on non-local magic found using the replica method in Ref.~\cite{Cao:2024nrx}.
Thus, the non-local magic saturates the upper bound.

\begin{figure}[!ht]
    \centering
    \includegraphics[width=0.9\columnwidth]{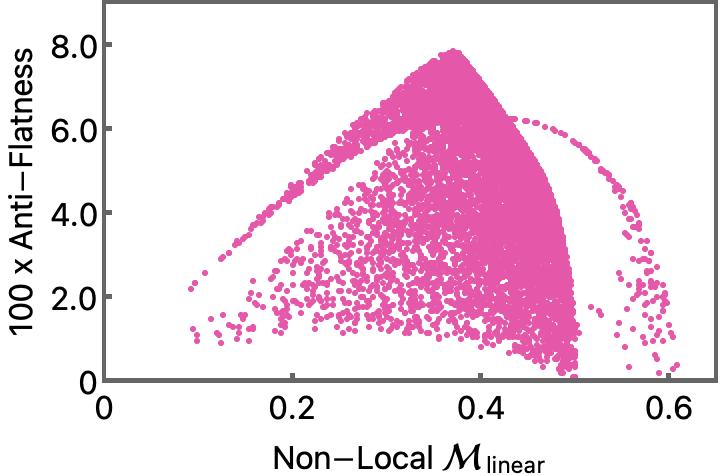}\\
 \vspace{0.5cm}
    \includegraphics[width=0.9\columnwidth]{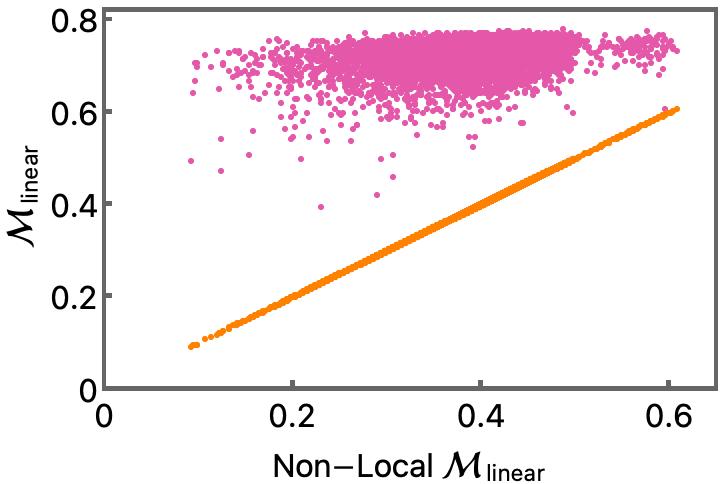}\\
    \caption{
    The upper panel shows the anti-flatness versus the non-local linear magic 
    for 10K
    two-qutrit states randomly selected from the SU(9) Haar measure.
    The lower panel shows the total linear versus the non-local linear magic.
    }
    \label{fig:2qutritsamplingMag}
\end{figure}

\begin{figure}[!ht]
    \centering
    \includegraphics[width=0.98\columnwidth]{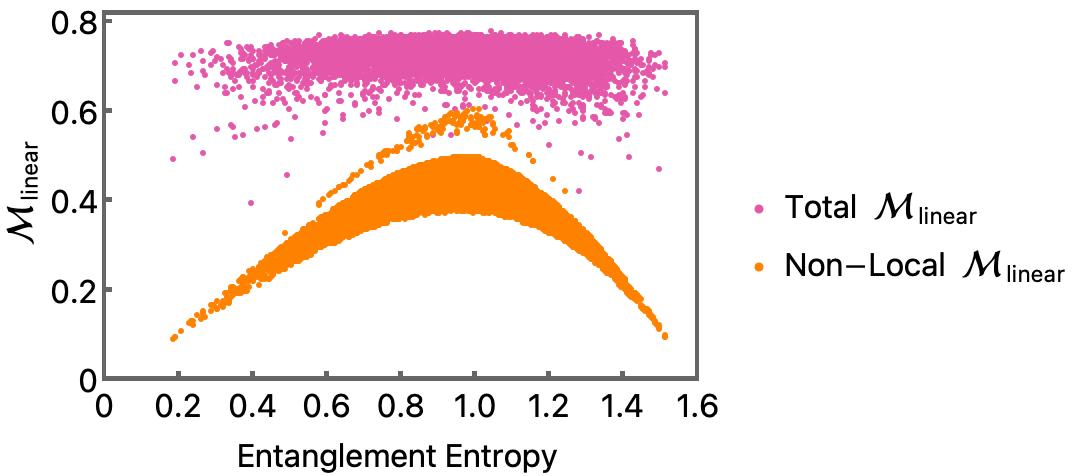}
    \caption{
    The total and non-local linear magic versus the entanglement entropy
    for 10K
    two-qutrit states randomly selected from the SU(9) Haar measure.
    }
    \label{fig:2qutritsMagVMag}
\end{figure}

In the case of $d=3$, for systems of two qutrits, relations between the various measures of quantum complexity are more complex.
A general 2-qutrit wavefunction is of the form
$|\psi\rangle = \hat U_{SU(9)} |00\rangle$,
and as discussed previously, 
the quantities considered for the 2-qubit system are generalizable.
The concurrence and generalized-concurrence 
of a n-qutrit system 
are found by forming the 
reduced density matrix for each qutrit, $\hat\rho_i$, 
and computing its eigenvalues, $\lambda_{i1,i2,i3}$.
The concurrence, $C$,  
for $\hat\rho_i$ is determined by four times the sum of products of two eigenvalues, while the generalized-concurrence,
$G$,
is the product of the three eigenvalues.  
These contributions are  summed over each qutrit, 
i.e.,
\begin{align}
    C & =  4 \sum_i 
    \left(\lambda_{i1}\lambda_{i2}+\lambda_{i1}\lambda_{i3}+\lambda_{i2}\lambda_{i3}\right)
    \  ,\ \nonumber\\ 
    G & = \sum_i \lambda_{i1}\lambda_{i2}\lambda_{i3}
     \ .
\end{align}
The $n$-tangles are formed from matrix elements of $n$ insertions of the SO(3) generators~\cite{Chen_2012}, 
$\hat J_i^n$, where,
\begin{align}
    J_1 & =  
    \begin{pmatrix}
    0&0&0 \\   0&0&-i \\ 0&i&0
    \end{pmatrix}
    \  ,\  
        J_2 \ =\  
    \begin{pmatrix}
    0&0&i \\   0&0&0 \\ -i&0&0
    \end{pmatrix}
    \  ,\  
        J_3 \ =\  
    \begin{pmatrix}
    0&-i&0 \\   i&0&0 \\ 0&0&0
    \end{pmatrix}
\  ,
\label{eq:so3gens}
\end{align}
and averaging over the squared-magnitude, i.e., 
for the 4-tangle,
\begin{eqnarray}
    \tau_4 & = {1\over {\cal N}_4} & \sum_i\ \sum_{a\ne b\ne c \ne d}\ 
    |\langle\psi|\ \hat J_{i,a}  \hat J_{i,b}  \hat J_{i,c}  \hat J_{i,d} |\psi\rangle|^2
    \ ,
\end{eqnarray}
where  ${\cal N}_4$ is the number of contributions to the sum.
This is the generalization of the $n$-tangles for n-qubit systems.
It is also the case that the anti-flatness and non-local magic are no longer uniquely related,
and in fact there are two distinct structures relating the two.
The concurrence and 2-tangle remain uniquely related, but not so for 
the generalized concurrence. 
\begin{figure}[!ht]
    \centering
    \includegraphics[width=0.9\columnwidth]{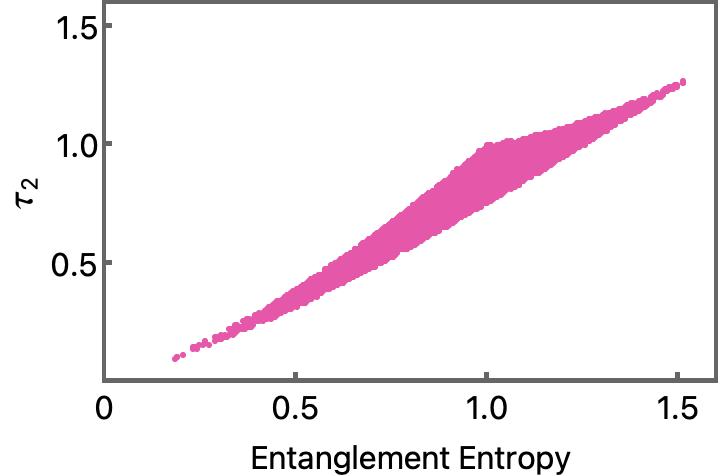  } \\
 \vspace{0.5cm}
    \includegraphics[width=0.9\columnwidth]{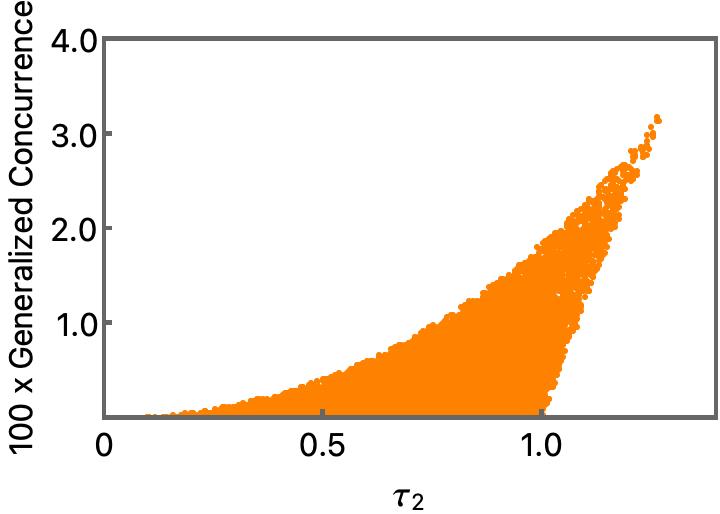  }
    \caption{
    The upper panel shows the 2-tangle (same as the concurrence) versus the entanglement entropy for 
    10K two-qutrit states randomly selected from the SU(9) Haar measure.   
    The lower panel shows the generalized concurrence versus the 2-tangle 
    (the concurrence and the 2-tangle are found to be equal).
    }
    \label{fig:2qutritsamplingEE}
\end{figure}
Figures~\ref{fig:2qutritsamplingEE}, \ref{fig:2qutritsamplingMag} 
and \ref{fig:2qutritsMagVMag}
show the correlations among measures of entanglement, magic and non-local magic.
The relation between anti-flatness and non-local magic that is present for qubits does not hold for qutrits,
and Fig.~\ref{fig:2qutritsamplingMag} suggests that there are two distinct structures constraining them.
This is reinforced by 
Fig.~\ref{fig:2qutritsMagVMag},
which shows two different regions of non-local magic when displayed against the entanglement entropy.

\section{Quantum Complexity Flow}
\label{sec:M2flow}
\begin{figure}[ht!]
    \centering
\includegraphics[width=0.95\linewidth]{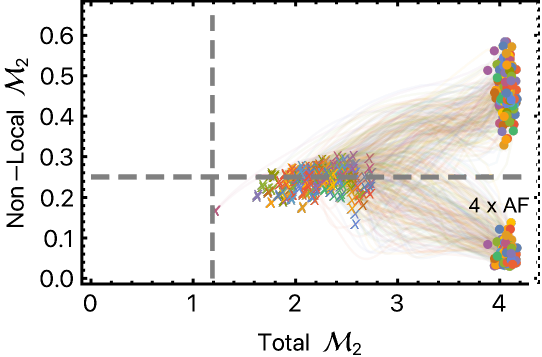}\\
 \vspace{0.5cm}
\includegraphics[width=0.95\linewidth]{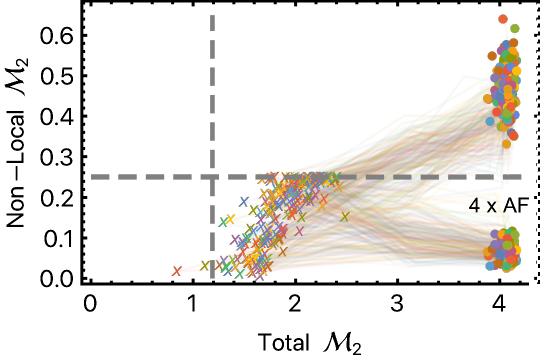}
\caption{
The renormalization flow of non-local magic, anti-flatness and total magic, ${\cal M}_2$ for 
Gaussian (upper panel) and Hadamard smearing (lower panel).
A set of 200 6-qubit Haar-random states was selected and flowed by successive Gaussian smearing 
(from 0 to $2^6$ applications of a width=1 Gaussian profile)
or Hadamard transform truncations over a finite interval.
The dashed gray lines correspond to the 2-qubit upper bounds on non-local magic and total magic.
}
    \label{fig:FlowM2}
\end{figure}

\noindent
This appendix is to provide  more detail supporting the discussion in the main text.
Figure~\ref{fig:FlowM2} shows the flow of non-local magic and total magic for the flow-parameters discussed in Sec.~\ref{sec:Flow}.
Unlike the non-local measures of magic, the total magic, the sum of local magic and non-local magic tends toward, but remains larger than,  2-qubit values for the selected flow parameters.
This is likely due to residual local rotations of the six qubits.
For flow parameters that are more than an order of magnitude larger, the total magic better recovers the permissible 2-qubit values. 
From this we conclude that non-local measures of magic converge much faster to the corresponding 
2-qubit values than the global measures.

\section{Two-Qubit Stabilizer States}
\label{app:stabs}
\noindent
For two qubits (with $d=4$), there are four stabilizer operators for each of the 
sixty stabilizer states given in Table~\ref{tab:TwoQstabs}.
Thirty-six of these states are tensor products formed from one-qubit stabilizers, 
while the remaining twenty-four are entangled states.
\begin{table*}
\caption{The complete set of sixty two-qubit stabilizer states. For notational purposes, we identify $|0\rangle=|\uparrow\rangle$ and $|1\rangle=|\downarrow\rangle$. The left set are from the tensor product of one-qubit stabilizer states, while the right set are entangled states. They are (generally) unnormalized, and require coefficients of either 1 or ${1\over\sqrt{2}}$ or ${1\over 2}$.}
\label{tab:TwoQstabs}
\centering
\begin{tabular}{c|cccc||c|cccc} 
\hline\noalign{\smallskip}
state & $|00\rangle$ & $|01\rangle$ & $|10\rangle$ & $|11\rangle$ & state & $|00\rangle$ & $|01\rangle$ & $|10\rangle$ & $|11\rangle$ \\
\noalign{\smallskip}\hline\noalign{\smallskip}
1 & 1 & 1 & 1 & 1    & 37 & 0 & 1 & 1 & 0 \\
2 & 1 & -1 & 1 & -1  & 38 & 1 & 0 & 0 & -1\\
3 & 1 & 1 & -1 & -1  & 39 & 1 & 0 & 0 & 1 \\
4 & 1 & -1 & -1 & 1  & 40 & 0 & 1 & -1 & 0\\
5 & 1 & 1 & $i$ & $i$    & 41 & 1 & 0 & 0 & $i$\\
6 & 1 & -1 & $i$ & $-i$  & 42 & 0 & 1 & $i$ & 0 \\
7 & 1 & 1 & $-i$ & $-i$  & 43 & 0 & 1 & $-i$ & 0 \\
8 & 1 & -1 & $-i$ & $i$  & 44 & 1 & 0 & 0 & $-i$\\
9 & 1 & 1 & 0 & 0    & 45 & 1 & 1 & 1 & -1\\
10 & 1 & -1 & 0 & 0  & 46 & 1 & 1 & -1 & 1\\
11 & 0 & 0 & 1 & 1   & 47 & 1 & -1 & 1 & 1\\
12 & 0 & 0 & 1 & -1  & 48 & 1 & -1 & -1 & -1\\
13 & 1 & $i$ & 1 & $i$   & 49 & 1 & $i$ & 1 & $-i$\\
14 & 1 & $-i$ & 1 & $-i$ & 50 & 1 & $i$ & -1 & $i$\\
15 & 1 & $i$ & -1 & $-i$ & 51 & 1 & $-i$ & 1 & $i$\\
16 & 1 & $-i$ & -1 & $i$ & 52 & 1 & $-i$ & -1 & $-i$ \\
17 & 1 & $i$ & $i$ & -1  & 53 & 1 & 1 & $i$ & $-i$ \\
18 & 1 & $-i$ & $i$ & 1  & 54 & 1 & 1 & $-i$ & $i$ \\
19 & 1 & $i$ & $-i$ & 1  & 55 & 1 & -1 & $i$ & $i$ \\
20 & 1 & $-i$ & $-i$ & -1& 56 & 1 & -1 & $-i$ & $-i$   \\
21 & 1 & $i$ & 0 & 0   & 57 & 1 & $i$ & $i$ & 1 \\
22 & 1 & $-i$ & 0 & 0  & 58 & 1 & $i$ & $-i$ & -1 \\
23 & 0 & 0 & 1 & $i$   & 59 & 1 & $-i$ & $i$ & -1 \\
24 & 0 & 0 & 1 & $-i$  & 60 & 1 & $-i$ & $-i$ & 1 \\
25 & 1 & 0 & 1 & 0  \\
26 & 0 & 1 & 0 & 1  \\
27 & 1 & 0 & -1 & 0  \\
28 & 0 & 1 & 0 & -1  \\
29 & 1 & 0 & $i$ & 0  \\
30 & 0 & 1 & 0 & $i$  \\
31 & 1 & 0 & $-i$ & 0  \\
32 & 0 & 1 & 0 & $-i$  \\
33 & 1 & 0 & 0 & 0  \\
34 & 0 & 1 & 0 & 0  \\
35 & 0 & 0 & 1 & 0  \\
36 & 0 & 0 & 0 & 1  \\
\noalign{\smallskip}\hline
\end{tabular}
\end{table*}

\clearpage
\bibliographystyle{sn-aps}
\addcontentsline{toc}{section}{References}  
\bibliography{biblio,bib_NeutMagicpaper}

\end{document}

%% file: Section_Introduction.tex
\section{Introduction}
\label{sec:intro}


Advances in quantum information science (QIS), quantum computing, 
communication and simulation, 
that are now being integrated into nuclear physics and high-energy physics research,
are changing how we understand, describe and compute systems of fundamental and composite particles.
While the role of quantum correlations in the structure and dynamics of matter has long been appreciated,
the re-interpretation of these correlations in terms of shared information
reveals new aspects of physical phenomena 
while allowing for direct connections with computation.
This shift in perspective has the potential to disruptively accelerate numerical simulations of physical systems, 
including in extreme regimes inaccessible to classical computing or experiment, and to advance searches for new particles, symmetries and phases of matter.

The foundations of QIS began with 
the observations of Einstein-Podolsky-Rosen (EPR)~\cite{Einstein:1935rr}
in the 1930s
and their objection to violations of local realism in quantum mechanics,
and 
with
the discovery of inequalities by Bell~\cite{Bell:1964kc} in 1964 to 
distinguish quantum from classical correlations,
that were
made experimentally
verifiable
by Clauser, Horne, Shimony and Holt (CHSH)~\cite{Clauser:1969ny} in 1969. 
First experimental tests of local hidden-variable theories were made by Clauser and Freedman~\cite{Freedman:1972zz} in 1972.
These
were solidified by experimental confirmations of the violation of Bell's inequalities
by Aspect~\cite{Aspect:1981prl,Aspect:1982fx,Aspect:1982new} and others in the 1970s and 1980s, 
consistent with the predictions of quantum mechanics and inconsistent with nature exhibiting 
local realism (see Refs.~\cite{Clauser:1978ng,aspect2004}).
In the 1980s, Feynman~\cite{Feynman1982,Feynman1986} and Manin~\cite{Manin1980,Manin2007} 
independently proposed that quantum systems could simulate aspects of physics more efficiently than classical ones, 
which followed works by Landauer~\cite{Landauer1961} and Bennett~\cite{Bennett1973,Bennett1982} 
exploring the reversibility in computing and the 
thermodynamic limits of information.
These theoretical advances 
gave rise to a series of remarkable breakthroughs in the 1990s, including
the development of Shor's quantum algorithm for 
integer factorization~\cite{Shor1994,Shor1997}, 
demonstrating that a quantum computer could break RSA encryption.
The discovery of quantum error correction (QEC) by Shor~\cite{Shor1995}, Shor and Calderbank~\cite{CalderbankShor1996} and Steane~\cite{Steane1996a,Steane1996b}
established a path toward robustly preserving  quantum information in
physically-realizable environments,
and paved the way for the long-perceived "paradoxes" of quantum mechanics 
to become functional resources for computing, sensing and communications.
In 1997, Preskill~\cite{Preskill1998}  presented the case for reliable quantum computers, 
establishing the accuracy threshold theorem,
corresponding to a lower bound on the error rate per quantum gate below which 
arbitrarily long quantum computations can be performed reliably.
Combined, these 
provided significant practical motivations for developing 
large-scale quantum computation and simulation, and
more generally, quantum technologies.
\\

In the future, 
quantum computers are expected 
to enable efficient and robust
real-time simulations~\cite{Lloyd1996}
of dynamical processes in fundamental physics, 
such as the high-energy collisions of nucleons and nuclei, 
transport in de-confined phases of QCD, 
and of the 
flavor evolution of neutrinos in supernovae, 
with a complete quantification of uncertainties,
tasks that are classically intractable due to exponential complexity growth 
and sign problems 
(for reviews and overviews, see Refs.~\cite{NSACQIS2019,Bauer:2022hpo,Beck:2023xhh,Bauer:2023qgm}). 
After a decade of advancing Noisy Intermediate Scale Quantum (NISQ) 
quantum computers~\cite{Preskill:2018jim},
recent demonstrations of logical qubits with fidelities greater than the physical qubits indicate that quantum computers with tens or hundreds of logical qubits can be expected to become available for
simulating fundamental physics in the near future.
Embedded in large high-performance classical computing environments, 
quantum computers will be ``strongly-coupled'' to classical computing throughout the course of a quantum simulation. 
As such, and given the superior relative performance of classical computing 
for classically efficient aspects of problems, 
understanding which elements of simulations require quantum computation, and which can be accomplished with classical computing in a timely fashion, will play a decisive role in minimizing time to solution, quantum computational resources, and errors due to noise.
This strategy involves quantifying the difficulty faced by classical computers,
identification of simple, classically-tractable approximations, 
and optimal routes to full quantum solutions.

Theoretical and numerical techniques have been developed to characterize the complexity of quantum states and to assess the challenges they pose to classical computation, with significant acceleration in recent years.
These advances have been enabling re-interpretations of traditional
many-body
techniques, 
in terms of information and resource theories, 
and, at the same time, are providing new aspects of classical efficiency and directions for systematic expansions around simple states.
Mean-field or Gaussian states, for example, have been long used in nuclear physics as starting points to include correlations, while the modern tools of tensor networks offer controlled, low-entanglement expansions around tensor-product states.
%
%
Another notable class of states in the Hilbert space, that was only recently introduced in the many body-physics area~\cite{Liu:2020yso}, is that of stabilizer states. As first pointed out by Gottesmann and Knill~\cite{Gottesman:1998hu}, these states are special in that they can be prepared efficiently with classical computing~\cite{Gottesman:1998hu,Aaronson_2004}, while being able to support high (even maximal) large-scale entanglement.
A currently promising line of exploration in this context is the development of techniques built upon large-scale entangled stabilizer states, providing efficiently preparable initial states for quantum computers.

Deviations from these "simple" classical states provide measures of complexity of a quantum state, which, together, evaluate the difficulty projected to be encountered by classical computation and the need for quantum computers.
{\it Non-stabilizerness}, or {\it quantum magic}~\footnote{In this review we will use the terms {\it non-stabilizerness} and {\it (quantum) magic} interchangeably}~\cite{Bravyi:2004isx}, 
for example, is necessary to generate complex, non-regular patterns of entanglement, while entanglement is needed to introduce non-stabilizerness in non-local degrees of freedom. 
Therefore, it is their interplay, captured by {\it non-local magic}~\cite{Cao:2024nrx}, that drives the need for quantum computers at scale.
The rapidly-developing tools to evaluate these (combined) aspects are now being used to 
determine the complexity structure of many-body systems important to nuclear and high-energy physics,
and formulate simulation strategies.
\\

From a computer science perspective, much is known about the complexity classification of general quantum many-body systems. 
For instance, it is known that real-time evolution 
from an initial wavefunction resides in {\bf BQP}, 
the class of bounded-error problems that can be solved efficiently with quantum computers~\cite{Lloyd1996}.
This selects forefront problems in nuclear and high-energy physics that cannot be addressed with classical computing,
as candidates for focus with the next generation of quantum computers. 
Interestingly, 
the computability of scattering amplitudes in 
$\lambda\phi^4$ scalar field theory has been examined in detail~\cite{Jordan2012,Jordan2014,Jordan2014Fermions},
where it was shown to be {\bf BQP}-complete~\footnote{This means that not only can the scattering amplitudes be computed efficiently using quantum computers, but 
that all {\bf BQP} systems can be mapped to the scattering amplitudes in $\lambda\phi^4$ with polynomial-scaling quantum resources.}.

On the other hand, preparing ground states of arbitrary Hamiltonians with two-body and higher interactions has been shown to be {\bf QMA}-complete~\cite{Kitaev2002,Kempe2006,Bravyi2005Commuting,Osborne2012},
{\it a priori} placing them out of reach of efficient preparation, even with quantum computers.
This classification, however, is defined by the system scaling of the most general case, while physically-relevant systems are typically far from generic, as they often exhibit high degrees of symmetries, and are not necessarily asymptotically large 
({\it e.g.}, finite nuclei). 
Moreover, ground states of Hamiltonians found in Nature tend to feature significantly less complexity than typical random quantum states.
In this context, deeper understandings and precise quantifications of the various aspects of quantum complexity in physically-relevant systems are likely to guide and accelerate ground-state searches,
providing 
improved predictive capabilities of classical modeling, 
and expanding the reach of quantum simulations.

Further, establishing sets of states with known quantum complexities, to provide a graded set of ``challenges'' for compute environments, including machine learning and artificial intelligence (ML/AI), 
can provide performance metrics not only for the hardware, but for the application software and simulation workflows. This role is becoming clearer with recent observations and results, see, {\it e.g.}, 
Ref.~\cite{Levine:2019gtf,Passetti:2022ilw,Denis:2023dww,Yang:2024yxu,Jreissaty:2025qip,Paul:2025uew,Lu:2026pgr,Sinibaldi:2025cst,Keeble:2026rto}. The structure of nuclei has been identified as one of the possible sets, where the fidelity in learning wavefunctions with Restricted Boltzmann Machines (RBMs) has been found to be correlated with the magic in the wavefunction~\cite{Keeble:2026rto}.
The efficiency of a combined hybrid classical-quantum computing environments delivering physically relevant results for any given system in the foreseeable future is something that will need to be determined by ``hands-on'' implementations using available hardware. This includes evaluating the effectiveness of fault-tolerant and error-corrected implementations.

At present, the non-local complexity structures of quantum many-body systems relevant to nuclear and particle physics  
remain poorly characterized and understood.
Early steps to elucidate these structures have been taken, including
the discovery of connections between fluctuations in entanglement~\cite{Beane:2018oxh} and magic~\cite{Robin:2024oqc,Robin:2025ymq} in scattering processes
and global symmetries of the theory.
There have been demonstrations
that leveraging~\cite{Papenbrock:2003bj,Papenbrock:2003az,Gorton:2024hbb} or re-arranging~\cite{Legeza:2015fja,Robin:2020aeh,Tichai:2022bxr,Robin:2023pgi,Hengstenberg:2023ryt} the entanglement structure 
can improve the convergence of low-energy effective model spaces
of multi-nucleon systems, and aid in identifying underlying structures ({\it e.g.}, shell structures)~\cite{Robin:2020aeh,Tichai:2022bxr} that enable accelerated simulation.
Such analyses have the potential to reveal new small expansion parameters for perturbative analyses, and are leading
to new algorithms for both classical and quantum simulation.
Recent observations of quantum complexity barriers in the evolution from a 
highly non-equilibrium state to equilibrium in gauge theories~\cite{Ebner:2025pdm} 
are consistent with analogous observations in other science domains~\cite{aaronson2014C}. 
Essentially, a relatively "simple" initial state of a many-body system evolves towards another simple equilibrium state. Along the path, however, the complexity of the system exhibits a transient increase before decreasing again, providing a computational barrier as the internal degrees of freedom re-arrange themselves.
This has implications for the transition from the quark-gluon plasma (a low-viscosity liquid), 
the cross-over transition in finite-temperature QCD (at zero baryon number), and the electroweak phase transition where the matter-antimatter asymmetry could be generated.
This has also been seen across the phase diagram of the Lipkin-Meshkov-Glick (LMG) 
model~\cite{LIPKIN1965188},
a model of fermions (or spins) with permutation-invariant interactions, 
in which simple highly-entangled collective states emerge from a phase of high complexity~\cite{Robin:2025wip}.
Explicit simulations of the evolution of the quark and gluon fields during fragmentation and hadronization 
in high-energy hadron collisions are expected to exhibit a similar complexity barrier, 
as suggested in recent work on complexity in string breaking, {\it e.g.}, Refs.~\cite{Grieninger:2025rdi,Grieninger:2026bdq}. 
It is hoped that improved understandings of these processes could be integrated into models 
used in the discovery of new physics in high-energy colliders, such as the LHC at CERN 
or the future EIC at Brookhaven National Laboratory.
It has also been suggested that analogous understandings of complexity in systems of dense neutrinos, 
created during supernova, may aid in improving simulations of these systems~\cite{Chernyshev:2024kpu}.
With regard to the static structure of systems, recent results suggest that quantum complexity and correlations may provide new windows into nature of nuclei, particularly those close to instability.
Recent studies in the nuclear shell model indicate that while classical measures of nuclear deformation become small and vanish toward the region of shape co-existence, quantum measures of complexity remain, 
indicating non-trivial structure in the nuclear wavefunction that is not reflected in classical expectation values~\cite{Brokemeier:2024lhq}.  
Such effects are expected to be magnified in the reactions of nuclei.
Finally, the Higgs sector of the Standard Model, and specifically the symmetry structures of minimally-extended Higgs sectors,  
are potentially constrained by requirements of minimizing or maximizing fluctuations in entanglement in scattering processes~\cite{Carena:2023vjc,Carena:2025wyh}, for example, in 2-Higgs doublet models~\cite{PhysRevD.15.1958,PhysRevD.15.1966}.
Interestingly, maximizing the entanglement power is found to lead to enhanced symmetries, in a way that is analogous to the enhanced structures required in error-correction codes~\cite{Carena:2025wyh}.

On the formal side, 
there are deep connections between quantum complexity and gravity that are important to have in mind (for a recent review, see Ref.~\cite{Baiguera:2025dkc}).
The relationship between quantum gravity and quantum complexity, 
particularly within the AdS/CFT correspondence~\cite{Maldacena:1997re}, 
suggests that the geometric evolution of spacetime is a physical manifestation of computational growth,
see {\it e.g.}, Refs.~\cite{Susskind:2014rva,Stanford:2014jda,Brown:2015bva,Brown:2015lvg,Chapman:2021jcl}. In this framework, 
the {\it Complexity Equals Volume} 
and {\it Complexity Equals Action} 
conjectures propose that the increasing complexity of a quantum state on the boundary is dual to the gravitational action of the bulk, 
{\it e.g.}, Refs~\cite{Ryu:2006bv,Carmi:2016wjl,Bouland:2019pzu}. 
Conversely, 
quantum complexity provides an explanation of why black hole interiors appear to evolve 
after they have reached thermal equilibrium, 
connecting the expansion of space with Hilbert space exploration, {\it e.g.}, Ref.~\cite{Cao:2016fxf}.
This suggests that spacetime  may not be fundamental, 
but an emergent property due to the complexity of entanglement 
and information processing.
Tools developed in exploring the connections between gravity and complexity are of utility in a broader context, including in exploring the complexity in fundamental physics, see {\it e.g.}, Ref.~\cite{Cao:2024nrx}.
\\

This is a remarkably exciting period for fundamental physics, 
fueled by a rapidly improving  understanding of entanglement and complexity in quantum many-body systems
in tandem with analogous advances in our ability to control non-local aspects of such systems in the laboratory.
In this review, we cover research in key areas of nuclear physics and high-energy physics phenomenology, 
where advances 
in QIS and quantum computing are anticipated to provide
new theoretical insights, research directions and simulation capabilities, 
to further our ability to understand and reliably predict the behavior of matter in the diverse environments found in our universe.

%% file: Section_Complexity.tex
\section{Many-Body Quantum Complexity}
\label{sec:QC}

Simulations of many-body systems and field theories involve the development of methods and algorithms that aim at predicting the properties and behaviors of these systems as precisely and accurately as possible.
While the concept of "complexity" plays a central role in performing this task, the term has often been employed in a subjective manner in the context of many-body physics, lacking clear and precise definitions. It is indeed commonly understood that the more complex a state or process is, the more challenging it is to describe or simulate.
Yet, the precise meaning of this, {\it i.e.} what really distinguishes simplicity from complexity in this context, is often unclear.
(Quantum) information theory~\cite{cover2006elements,nielsen2010quantum}, on the other hand, provides a framework for defining and quantifying aspects of complexity in more rigorous ways, 
that are in one-to-one correspondence with given resources, and thus directly connect with computational resource requirements.

\subsection{Quantum State Complexity}
\label{subsec:Q_State_Complex}

The complexity of a given physical state, for example, can be quantified by the amount of information that is needed to represent it, that is, how many bits. 
If a state can be described by a number of bits
that scales polynomially with the number of degrees of freedom $N$ and the desired accuracy~\footnote{Specifically, the number of bits is required to scale polynomially in $\log(1/\epsilon)$, where $\epsilon$ denotes the desired approximation error.}, in other words if that state admits an {\it efficient} classical representation, then one can say that the state is "simple". 
On the other hand, if the state requires an exponentially-scaling number of bits, then it can be said to be "complex".

The state of a classical system, if known, can always be specified using an amount of information that scales (at most) linearly with $N$. This is because classical systems can only exist in one single configuration at a time.
Quantum systems, on the other hand, can exist in coherent superpositions of exponentially many classical configurations, and thus a full description of an arbitrary quantum state, will, in general, require an amount of classical information that increases exponentially with $N$. 
This {\it a priori} exponential complexity arises from intrinsically quantum features such as superposition and entanglement, which allow information to be distributed non-locally across many degrees of freedom.
Ground or low-lying states of physically-relevant many-body quantum systems, however, often possess special properties, such as symmetries and particular underlying structures, which set them apart from random quantum states, and, in some cases, allows for compression and efficient descriptions.\\

Ground states of low-dimensional gapped and (geometrically) local Hamiltonians, for example, are known to exhibit entanglement properties that scale with the size of the boundary of the sub-region of interest, as opposed to the volume~\cite{Hastings:2007iok,RevModPhys.82.277}. Such entanglement area laws
enable efficient representations of these states with tensor network (TN) techniques~\cite{Orus:2013kga,RevModPhys.93.045003,Banuls:2022vxp}. 
In these frameworks, quantum states are represented by networks of connected tensors where the connections represent the amount of entanglement between degrees of freedom. In systems where the correlation length decreases exponentially, these methods allow for powerful entanglement-based truncations, to suppress unimportant information and compress the state while maintaining a good description of entanglement properties.

A well known example is the density matrix renormalization group (DMRG), 
which was proposed by White in the early 1990's~\cite{PhysRevLett.69.2863} and was later re-interpreted 
as one-dimensional (1D) TN, or matrix product state (MPS)~\cite{McCulloch_2007,Schollwock_2011}. 
In principle, any arbitrary quantum state $\ket{\Psi}$ can be written as an MPS by introducing auxiliary degrees of freedom via successive singular value decompositions (SVDs), or Schmidt decompositions,~\cite{Perez-Garcia:2006nqo} as
\begin{align}
&\ket{\Psi} \ = \ \sum_{i_1, i_2, ..., i_N}\  C_{i_1, i_2, ..., i_N} \ket{i_1 i_2 ... i_N}  
\nonumber \\
& \xrightarrow[\text{SVDs}]{} \sum_{i_1, i_2, ..., i_N}
Tr(A^{[1] i_1}A^{[2] i_2} ... A^{[N] i_N}) 
\ \ket{i_1 i_2 ... i_N}  \; ,
\end{align}
with
\begin{align}
& Tr(A^{[1] i_1}A^{[2] i_2} ... A^{[N] i_N}) \nonumber \\
& = \sum_{a_1,...,a_{N-1}} A^{[1]i_1}_{a_1} A^{[2]i_2}_{a_1 , a_2}... A^{[N-1] i_{N-1}}_{a_{N-2} , a_{N-1}} A^{[N] i_N}_{a_{N-1}} \; ,
\end{align}
where the indices $i_k$ (resp. $a_k$) denote a basis for the physical (resp. auxiliary) degrees of freedom.
The dimensions of the tensor $A^{[k] i_{k}}_{a_{k-1} , a_{k}}$  set the amount of the entanglement between neighboring sites $k-1$ and $k$, and are known as {\it bond dimensions}.
In practice these dimensions are kept small by imposing a truncation on the spectrum of singular values and keeping only the largest ones. 
It was shown that the error incurred in the fidelity of the quantum state is of the order of the sum of the discarded squared singular values~\cite{PhysRevB.73.094423}. 
When the spectrum decays rapidly, which is the case in 1D systems with gapped local Hamiltonians, 
the loss of information decreases at the same rate.

Extensions of MPS to more expressive ansatz and/or higher-dimensional systems include tree-tensor networks~\cite{PhysRevA.74.022320}, projected entangled pairs states~\cite{Verstraete:2004cf}, or MERA~\cite{Vidal:2008zz}, as well as other variants. 
For more details on TNs we refer the reader to recent reviews, 
{\it e.g.}, Refs.~\cite{RevModPhys.93.045003,Banuls:2022vxp}.
\\

While TNs are based on the fact that low-entangled states can be efficiently representable,
highly-entangled states are, in fact, not necessarily hard to capture.
This finding
goes back to Gottesman and Knill, who, in the late 1990s, showed that a certain class of "stabilizer states", 
which can be strongly entangled in a collective, multipartite and long-range manner, can nevertheless be easy to describe~\cite{Gottesman:1998hu}. 

Mathematically, the stabilizer formalism is centered around the generalized Pauli group, which, for an $N$-qubit system, is defined as
\begin{equation}
\mathcal{G}_N = \{ \varphi \, \hat{\sigma}^{(1)} \otimes \hat{\sigma}^{(2)} \otimes ...\otimes  \hat{\sigma}^{(N)} \} \; ,
\label{eq:Pauli_group}
\end{equation}
where $\hat{\sigma}^{(k)} \in \{ \mathds{1}, \hat{\sigma}^{(k)}_x, \hat{\sigma}^{(k)}_y, \hat{\sigma}^{(k)}_z \}$ denotes a Pauli operator (or identity) acting on qubit $k$ and $\varphi  \in \{ \pm 1, \pm i \}$ is a phase.
The stabilizer group of a quantum state $\ket{\Psi}$, denoted as $\mathcal{S} (\ket{\Psi})$, is the sub-group of $\mathcal{G}_N$ which stabilizes $\ket{\Psi}$, i.e.,
\begin{equation}
\hat{P} \ket{\Psi} = \ket{\Psi} \; , \forall \hat{P} \in  \mathcal{S} (\ket{\Psi}) \; ,
\end{equation}
and is Abelian.
Stabilizer states have stabilizer groups with exactly $|\mathcal{S} (\ket{\Psi})| = 2^N$ elements. 
They are thus completely and uniquely specified by their stabilizer group, which, in turn, is specified by $N$ linearly independent generators. 
This property underlies the low-cost representation of stabilizer states.
Since each stabilizer generator can be specified by $2N +1$ bits,
any stabilizer state can be specified with $N(2N+1) = O(N^2)$ bits.

Moreover, as Clifford operations preserve the generalized Pauli group, 
they act as a basis transformation between stabilizer groups, and thus between corresponding stabilizer states.
Any stabilizer state can be created from the trivial state 
$\ket{0}^{\otimes N} = \ket{0} \otimes \ket{0} ... \otimes \ket{0}$ with Clifford operators.
The two above results form the core of the Gottesman-Knill theorem~\cite{Gottesman:1998hu,Aaronson:2004xuh}.

From a many-body point of view, stabilizer states exhibit regular, highly-structured entanglement patterns, which makes them closely related to graph states.
Such graph states are known to encompass forms of multipartite entanglement~\cite{PhysRevA.69.062311}, and possess intuitive representations as mathematical graphs, directly representing the physical degrees of freedom of the physical system and their entanglement structure~\cite{Hein2006GraphStates}

\begin{figure}[ht]
\centering
\includegraphics[width=\columnwidth]{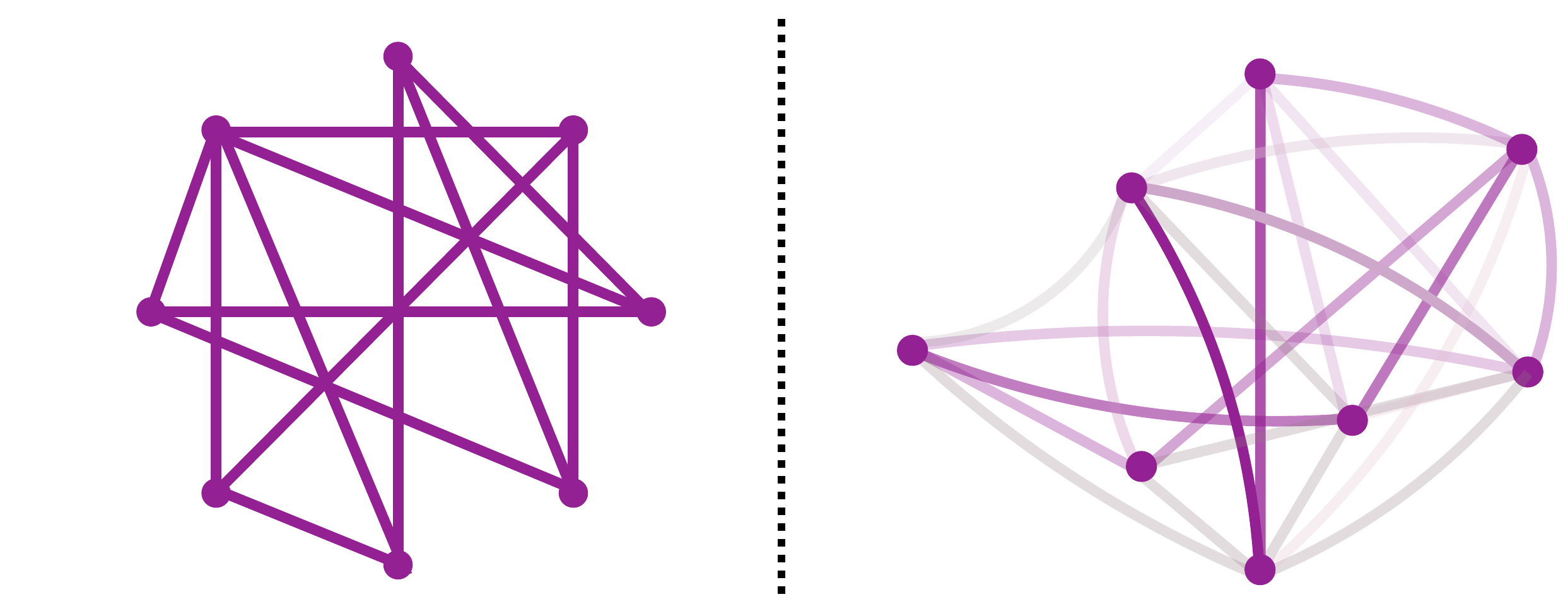}
\caption{Left: 8-qubit graph (LC equivalent to a stabilizer) state with volume-law entanglement entropy.  The vertices represent the qubits and edges represent entanglement between them created by the CZ gates in Eq.~\eqref{eq:graph0}. Right: Arbitrary 8-qubit state (schematic) with entanglement and magic. The magic creates complex (non-regular) patterns of entanglement. }
\label{fig:graph-arb-state}
\end{figure}

A $N$-qubit graph state corresponding to a given graph with set of edges $E$ and set of vertices $V=\{ i \}$, can be written as
\begin{equation}
    \ket{G} = \prod_{(i,j) \in E} \mathrm{CZ}_{ij} |+\rangle^{\otimes N} \; ,
\label{eq:graph0}
\end{equation}
where $(i,j)$ label qubits connected by an edge, CZ is the controlled-Z gate, and $\ket{+} = (\ket{0} + \ket{1})/\sqrt{2}$ is a single-qubit superposed state, eigenvector of the $\hat{\sigma}_x \equiv \hat X$ Pauli operator. 
Any stabilizer state is then equivalent to a graph state, up to local Clifford operations~\cite{Schlingemann:2003ika,Grassl,Hein2006GraphStates} (LC equivalence).
This is illustrated on the left panel of Fig.~\ref{fig:graph-arb-state}.

\begin{figure}[ht]
\centering
\includegraphics[width=0.5\columnwidth]{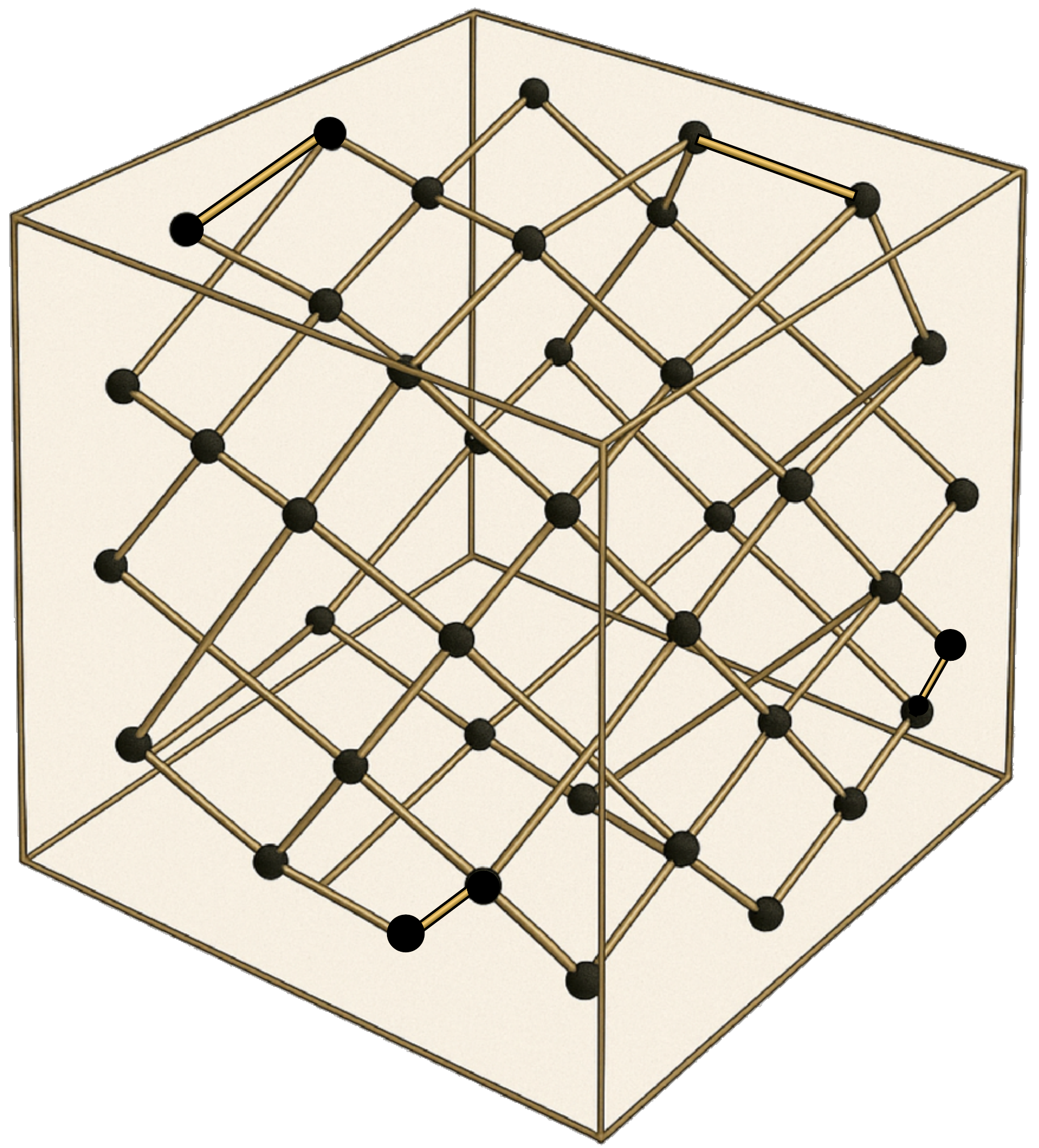}
\caption{Artist view of the lattice of stabilizer states mapping a finite many-body Hilbert space.}
\label{fig:stab-lattice}
\end{figure}

A single-qubit system has six stabilizer states corresponding to the six eigenstates of the Pauli operators. For two-qubit systems there are 60 stabilizer states comprising 36 tensor products of one-qubit stabilizer states, and 24 entangled stabilizers including the maximally entangled Bell states. They are listed in appendix~\ref{app:stabs}.
More generally, the number of stabilizer states grows doubly-exponentially as $2^{(1/2 + o(1))N^2}$~\cite{PhysRevA.70.052328}.
Stabilizer states can thus be seen as providing some sort of non-orthogonal discrete lattice, or overcomplete basis of the many-body Hilbert space, as illustrated schematically in Fig.~\ref{fig:stab-lattice}. 
A useful property for qubit systems is that the set of stabilizer states forms a complex projective 3-design, 
{\it i.e.}, averaging over them reproduces up to the 
third
moment of the Haar-random state distribution~\cite{Dankert:2009yux,Kueng:2015mts,Zhu:2017psv}. 
For qudit systems with dimension equal to the power of a prime number, stabilizer states form a 2-design, reproducing Haar averages up to 
second
order only~\cite{Kueng:2015mts}.

\begin{figure}[ht]
\centering
\includegraphics[width=.85\columnwidth]{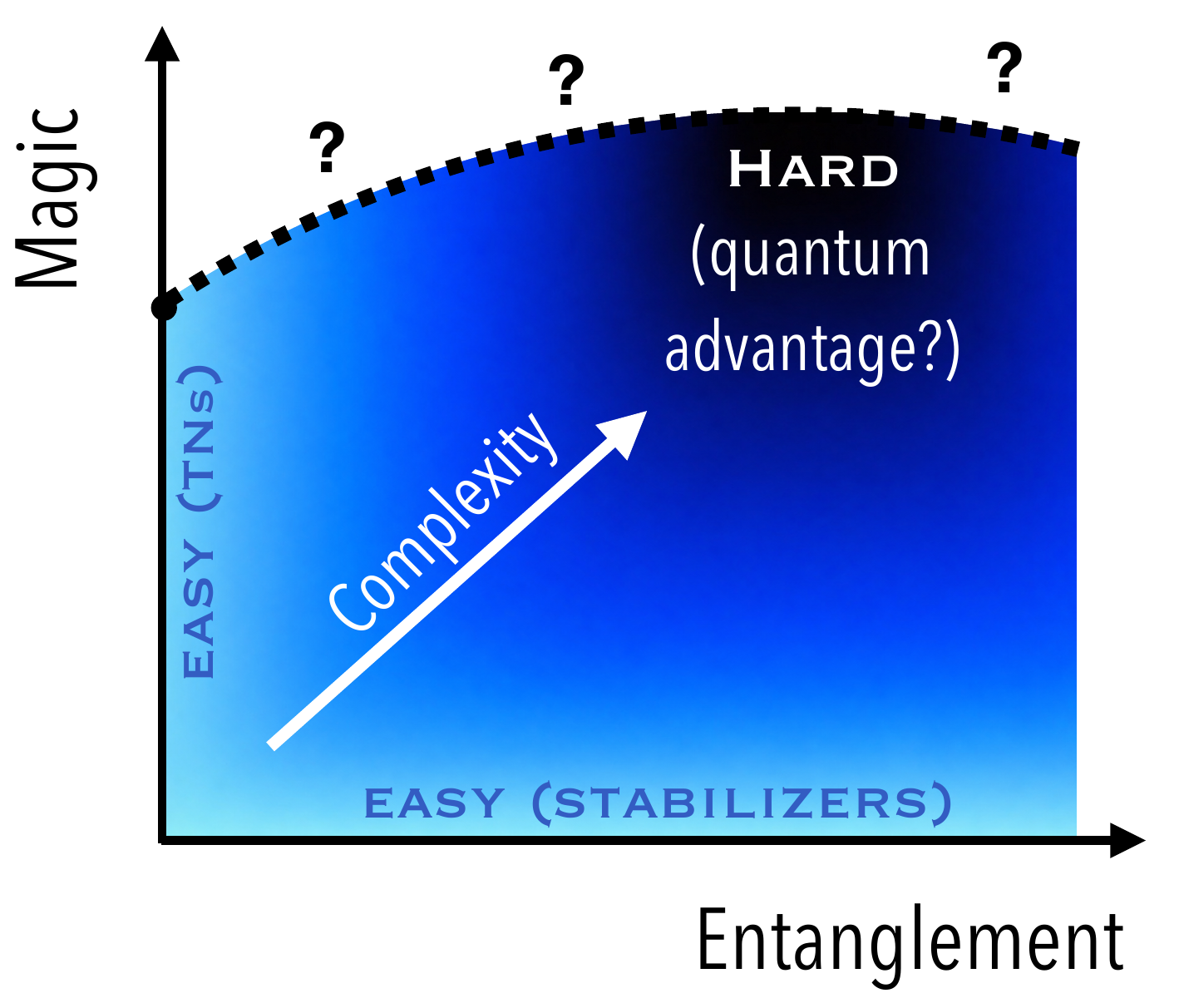}
\caption{Entanglement-magic diagram (schematic) and hardness of classical simulations. 
The light blue regions denote regions of the Hilbert space that are easy to represent, while the darker regions represent states with higher complexity. 
States with low (area law) entanglement (near the $y$ axis) may be described efficiently with tensor networks (TNs) while states with low magic and arbitrary entanglement entropy (near the $x$ axis) may be represented with (low-rank superposition of) stabilizer states.
The dashed line shows that the magic profile depends on the entanglement content. 
For example, states with low-entanglement cannot encompass maximal magic. 
The shape shown here is based on calculations involving few-qubit random states 
(see appendix~\ref{sec:TwoQudits} and Ref.\cite{Iannotti:2025lkb}), but the exact form of that boundary is unknown for general systems. 
Inspired by a figure from Ref.~\cite{Hamma_IQuS_YouTube}.}
\label{fig:magic-entang-diagram}
\end{figure}

In order to attain non-stabilizer states, and fill the gaps in the Hilbert space, it is necessary to introduce some amount of {\it non-stabilizerness}.
This is the extra ingredient, also known as {\it magic}~\cite{Bravyi:2004isx}, which is generated by non-Clifford operators, and is required to induce complex, non-regular entanglement patterns, see right panel of Fig.~\ref{fig:graph-arb-state}.
In general, a state that exhibits both high entanglement and high non-stabilizerness will require exponentially-scaling amounts of classical information to be specified. This is depicted on the "complexity phase diagram" in Fig.~\ref{fig:magic-entang-diagram}.
\\

Another class of easily simulable states, with a long history in many-body physics, is that of Gaussian states. 
These correspond to ground and thermal states of quadratic (free) Hamiltonians, and are fully characterized by their first and second moments (one- and two-point correlation functions).
Consequently, Gaussian states admit an efficient classical description with $O(N^2)$ bits, where $N$ is the number of bosonic or fermionic modes.

In the bosonic case, examples of Gaussian states include coherent and squeezed states, as well as any state obtained from the vacuum by application of a Gaussian unitary, {\it i.e.} a unitary that is generated from an operator which is quadratic in the canonical variables ($\hat x, \, \hat p$) or harmonic oscillator ladder operators ($\hat b, \, \hat{b}^\dagger$)~\cite{Weedbrook:2011wxo}.

Fermionic Gaussian states (FGS) are naturally described using fermionic (Dirac) creation and annihilation operators $a^\dagger_k, a_k$ , or equivalently, their Hermitian combinations, Majorana operators 
\begin{equation}
    \gamma_{2k-1} = a_k + a^\dagger_k \; , \ \ \gamma_{2k} = i (a_k - a^\dagger_k) \; .
\end{equation}
Pure FGS $\ket{\Psi_{FGS}}$ can be obtained from the vacuum $\ket{0}$ by acting with fermionic Gaussian unitaries  (FGU)
\begin{eqnarray}
    U_{FG} = \exp \left( \frac{1}{4} \sum_{k,l} U_{kl} \gamma_k \gamma_l \right) \; ,
\end{eqnarray}
{\it i.e.}, operators that are quadratic in Dirac or Majorana operators.
Consequently they are free fermionic states and thus can be fully characterized by their one-body correlation functions, 
\begin{align}
    \mathcal{C}_{kl} &= \langle \Psi_{FGS} | c^\dagger_k c_l| \Psi_{FGS} \rangle \; , \\
    \kappa_{kl} &= \langle \Psi_{FGS} | c_k c_l| \Psi_{FGS} \rangle \; ,
\end{align}
compactly encoded into the $2N \times 2N$ covariance matrix
\begin{equation}
    \Gamma_{kl} = - \frac{i}{2} \langle \Psi_{FGS} | [ \gamma_k, \gamma_l ]| \Psi_{FGS} \rangle \; .
\label{eq:covariance_matrix_fermion}
\end{equation}
Wick's theorem~\cite{fetter2003quantum} then allows for the reduction of 
any $M$-body correlation function as products of $\mathcal{C}$'s, $\kappa$'s, and $\kappa^\dagger$'s.
In the context of nuclear, and more generally many-body physics, $\mathcal{C}$ and $\kappa$ are known as normal and pairing densities, and are encountered, 
{\it e.g.}, in Bardeen-Cooper-Schrieffer (BCS) or Hartree-Fock-Bogoliubov (HFB) formalisms of superconductivity or superfluid pairing~\cite{ring2004nuclear}. 

Gaussian states can exhibit extensive entanglement,
however since all correlations are fully determined by second moments or covariance matrices, this entanglement is structurally constrained. 
For example, Ref.~\cite{Bianchi:2021lnp} showed that random FGS do not saturate the Page curve, while Ref.~\cite{Froland:2024wxj} showed that Gaussian states are not able to capture entanglement signatures of thermalization. Consequently, some amount of {\it non-Gaussianity} is needed to explore the full Hilbert space.

\subsection{Quantum Computation and Resource Theories}

The above notions of many-body complexity -- entanglement, non-stabilizerness, and non-Gaussianity -- can be understood as genuine quantum information resources, and are therefore naturally described within the framework of quantum resource theories~\cite{Chitambar:2018rnj}, as used, {\it e.g.}, in quantum computing. In such frameworks, one defines a set of “free” operations that are efficiently simulable classically, and identifies non-free states whose resource content enables universality and quantum advantage.
\\

The resource theory based on Clifford+T gates, for example, is one of the most widely examined. 
In this setting, the Clifford operations, which can generate highly-entangled stabilizer states, are the free operations~\cite{Aaronson:2004xuh}.
Precisely, the Clifford group is generated by a set of three gates: the two single-qubit gates  H, S and the two-qubit CNOT gate, which are responsible for creating superposition, complex relative phases, and entanglement, respectively~\cite{Gottesman:1998hu}. 
Stabilizer computations based on these gates are also well suited for fault-tolerant implementations, as H, S  and CNOT gates can be implemented efficiently in many stabilizer quantum error-correcting codes such as the surface code~\cite{Kitaev:1997wr,Fowler:2012hwn}, or 2D color codes~\cite{Landahl:2011den,Bombin:2015tpp}.
In contrast, non-stabilizerness, which is necessary for achieving universality,
arises from the inclusion of non-Clifford operations, such as the single-qubit T gate (or equivalently, magic state injection)~\cite{Bravyi:2004isx}.
These operations render classical simulation inefficient and require more costly fault-tolerant implementations, typically relying on techniques such as magic state distillation~\footnote{In 3D color codes, 
the T-gate or CZZ-gate can be transversal while the H-gate is not~\cite{bombin2015a,PhysRevA.91.032330,Vasmer_2019}.}.
\\

Gaussian states are naturally described within the framework of a resource theory in which Gaussian operations are taken as free, and non-Gaussianity constitutes the resource. In this setting, Gaussian states and operations can be efficiently simulated classically, while non-Gaussian elements are required for achieving quantum advantage and universality~\cite{PhysRevA.97.052317}.
Such structures arise across a variety of physical platforms. In bosonic systems, Gaussian operations are implemented in continuous-variable architectures, including quantum optical systems, microwave superconducting cavities, and optomechanical platforms~\cite{Weedbrook:2011wxo,Blais2021,Aspelmeyer2014}.
In fermionic systems, the analogous class of efficiently simulable operations is given by (nearest-neighbor) matchgate circuits, which correspond to fermionic Gaussian dynamics under Jordan-Wigner mappings~\cite{Terhal:2001aml}. These arise, for example, in systems of non-interacting fermions, including certain regimes of cold atoms in optical lattices and topological platforms based on Majorana modes~\cite{Bloch2008,Nayak2008,Alicea2012}. In both cases, the inclusion of non-Gaussian operations is necessary to go beyond classically simulable regimes.
In the fermionic case, this can be done by extending the range of matchgates with SWAP gates.
\\

Entanglement can similarly be viewed as the relevant resource in measurement-based quantum computation (MBQC, or one-way quantum computing)~\cite{PhysRevLett.86.5188,Raussendorf:2003rug,Raussendorf:2012tof}. In this setting, the free operations consist of adaptive local measurements and classical communication, while the resource is a highly entangled universal graph state, such as the two-dimensional cluster state.
In this sense, operations restricted to local measurements and classical communication alone remain classically simulable, while entanglement enables universality.
\\

\color{black}
\subsection{Measures of Complexity}
\label{sec:measures_complexity}

A wide range of measures have been developed to characterize and quantify various aspects of complexity, from the early notions of bipartite entanglement, to concepts of multipartite entanglement, non-stabilizerness, non-Gaussianity, and interplays between these aspects.
By probing the complexity structures of quantum states and dynamical processes, these measures serve as diagnostics for the hardness of classical simulations, and relate, to varying degrees, to quantum resource costs.
Below, we summarize a selection of such measures, focusing on those most relevant to the subsequent reviewed studies in 
nuclear and high-energy physics, with an emphasis on modern notions of complexity beyond bipartite entanglement.

\subsubsection{Entanglement}

Entanglement, or the fact that information can be distributed among several subsystems of a larger one, can be quantified by partitioning a quantum state in various ways and evaluating how much information about the whole is inaccessible by examining individual parts.

In the case of a bipartite system in a pure state 
\begin{equation}
    \ket{\Psi_{AB}} = \sum_{i_A=1}^{d_A} \sum_{\mu_B=1}^{d_B} C_{i_A, \mu_B}  \ket{i}_A \otimes \ket{\mu}_B \; ,
\label{eq:bipartite_pure_state}
\end{equation}
the von Neumann entanglement entropy quantifies this by measuring the loss of information when tracing out one of the two subsystems.
For example, the von Neumann entropy associated with the reduced density matrix $\rho_A = \mbox{Tr}_B (\ket{\Psi_{AB}} \bra{\Psi_{AB}})$ of subsystem $A$ is given by
\begin{equation}
S(\rho_A) = - \mbox{Tr} (\rho_A \log \rho_A ) \; ,
\end{equation}
and $S(\rho_A) = S(\rho_B)$.
The entropy $S(\rho_A)$ satisfies various fundamental properties including non-negativity, invariance under local unitaries, and monoticity on average under local operations and classical communications (LOCC)~\cite{nielsen2010quantum}.
It vanishes when $\rho_A$ is pure, in which case A and B are unentangled and the bipartite states can be written as a product state $\ket{\Psi_{AB}} = \ket{\phi_A} \otimes \ket{\chi_B}$. 
On the other hand, $S(\rho_A)$ reaches its maximum value
$S(\rho_A)_{\mathrm{max}} = \log (\mathrm{min} (d_A, d_B))$
when $\rho_A$ is maximally mixed, in which case $\ket{\Psi_{AB}}$ is said to be maximally entangled.

Other measures of pure state bipartite entanglement include the R\'enyi entropies 
\begin{equation}
    S_\alpha(\rho_A) = \frac{1}{1-\alpha} \log \mathrm{tr} ( \rho_A^\alpha ) \; .
\end{equation}
This definition approaches the von Neumann entropy $S(\rho_A)$ for $\alpha \rightarrow 1$. The 2-R\'enyi entanglement entropy has the advantage of being practically computable, for example in tensor network approaches or Monte-Carlo methods~\cite{Hastings:2010zka}, and is also directly measurable in experiments~\cite{Islam:2015mom}.

The entanglement spectrum, given by the eigenvalue spectrum $\{ \lambda_k \}$ of $\rho_A$, is often used to obtain more detailed information about the bipartite entanglement structure of $\ket{\Psi}_{AB}$.
Specifically, the spectrum $\{ \lambda_k \}$ is directly related to the Schmidt decomposition of $\ket{\Psi}_{AB}$, 
or singular-value decomposition (SVD)
of $C_{i_A,\mu_B}$, which provides a compressed representation of that state, as
\begin{align}
    \ket{\Psi}_{AB} 
    = \sum_{k=1}^{r} \sqrt{\lambda_k} \, \ket{u_k}_A \otimes \ket{v_k}_B \; ,
\label{eq:Schmidt_decompo}
\end{align}
where $r$ is the Schmidt rank, equal to the number of non-zero eigenvalues $\lambda_k$.
When the number of significant eigenvalues is small, the state is highly compressible and can be approximated to high fidelity by truncating the expansion in Eq.~\eqref{eq:Schmidt_decompo}. This kind of entanglement-based truncation is at the heart of tensor network methods~\cite{Banuls:2022vxp}, as discussed above in section~\ref{subsec:Q_State_Complex}.
\\

Quantifying bipartite entanglement in mixed states is significantly more challenging, due to more complex criteria for separability. Despite this, computable measures have been advanced such as the concurrence and entanglement of formation for two-qubit states~\cite{Wootters:1997id}. Further, the (logarithmic) negativity~\cite{Vidal:2002zz}, based on the partial transpose of the reduced density matrix, quantify the amount of (distillable) entanglement between two sub-parts AB of a larger closed system ABC. Precisely, the negativity is defined as
\begin{equation}
\mathcal{N}(\rho_{AB}) = \frac{|| \rho_{AB}^{T_A} -1 ||_1}{2} \; ,
\end{equation}
where $\rho_{AB}^{T_A}$ is the partial transpose of $\rho_{AB}$. The negativity $\mathcal{N}(\rho_{AB})$ coincides with the absolute value of the sum of negative eigenvalues of $\rho_{AB}^{T_A}$, and vanishes when A and B are unentangled. 
The logarithmic negativity $E_\mathcal{N}(\rho_{AB}) = \log (|| \rho_{AB}^{T_A} ||_1)$ provides a bound on distillable entanglement~\cite{Vidal:2002zz}.

The mutual information defined as
\begin{equation}
I(A:B) = S(\rho_A) + S(\rho_B) - S(\rho_{AB}) \; ,
\label{eq:MI}
\end{equation}
is also commonly used as it is often straightforward to compute, although this measure captures both classical and quantum correlations.

Interactions in many-body systems often give rise to rich and intricate multi-particle correlations. Understanding the nature of multipartite entanglement is therefore crucial for establishing connections to physical phenomena, such as the emergence of collectivity and new degrees of freedom. 
Defining and quantifying multipartite entanglement, however, constitute significantly greater tasks than for the bipartite case. Because of the many and inequivalent types of multipartite entanglement, and the difficulty of capturing genuine multi-body entanglement, there is no unique measure that universally captures all relevant correlations.

The $n$-tangles~\cite{PhysRevA.63.044301}, for example, can provide a measure of $n$-partite entanglement in a $N$-qubit state $| \Psi \rangle$ ($N\geq n$):
\begin{equation}
\tau_{(n)}^{(i_1...i_n)} = \left| \langle \Psi | \sigma_y^{(i_1)} \otimes ... \otimes \sigma_y^{(i_n)}     |\Psi^* \rangle \right| \; ,
\label{eq:ntangles}
\end{equation}
where $\sigma_y^{(i_k)} $ is the $y$ Pauli operator acting on qubit $i_k$.
The $2$-tangle coincides with the squared concurrence, while the $3$-tangle provides a measure of genuine 3-qubit entanglement~\cite{PhysRevLett.78.5022,Wootters:1997id,PhysRevA.61.052306,Osborne:2002vcf}.
For $n\geq 4$, however,  the $n$-tangles may include lower-rank contributions, and thus provide a measure of non-irreducible $n$-qubit entanglement. 
Moreover $n$-tangles for odd values of $n >3$ are more difficult to define~\cite{PhysRevA.63.044301,Li:2011ayg}.

The $n$-tangles may be complemented by, for example, the Quantum Fisher Information, which is known to be a witness for genuine $n$-qubit entanglement~\cite{Hyllus:2012ufd,Toth:2012lpv,Strobel:2014npf}.
\\

For a comprehensive review of entanglement measures in pure and mixed states, we refer the reader to {\it e.g.}, Ref.~\cite{Horodecki:2009zz}, 
or more recently Ref.~\cite{Ma:2023ecg} on multipartite entanglement.

\subsubsection{Non-Stabilizerness (Quantum Magic)}

Similarly to entanglement, various measures to quantify the magic in a quantum state have been developed. 
For example, the minimum relative entropy of magic~\cite{Emerson:2013zse}, or robustness of magic~\cite{PhysRevLett.115.070501,PhysRevLett.118.090501,Heinrich_2019} characterize the minimum distance between a quantum state and the nearest stabilizer state.
Other measures, such as the mana~\cite{Emerson:2013zse} and thauma~\cite{PhysRevLett.124.090505}, 
stabilizer extent~\cite{Bravyi2019simulationofquantum}, stabilizer norm~\cite{PhysRevA.83.032317} and stabilizer nullity~\cite{Beverland:2019jej}, are closely related to the minimum number of stabilizer states required to expand the quantum state of interest, or stabilizer rank. 
Recently, more easily computable measures of magic which do not require explicit minimization procedures, have been introduced, such as the stabilizer R\'enyi entropies (SREs)~\cite{Leone:2021rzd} and the Bell magic~\cite{PRXQuantum.4.010301}, which have been shown to be measurable in quantum computing experiments~\cite{Oliviero_2022,PRXQuantum.4.010301,Bluvstein:2023zmt}, and efficiently calculable for MPS~\cite{Haug:2022vpg,Haug:2023hcs,Tarabunga:2024ugl,Lami:2023naw}.
\\

Since their introduction a few years ago, SREs, in particular, have become a central tool for characterizing non-stabilizerness in various quantum systems, including those relevant to nuclear and high-energy physics, as discussed throughout this review. For that reason, we will primarily focus on those, while briefly exploring other measures.

\begin{figure}[ht!]
\centering
\includegraphics[width=\columnwidth]{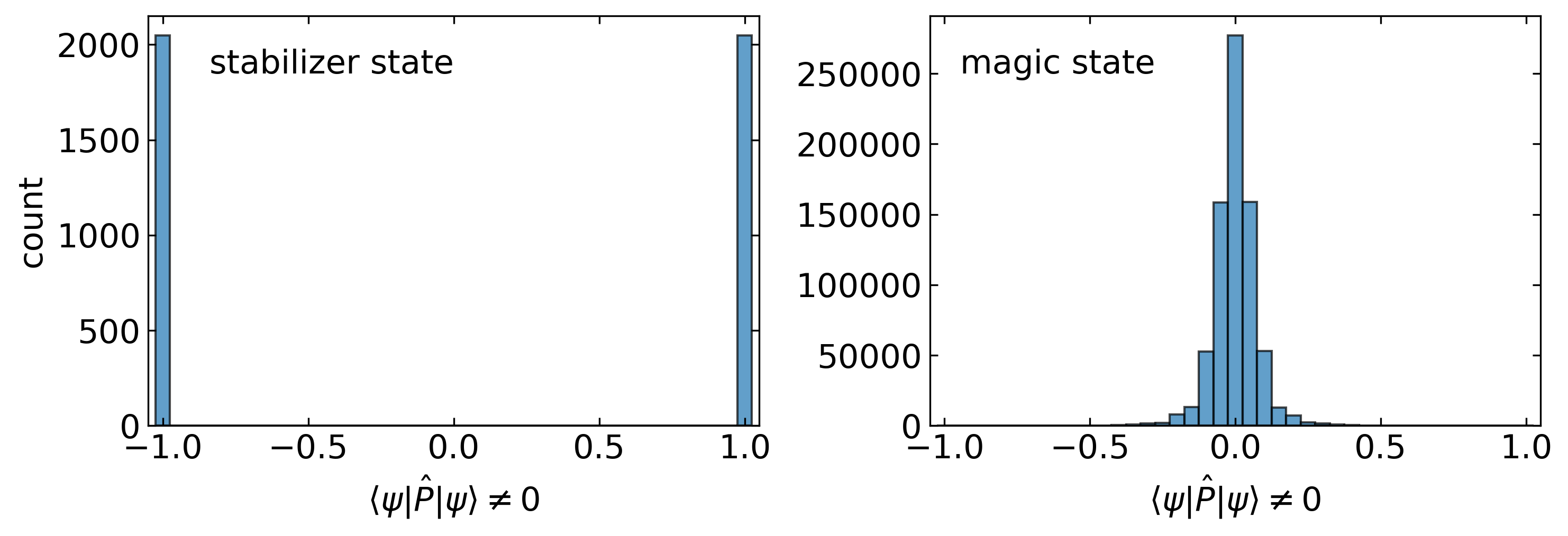}
\caption{Pauli distribution of a 12-qubit stabilizer state (left) 
and an arbitrarily selected state with non-zero magic
(right). Only non-zero Pauli string expectation values are shown. }
\label{fig:stab_state_dist}
\end{figure}

The derivation of SREs for a $N$-qubit pure state $\ket{\Psi}$ is based on
a general expansion of the corresponding density matrix into tensor products of Pauli operators, or Pauli strings, $\hat P$
\begin{equation}
    \hat{\rho} = \ket{\Psi} \bra{\Psi} = 
    \frac{1}{d} \sum_{\hat P \in \widetilde{\mathcal{G}}_{N}} \langle \Psi |\hat{P} | \Psi \rangle \, \hat{P} 
    \; ,
\end{equation}
where $d=2^N$, 
and $\widetilde{\mathcal{G}}_{N}$ is the subgroup of the Pauli group $\mathcal{G}_{N}$ defined in Eq.~(\ref{eq:Pauli_group}) with phases $\varphi = +1$.
A central element of the derivation is that the quantity 
\begin{equation}
    \Xi_P \equiv   \frac{\langle \Psi |\hat{P} | \Psi \rangle^2 }{d} \; ,
\end{equation}
is a probability distribution~\cite{Leone:2021rzd}, and that $\ket{\Psi}$ is a stabilizer state if and only if~\cite{zhu2016clifford}
\begin{equation}
\Xi_P  =
\begin{cases}
1/d & \text{for $d$ commuting Pauli strings } \hat P \in \widetilde{\mathcal{G}}_{N} \; , \\
0 & \text{for the remaining } d^2-d \text{ strings} \; .
\end{cases}
\end{equation}
Consequently, 
the stabilizer $\alpha$-R\'enyi entropies~\cite{Leone:2021rzd},
\begin{equation} 
\mathcal{M}_{\alpha}(\ket{\Psi})= -\log d + \frac{1}{1-\alpha} \log 
\left( \sum_{\hat{P} \in \widetilde{\mathcal{G}}_{N}} \Xi_P^{\alpha} \right) \; ,
\label{eq:SRE_def}
\end{equation}
provide a measure of non-stabilizerness (magic) in $\ket{\Psi}$. 
The constant offset, $- \log d$, is introduced to ensure that the SREs vanish for stabilizer states. 
The SREs thus capture the deviation between the Pauli distribution of $\ket{\Psi}$ and the characteristic distribution of stabilizer states, which is uniform on their stabilizer group, see Fig.~\ref{fig:stab_state_dist}.

Different values of $\alpha$ probe different aspects of magic.
Specifically, SREs with $\alpha > 1$ provide a measure of the distance to the nearest stabilizer state, while those with $\alpha < 1$ relate to the stabilizer rank of $\ket{\Psi}$~\cite{Haug:2024ptu}.
Additionally, it was demonstrated that SREs with $\alpha \geq 2$ are magic monotones for pure states, in contrast to $\alpha < 2$~\cite{Leone:2024lfr,Haug:2023hcs}.
For this reason, the stabilizer $2$-R\'enyi entropy 
\begin{equation} 
\mathcal{M}_{2}(\ket{\Psi})= - \log 
\left( \frac{1}{d} \sum_{\hat{P} \in \widetilde{\mathcal{G}}_{N}} \langle \Psi |\hat{P} | \Psi \rangle^{4} \right) \; ,
\label{eq:2SRE}
\end{equation}
or its linear version $\mathcal{M}_{\mathrm{lin}}(\ket{\Psi})$ with
\begin{equation} 
\mathcal{M}_{2}(\ket{\Psi})= - \log 
\left( 1- \mathcal{M}_{\mathrm{lin}}(\ket{\Psi}) \right) \; ,
\label{eq:linSRE}
\end{equation}
which has been shown to be a strong monotone~\cite{Leone:2024lfr},
are often preferred measures of magic for applications.
Figure~\ref{fig:1q_Bloch_magic} shows $\mathcal{M}_{2}(\ket{\Psi})$ for a single qubit
on the Bloch sphere.

As mentioned above, an appealing feature of SREs is that they do not involve optimization procedures, making them significantly more tractable than many previously proposed measures. In particular SREs can be computed efficiently in MPS~\cite{Haug:2022vpg,Haug:2023hcs,Lami:2023naw,Tarabunga:2024ugl}. For general $N$-qubit states, although their definition involves $4^N$ Pauli string expectation values, SREs can be computed using replica methods, which are well-suited for sparse wave functions~\cite{Tarabunga:2023xmv}, or via Markov-Chain-Monte-Carlo (MCMC)
techniques~\cite{Tarabunga:2023ggd,Brokemeier:2024lhq,Sinibaldi:2025cst,Sierant:2026jru}.
SREs can also be straightforwardly generalized to pure states of high-dimensional qudit systems, using the formalism of qudit stabilizer states~\cite{Howard:2013wxp,Turkeshi:2024pnj}.
\\

Similarly to entanglement, defining measures of magic for mixed states is, however, more difficult. Definition of mixed-state SREs have been proposed that resemble the definition of SRE for pure states, through proper (re-)normalization of the probability distribution $\Xi_P$~\cite{Leone:2021rzd}
but these apply only to a subset of mixed stabilizer states (states in the stabilizer polytope)~\cite{Leone:2021rzd,Tarabunga:2025wym}.
Recently, genuine magic witnesses for mixed states derived from SREs have been proposed and efficiently measured on trapped-ion quantum computers~\cite{Haug:2025ncl}.

A measure of non-stabilizerness that is applicable to both pure and mixed states is the 
Robustness of Magic (RoM)~\cite{PhysRevLett.115.070501,PhysRevLett.118.090501,Heinrich_2019}. 
It is defined by the minimum distance to the surface defined by the stabilizer states,
\begin{align}
    R(\hat{\rho})  & =   \min_{\bm x} \left\{
    \ ||\bm x||_1 \ \ \  \bigg\rvert \ \ \ \hat{\rho} = \sum_{i}
     x_i\hat{\rho}_{s_i}\right\} 
     \ ,
\label{eq:RoMdef}
\end{align}
given by the 1-norm of the coefficients of the stabilizer density matrices $\hat{\rho}_{s_i}$.
Elements of the stabilizer polytope have $x_i>0$ and $R(\hat{\rho})=1$, 
while non-stabilizer states have some $x_i<0$, and hence $R(\hat{\rho})>1$.
As  $\sum_i x_i=1$, $R(\hat{\rho})$ measures the amount of negativity in the 
expansion coefficients,
\begin{eqnarray}
||\bm x||_1 = 1 + 2 \sum\limits_{i \ \forall\  x_i<0} |x_i|
\ ,
\end{eqnarray}
and thus provides a measure of the difficulty for classical simulation
(more specifically, $R(\hat{\rho})-1$), 
because of the sign problem encountered in sampling (from the non-positivity of the $x_i$).
The optimization involved in computing the RoM makes this measures however difficult to compute in practice for large subsystems.

\subsubsection{Entanglement-Magic Interplay}

\begin{figure}[ht!]
\centering
\includegraphics[width=\columnwidth]{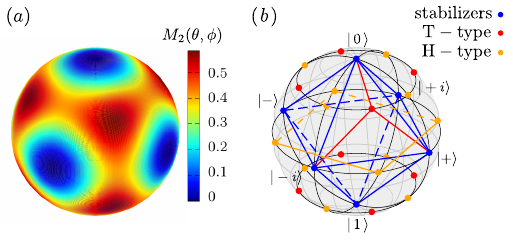}
\caption{
(a) 
$\mathcal{M}_2 (|\psi\rangle)$  of a single qubit state $\ket{\psi} = \cos{(\theta/2)} \ket{0} + \mathrm{e}^{i \phi} \sin(\theta/2) \ket{1}$ on the Bloch sphere.
(b) Special states on the Bloch sphere: 
Stabilizer states $\ket{0}, \ket{1}, \ket{+}, \ket{-}, \ket{+i}, \ket{-i}$ (blue points) occupy the vertices of an octahedron, T-type states $\ket{\mathrm{T}}$ (red points) possess maximal magic $\mathcal{M}_2 = \log_2(3/2)\simeq 0.585$, while H-type states $\ket{\mathrm{H}}$ are associated with the most probable value of $\mathcal{M}_2 = \log_2(4/3)\simeq 0.415$~\cite{Szombathy:2024tow}. 
[Figure from Ref.~\cite{Szombathy:2024tow} provided by Dominik Szombathy and used with permission from the authors under {\it Creative Commons Attribution 4.0 International license}~\cite{cc_by_4.0}.]
}
\label{fig:1q_Bloch_magic}
\end{figure}

Like entanglement alone, magic alone does not provide a full characterization of quantum complexity. This is because magic can be contained in local degrees of freedom, which can thus be eliminated or created via local unitary operations.
For example, the maximal stabilizer 2-R\'enyi entropy for a single qubit is $\mathcal{M}_2^{N=1, \, \mathrm{max}} = \log{(3/2)}$, which is attained for the T-type states~\cite{Szombathy:2024tow}, introduced in Ref.~\cite{Bravyi:2004isx}, and represented by the red points with Bloch vectors $(\pm 1, \pm 1, \pm 1)/\sqrt{3}$ in Fig.~\ref{fig:1q_Bloch_magic}. 
\\
Using the additivity property of $\mathcal{M}_2$~\cite{Leone:2021rzd} means that the maximal stabilizer 2-R\'enyi entropy for an $N$-qubit tensor product state is $\mathcal{M}_2^{N, TP, \, \mathrm{max}} = N \, \log(3/2)$. 
Thus, even a tensor product state can follow a volume-law scaling of magic, 
although 
it is bounded above by
$\mathcal{M}_2^{N, \, \mathrm{max}} = \log ((2^N+1)/2)$~\cite{Leone:2021rzd}, 
and, as discussed below, 
does not reach the maximum attainable value in entangled states. 
\\

Thus it is the {\it non-local magic}, {\it i.e.}, the magic which resides in the non-local degrees-of-freedom, and thus requires both entanglement and magic to exist,  that makes a quantum state truly complex.
The concept of non-local magic was introduced in Ref.~\cite{Cao:2023mzo,Cao:2024nrx}, where it was defined as the minimum magic over tensor-product local unitaries.
Specifically, given an $n$-partition of the state $\ket{\Psi}$ and a chosen measure of magic $M$ ({\it e.g.}, SRE or RoM), the $n$-partite non-local magic is given by
\begin{equation}
M^{(n,\, NL)} \left( |{\Psi}\rangle \right) = \min_{ U_{A_1} \otimes ... \otimes U_{A_n} } 
M\left( U_{A_1} \otimes ... \otimes U_{A_n} |{\Psi}\rangle \right) \; .
\label{eq:NLMdef}
\end{equation}
The minimization over local unitaries $\otimes_i U_{A_i}$ eliminates the local magic contained in the individual subsystems $A_i$, and thus only leaves the magic that resides in the correlations between them. In this way, non-local magic provides a measure that is independent on the local basis.
Note, that this definition is tied to the adopted prescription for the local degrees of freedom. For example, if $\ket{\Psi}$ is a $N$-qubit state, the subsystems $A_i$ in Eq.~\eqref{eq:NLMdef} may be single qubits, or groups of qubits. In physical systems, the unitaries may be chosen to preserve the relevant symmetries and gauge invariances~\cite{Grieninger:2026bdq,Iannotti:2026haw,Keeble:2026}.
\\

The explicit minimization in Eq.~\eqref{eq:NLMdef} is generally untractable for systems with a  
large number of qubits.
In the bipartite case ($n=2$), 
however, 
an estimate of the non-local stabilizer 2-R\'enyi entropy $\mathcal{M}_2^{(2,\, NL)}$, based on the entanglement spectrum of the state, has been derived~\cite{Cao:2024nrx}.
Precisely, given a bipartite pure state $\ket{\Psi}_{AB}$ with entanglement spectrum $\{ \lambda_k \}_{k=1,...,r}$, one can form a new state through local change of Schmidt basis, as
\begin{equation}
    \ket{\Psi'}_{AB} =  \sum_{k=1}^{r} \sqrt{\lambda_k} \ket{s_k}_A \otimes \ket{s_k}_B \; ,
\end{equation}
where the $\ket{s_k}$ denote stabilizer states.
The reduced density matrix associated with subsystem $A$ (or $B$) is $\rho_A' = \sum_r \ket{s_k} \bra{s_k}$, 
which is a convex mixture of stabilizer states ({\it i.e.}, it belongs to the stabilizer polytope), and by definition has zero magic.  Therefore the local transformation $\ket{\Psi}_{AB} \rightarrow \ket{\Psi'}_{AB} $ preserves the entanglement spectrum while eliminating local magic. Consequently, the total magic of $\ket{\Psi'}_{AB}$ provides a good estimate to the non-local magic of $\ket{\Psi}_{AB}$.
Strictly speaking, it provides a close upper bound~\cite{Cao:2024nrx}. For example, for the stabilizer $2$-R\'enyi entropy
\begin{equation}
     \mathcal{M}_2^{(2, NL)}(\ket{\Psi}_{AB}) \leq \mathcal{M}_2(\ket{\Psi'}_{AB})  \; .
     \label{eq:NL_M2_upperbound}
\end{equation}
This is because local and non-local magic are not {\it a priori} independent, and so there might exists other local unitaries $U_A \otimes U_B$ which also eliminate local magic while yielding lower values of the total magic. 

In practice, however, it is found through comparisons with explicit minimization, that the upper bound is saturated for few-qubit systems.
This was shown analytically for 2-qubit systems in Ref.~\cite{Qian:2025oit}.
We have verified the equality numerically for systems with $N\le 6$ qubits. Further, Ref.~\cite{Busoni:2026lvp} derived a similar upper-bound based on {\it Schmidt attainment} for bipartite qudit systems, and showed that this bound is saturated for  2-qutrit and 2-qu5it systems, while it provides a good estimate for composite local dimensions such as qu4its.
\\

We note that related concepts of {\it long-range} or {\it mutual magic} have been proposed as analogous to the mutual information in Eq.~\eqref{eq:MI}, using measures of magic for mixed states. Although, similarly to mutual information, mutual magic is not a proper measure of quantum information, it captures, to some extent, aspects of non-local magic, 
see {\it e.g.}, Refs.~\cite{Fliss:2020yrd,Ellison:2020dkj,White:2020zoz,Tarabunga:2023ggd,Frau:2024zaa,Tarabunga:2025wym,Korbany:2025noe}.
\\

\begin{figure}[ht!]
\centering
\includegraphics[width=\columnwidth]{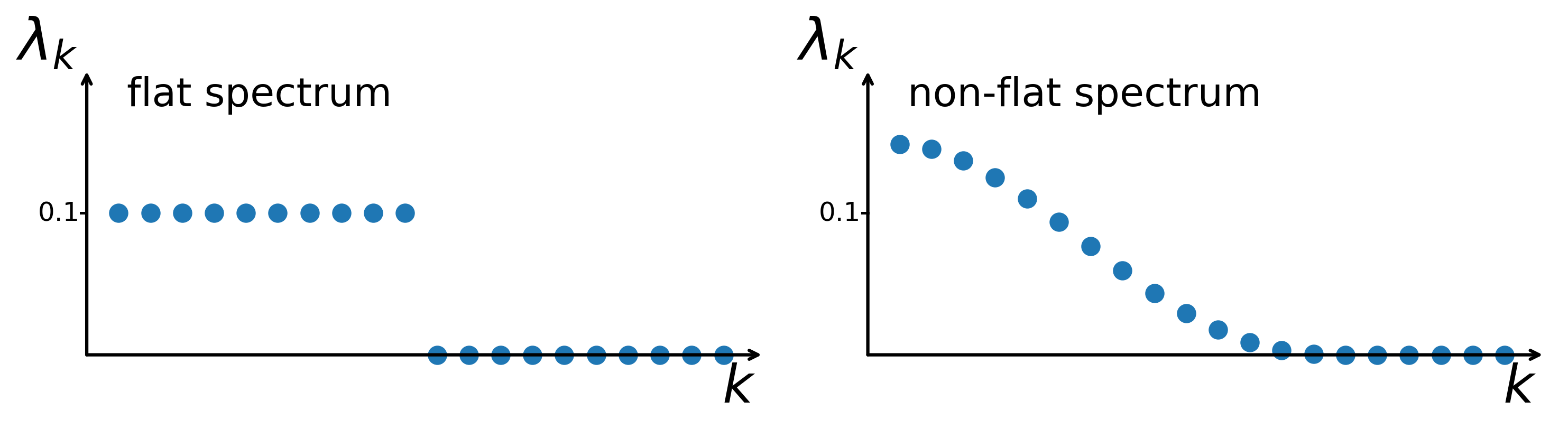}
\caption{Flat (left) versus non-flat (right) bipartite entanglement spectra. Both spectra yields the same von Neumann entanglement entropy $S \simeq 3.322$ (calculated here with $\log_2$).}
\label{fig:entang_spectrum_schematic}
\end{figure}

The above discussion indicates that magic and entanglement are directly connected and mutually influence each other. 
Entanglement allows for
non-local magic, while, in turn, non-local magic, imprints the entanglement spectrum.
The latter feature can be captured, to some extent, through the {\it anti-flatness} of the entanglement spectrum. 
Anti-flatness is defined 
as the variance of the reduced density matrix $\hat\rho_A$ for subsystem A~\cite{Tirrito:2023fnw}. That is,
\begin{eqnarray}
{\cal F}_A (\ket{\psi}_{AB}) & = & \langle\rho_A^2\rangle - \langle\rho_A\rangle^2
\ =\ 
{\rm Tr} (\rho_A^3) - \left({\rm Tr} \rho_A^2\right)^2 \; .
\label{eq:Antiflat}
\end{eqnarray}
When the entanglement spectrum is flat, 
{\it i.e.}, when the non-zero eigenvalues of $\hat\rho_A$ are equal (see Fig.~\ref{fig:entang_spectrum_schematic}), 
the two contributions in Eq.~\eqref{eq:Antiflat} cancel. 
This occurs if and only if $\ket{\psi}_{AB}$ is unentangled or has no magic~\cite{Tirrito:2023fnw}, {\it i.e.}, if it is a tensor product or stabilizer state. 
Since the entanglement spectrum (and thus the anti-flatness) is independent of the local basis, this means that the presence of non-local magic is a necessary and sufficient condition for anti-flatness of the entanglement spectrum~\cite{Cao:2024nrx}
\footnote{Exceptions of the sufficiency condition are found for some particular non-integer $\alpha$-SREs~\cite{Cao:2024nrx}.}.

The total linear magic $\mathcal{M}_{\mathrm{lin}}$, 
with $\mathcal{M}_2 = - \log(1-M_{\mathrm{lin}})$, is directly proportional to the anti-flatness averaged over Clifford orbits~\cite{Tirrito:2023fnw}
\begin{equation}
    \langle \mathcal{F}_A (\hat{\Gamma} \ket{\psi}_{AB}) \rangle_\mathcal{C} 
    = c(d,d_A) \ \mathcal{M}_{\rm lin} (\ket{\psi}_{AB}) \; ,
\label{eq:Cliffav_AF}
\end{equation}
where the left-hand side is the anti-flatness of $\hat{\Gamma} \ket{\psi}_{AB}$ averaged over 
Clifford unitaries $\hat{\Gamma} \in \mathcal{C}$, and the proportionality constant
$c(d,d_A)$ is given by 
\begin{equation}
    c(d,d_A) = \frac{(d^2 - d_A^2) (d_A^2 -1)}{(d^2-1) (d+2) \, d_A^2} 
    \; ,
    \label{eq:cfun}
\end{equation}
where $d_A$ is again the dimensionality of the Hilbert space of system A. 
The averaging over Clifford operators provides a quantity that is Clifford-invariant, while re-distributing magic among local and non-local degrees of freedom.

Further, Ref.~\cite{Cao:2024nrx} showed that anti-flatness provides a lower bound to bipartite non-local magic quantified by the trace distance, while an upper bound can be obtained from the entanglement entropy.
Additionally, numerical studies of up to 40-qubit systems~\cite{Keeble:2026} indicate the following bound for the linear non-local SRE:
\begin{equation}
    \frac{{\cal F}_A (\ket{\psi}_{AB})}{4} \leq \mathcal{M}_{\mathrm{lin}}^{2,NL} (\ket{\psi}_{AB}) \; .
\end{equation}
This lower bound has been found to be saturated for 2-qubit systems in Ref.~\cite{Robin:2025ymq} 
(see also Ref.~\cite{Busoni:2026lvp}). 
The equality however appears to be specific to 2-qubit systems, as it has not been verified for higher-dimensional 2-qudit systems~\cite{Busoni:2026lvp}.
\\

In summary, although entanglement and magic may appear  
uncorrelated when focusing on simple or global measures~\cite{Iannotti:2025lkb}, 
they are, in fact, deeply interconnected: entanglement is needed to distribute magic over non-local degrees of freedom,  while magic (precisely non-local magic) is required to generate complex entanglement patterns.
These interconnections are not captured by global magic measures or simple entanglement entropies, but become manifest, for example, through non-flat entanglement spectra, which capture, to some extent, the "response" of entanglement to magic~\cite{Tirrito:2023fnw}. 
Anti-flatness, however, only provides a lower bound to bipartite non-local magic, which is found to be unsaturated in many-qubit systems.
This signals the presence of higher-order effects, beyond the variance of the reduced density matrix, and 
understanding the nature of these effects is an intriguing question to pursue.
Beyond bi-partitions, it has also been shown that magic (precisely multipartite non-local magic) is necessary to reproduce the correct patterns of multipartite entanglement in the context of holography, see {\it e.g.}, Refs.~\cite{Cao:2024nrx,Nezami:2016zni}.
As non-local magic provides a measure of magic that is independent of the local basis, it can potentially provide a tool to probe these effects in laboratory experiments relevant to nuclear and high-energy physics, as discussed in sections~\ref{sec:ApplicationsLENP}, \ref{sec:QI_driven_nuclear_methods} and \ref{sec:ApplicationsHEP}.

\subsubsection{Studies in Random Quantum Circuits}

Random circuits are prototypical systems which can be used to study  
complexity growth, scrambling of information across degrees of freedom, and onset of chaotic behavior in the dynamics of the system.
In monitored tunable circuits, varying the amount of non-Clifford (T) gates or the rate of measurements, allows one to track how entanglement and magic develop over time, and to locate the transition between classically tractable and intrinsically quantum regimes.

\begin{figure}[ht]
\centering
\includegraphics[width=.79\columnwidth]{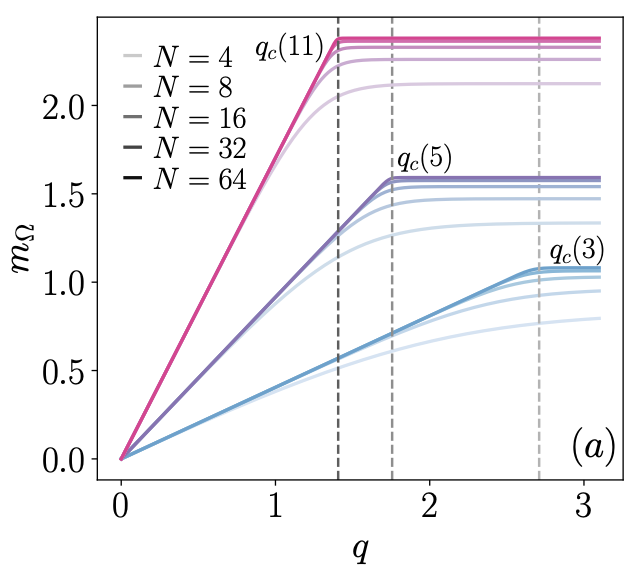}
\caption{GSE per qudit as a function of the qudit T-gate doping rate $q$, for different qudit local dimensions ($d_{\mathrm{loc}}=3,5,11$) and system sizes $N$. The value $q_c (d_{\mathrm{loc}})$ denotes the critical doping rate, at which the GSE density saturates. For $N\rightarrow \infty$ the GSE saturates to the Haar value $\log(d_{\mathrm{loc}})$. 
[Figure from Ref.~\cite{Magni:2025xbm} used with permission from the authors under {\it Creative Commons Attribution 4.0 International license}~\cite{cc_by_4.0}.]
}
\label{fig:magic_qudits_random}
\end{figure}

There is now a large number of studies in this area, and thus we will only mention a few of them here.
For example, studies in brick-wall Haar random circuits showed that magic grows more rapidly than entanglement 
and saturates at times that scale logarithmically in system size $t_{sat}^{mag} \sim \log(N)$, while entanglement grows ballistically with $t_{sat}^{mag} \sim N$~\cite{Turkeshi:2024pnj}.
By defining generalized stabilizer entropies (GSE) as a generalization of SREs to qudits~\cite{Turkeshi:2024pnj}
it was further found that qudits with higher dimensions converge faster to the Haar regime of maximal complexity~\cite{Magni:2025xbm}, as shown in Fig.~\ref{fig:magic_qudits_random}.

\begin{figure}[ht]
\centering
\includegraphics[width=\columnwidth]{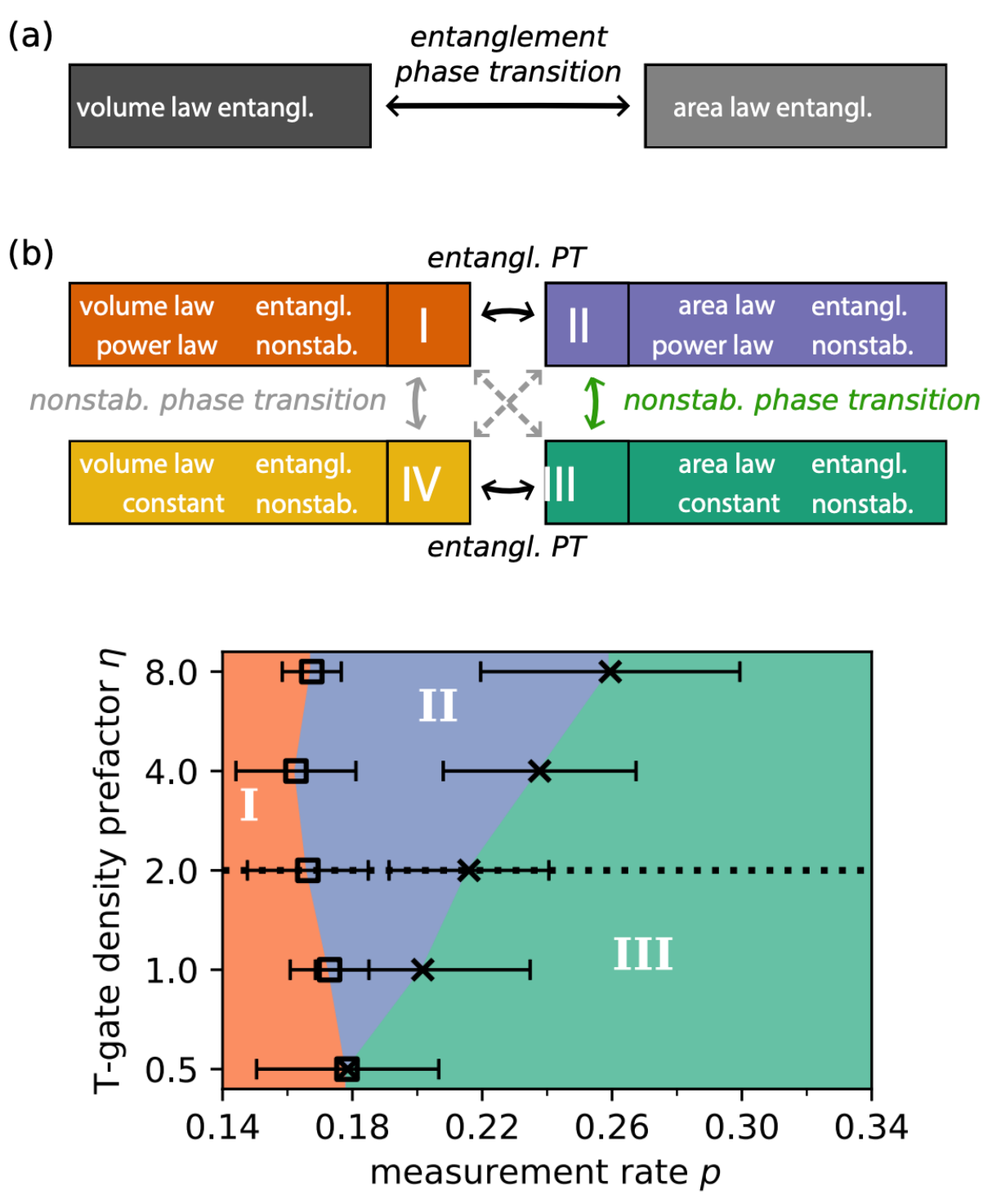}
\caption{The upper panel shows (a): phase transitions in entanglement; and (b): possible phase transitions in entanglement and nonstabilizerness. 
The lower panel shows the magic-entanglement phase diagram observed in Ref.~\cite{Fux:2023brx}, with the new transition between regions II and III. 
[Figure adapted from Ref.~\cite{Fux:2023brx} with permission from the authors under {\it Creative Commons Attribution 4.0 International license}~\cite{cc_by_4.0}.]
}
\label{fig:entang_magic_PT_random}
\end{figure}

Other studies investigated entanglement and magic phase transitions driven by T-gate magic injection and local measurements. 
In the case of entanglement, it is well established that unitary evolution leads to volume-law scaling, while local measurements drive the system into an area-law regime. This behavior is observed in both Haar-random and random Clifford circuits~\cite{Skinner:2018tjl,Li:2018mcv,Li:2019zju}, confirming that volume-law entanglement alone does not imply intrinsically quantum dynamics.
Recently Ref.~\cite{Fux:2023brx}, using 2-SRE to quantify magic, identified a new magic phase transition between power-law scaling and constant non-stabilizerness, controlled by the measurement rate. This is shown in Fig.~\ref{fig:entang_magic_PT_random}. Importantly, the critical measurement rate differs from that of the entanglement volume-to-area law transition, indicating that the underlying mechanisms driving these two transitions are distinct~\cite{Fux:2023brx}.
Note however that Ref.~\cite{Bejan:2023zqm}, which used another magic witness, did not find such a separation between entanglement and magic phase transitions, suggesting that the conclusion above may be sensitive to the choice of magic witness or the specific circuit setup. 
As this difference is not yet resolved~\cite{Fux:2023brx}, 
further studies are needed to fully understand the interplay between entanglement and magic.

\subsubsection{Entanglement and Magic of an Operator}
\label{subsubsec:power}

The amount of complexity that can be induced by a given unitary operator can be captured 
by its {\it entanglement power} and {\it magic power}. \\

The entanglement power (or entangling power) 
of a unitary operator $\hat U$ can be defined as the average entanglement generated by $\hat U$ when acting on tensor-product states~\cite{Zanardi:2001zza,BallardWu:2011}.
In the bipartite case, the entanglement entropy  $S$, 
such as the von Neumann or linear entropy, 
can be used define the entangling power of $\hat U$,
\begin{equation}
    \overline{S} (\hat U) = S \overline{ \left[ \hat U \left(\ket{\Phi}_A \otimes \ket{\Psi}_B \right) \right]} \; ,
\end{equation}
where $\overline{\phantom{ABC}}$ denotes the average over the tensor product states $\ket{\Phi}_A \otimes \ket{\Psi}_B$. Loosely speaking, the entangling power of $\hat U$ measures how far $\hat U$ is from being expressible as a tensor product of local operators $\hat{U}_A \otimes \hat{U}_B$.

In the two-qubit case, $\overline{S} (\hat U) $ can be computed exactly by averaging over the Bloch sphere angles for each qubit~\cite{Beane:2018oxh}. For larger systems, however, continuous averaging becomes computationally prohibitive.
Alternatively, in the case of the linear entropy $S_{\mathrm{lin}} = 1 - \mathrm{Tr}(\rho_A^2)$, $\overline{S} (\hat U) $ may be obtained by discrete averaging or sampling over tensor products of stabilizer states~\cite{Robin:2024oqc}, since stabilizer states form a projective 3-design for qubit systems~\cite{Dankert:2009yux,Kueng:2015mts,Zhu:2017psv}. 
Thus,
\begin{equation}
    \overline{S}_{\mathrm{lin}} (\hat U) = \frac{1}{\mathcal{N}_{stab}^{(A)}\,\mathcal{N}_{stab}^{(B)}} \sum_{i=1}^{\mathcal{N}_{stab}^{(A)}}  \sum_{j=1}^{\mathcal{N}_{stab}^{(B)}}  {S}_{\mathrm{lin}} \left[ \hat U \left( \ket{\phi_i}_A \otimes \ket{\psi_j}_B \right) \right]\; ,
\end{equation}
where $\mathcal{N}_{stab}^{(A)}$ and $\mathcal{N}_{stab}^{(B)}$ denote the number of stabilizer states for system A and B, respectively, while $\ket{\phi_i}_A$ and $ \ket{\psi_j}_B$ label their respective stabilizer states.
\\

Analogously, the magic power, or non-stabilizing power, of a unitary operator $\hat U$ was introduced in Ref.~\cite{Leone:2021rzd} as the average magic generated by $\hat U$ acting on the set of stabilizer (zero-magic) states. 
That is, given a 
measure of magic $M$, 
for example a SRE, 
the magic power $\overline{M}(U)$ is defined as
\begin{equation}
    \overline{M}(\hat U) = \frac{1}{\mathcal{N}_{stab}} \sum_{i=1}^{\mathcal{N}_{stab}} M\left(\hat{U} \ket{\Psi_i} \right) \; ,
    \label{eq:magic_power_def}
\end{equation}
where $\mathcal{N}_{stab}$ is the total number of stabilizer states, and $\ket{\Psi_i}$ denotes the stabilizer states for the full system.
Similarly, one can define the non-local magic power of $\hat{U}$~\cite{Robin:2025ymq}.

The linear SRE, 
$\overline{\mathcal{M}}_{lin} (\hat U)$, 
has been shown to provide a lower bound to the T-gate count $t(\hat U)$~\cite{Leone:2021rzd}, {\it i.e.}, the minimal number of T gates required to implement $\hat U$ in a quantum circuit,
\begin{equation}
    t(\hat U) \geq -\log \left[ d- (4+d)  \overline{\mathcal{M}}_{lin} (\hat U) \right] + \log(d+3) -2 \; .
\end{equation}
In addition,  maximal magic power is required to induce chaotic behavior~\cite{Leone:2021rzd}. 
These aspects have been further studied in the context of random circuits, 
see, {\it e.g.}, Ref.~\cite{Szombathy:2024tow}.
\\

In principle, the magic content of an operator can also be characterized by its action on Pauli strings. This is because a Clifford operator $\hat{\mathcal{C}}$ turns a single Pauli $\hat P$ 
into another one,
\begin{equation}
    \hat{\mathcal{C}} \,  \hat P \, \hat{\mathcal{C}}^\dagger \rightarrow \hat{P}' \; .
\end{equation}
while a non-Clifford $\hat U$ will lead to spreading into the Pauli basis,
\begin{equation}
    \hat{U} \,  \hat P \, \hat{U}^\dagger \rightarrow \sum_i a_i \hat{P}_i \; .
\end{equation}
Thus another useful measure, particularly in the context of quantum simulations in the Heisenberg picture, is the operator stabilizer entropy (OSE), 
defined  as an operator-space analog of the stabilizer R\'enyi entropy~\cite{Dowling:2024wvo}. Physically, this measure captures how well an evolved operator can be approximated 
by a small number of Pauli strings, 
and has been shown to provide a tight lower bound to the minimum number of T-gates in an associated quantum circuit.
\\

Alternative definitions of entangling and magic power, as well as other notions of operator complexity, have been defined in, {\it e.g.}, Refs.\cite{Seddon:2019ykh,Eisert:2021mjg,Bu:2022ozl}, with relations to quantum circuit resource requirements. 
These definitions, however, rely on optimizations over sets of operators making them difficult to compute in practice.

\subsubsection{Non-Gaussianity}

Although Gaussian states are a well-established concept in many-body physics, connections of (non-)Gaussianity to precise notions of many-body complexity and quantum resources have only been established more recently.
For completeness, we briefly mention a few selected measures of non-Gaussianity below, without, however, going into details, as these aspects have so far seen limited explorations in nuclear and high-energy physics.

In the case of bosonic, continuous variable systems, non-Gaussianity can be characterized via the Wigner function of the quantum state~\cite{Weedbrook:2011wxo}.
Specifically, if the latter takes a Gaussian form, then the state is said to be Gaussian. Furthermore, in the case of pure states, Gaussian states are the only ones with a non-negative Wigner function~\cite{Hudson:1974,10.1063/1.525607}.
Measures of deviations from Gaussianity have then been developed, based on the Wigner negativity or distance measures, see, {\it e.g.}, Refs.~\cite{PhysRevA.76.042327,PhysRevA.78.060303,PhysRevLett.106.200401,PhysRevA.87.062104,PhysRevA.90.013810,PhysRevA.97.062337,PhysRevLett.124.063605}.
Analogous to entanglement and magic powers, the non-Gaussianity generating power of a unitary operation, as defined in Ref.~\cite{PhysRevA.97.052317}, quantifies the maximal non-Gaussianity that this operation can induce on Gaussian input states, thereby providing a lower bound on the non-Gaussian resources required to implement it.

Similarly, various measures of fermionic non-Gaussianity have been advanced, to quantify the departure of a quantum state from the manifold of Gaussian states. These include, for example, Gaussian fidelity~\cite{Dias:2023zsg}, fermionic rank~\cite{Dias:2023zsg}, fermionic Gaussian extent~\cite{Dias:2023zsg,Reardon-Smith:2023vbm}, which involve optimization procedures.

Recently, the fermionic anti-flatness (FAF) was introduced~\cite{Sierant:2025fax}, 
\begin{equation}
   \mathcal{F}^F_m (\ket{\Psi}) = N -  \frac{1}{2}  \mathrm{Tr} \left[ (\Gamma^T \Gamma)^m \right] \; ,
\label{eq:FAF}
\end{equation}
where $\Gamma$ is the covariance matrix defined in Eq.~\eqref{eq:covariance_matrix_fermion}, and $m \geq 1$ is a chosen integer.
For a FGS, $(\Gamma^T \Gamma)  = \mathds{1}$, and thus, the FAF cancels due to the offset $-N$ in definition \eqref{eq:FAF}. The FAF can be interpreted, in a way, as quantifying the amount of information not captured by the one-body (two-point) correlation matrices, or covariance matrix.
This measure has also been shown to satisfy further good properties of faithfulness, or invariance under FGU, among others, and to be efficiently computable for MPS, and experimentally measurable~\cite{Sierant:2025fax}.
The FAF is invariant under fermionic mode (basis) transformation, and, in systems with good particle numbers, can be expressed entirely in terms of the occupation numbers in the natural basis. 
Other measures of non-Gaussian magic have been recently proposed, 
based on the entropy of the closest Gaussian state, which can be efficiently computed for pure states~\cite{Coffman:2025usz}.
\\

The behavior of the FAF in the dynamics of brick-wall Haar-random circuits has been studied~\cite{Sierant:2025fax}, starting from an initial fermionic Gaussian state. 
It becomes extensive already at constant circuit depth $t=O(1)$, 
and subsequently saturates to the Haar-random value on a timescale that grows logarithmically with system size $t \sim \log(N)$, similarly to non-stabilizerness, and much faster than the linear timescale $t\sim N$ required for entanglement entropy to saturate.
\\

Exploring the complexity aspects of fermionic Gaussian states (FGSs) and their extensions is an active area of research. A natural setting is provided 
by random nearest-neighbor matchgate circuits, which correspond, under the Jordan–Wigner mapping, to free-fermion (Gaussian) dynamics. Deviations from Gaussianity can be introduced and controlled by doping the circuit with non-Gaussian resources, such as SWAP gates, in analogy with T-doping in Clifford circuits. Such setups may offer insight into traditional many-body methods based on systematic expansions around Gaussian states, such as beyond-mean-field methods.
For example, it is known that random matchgate circuits exhibit slow entanglement growth $S \sim \sqrt{t}$, in contrast to the ballistic growth $S \sim {t}$ of random unitary circuits and random Clifford circuits,
and that this entanglement is very fragile under measurements, see, {\it e.g.}, Refs.~\cite{PhysRevB.93.134305,PhysRevX.7.031016,Paviglianiti:2025zdm}. 
Ref.~\cite{Paviglianiti:2025zdm} further studied the entanglement evolution in doped matchgates circuits, and found that an extensive number of non-Gaussian operations are needed to recover ballistic growth of entanglement and genuine volume law consistent with Page-typical (maximal) entanglement.

In contrast, fermionic Gaussian states exhibit non-stabilizerness comparable to that of generic Haar-random states~\cite{Collura:2024ida}, up to subleading corrections that grow only logarithmically with system size. 
The extent to which this magic is genuinely non-local, however, remains to be determined.
\\

Finally, measures to quantify non-Gaussianity in hybrid boson-fermion systems are now also being developed, 
see, {\it e.g.}, Ref.~\cite{Sarkis:2025cic}.
\\

There are other measures of complexity that we do not discuss in detail.
Two examples, that are surprisingly related to each other, are Krylov complexity and Nielsen complexity, which measure the complexity in unitary evolution.
A Krylov basis can be generated by repeated applications of the Hamiltonian to an initial state, 
$\{ |v_0\rangle, |v_1\rangle, ... \}$.  Starting from the initial states, Hamiltonian evolution generates a sum over these vectors 
with time-dependent amplitudes.  The Krylov Complexity is can be defined to be the average distance in the vector space.
The Nielsen complexity can be defined as the length of the minimum geodesic in the space to evolve from the initial to final states.
It has been shown that the trace of a matrix related to the Krylov complexity time average determines an upper bound on the Nielsen complexity~\cite{Craps_2024}.

%% file: Section_Info_Flow.tex
\subsection{Information and Renormalization Flow}
\label{sec:Flow}
%
The renormalization group plays a central role in quantum field theory, 
both in providing an understanding of the role of quantum fluctuations in the strengths of coupling constants probed at a given length scale, 
and in the relative importance of higher-dimension operators in an effective field theory (EFT).
Typically, contributions from fluctuations in quantum fields with momenta greater than the renormalization scale $\mu$, or a momentum cut off $\Lambda$, are ``integrated out'' of the theory, with the effects of those modes on low-energy observables  included through higher-dimension operators and the renormalization of dimension-4 or lower operators.
As such, while leaving observables at scales below $\mu$ or $\Lambda$ invariant (up to corrections that are power-law suppressed), 
reducing $\mu$ or $\Lambda$ eliminates the ability to compute observables probed at scales greater than
$\mu$ or $\Lambda$.  
Thus, the information in a low-energy EFT is less than that in the full field theory.
\\

Renormalization has been considered extensively in the area of quantum information in multiple contexts, 
{\it e.g.}, Refs.~\cite{Verstraete_2005,Vidal_2007,han2020U,Jefferson_2017,han2020U,Meurice_2022,Korbany_2025}.
It forms the basis for modern tools such as DMRG~\cite{PhysRevLett.69.2863}, MERA~\cite{Vidal_2007}, 
and  Real-Space Mutual Information (RSMI)~\cite{Gkmen_2021}, and more.
We will not review these (powerful) tools, 
but will focus on the impact of smearing lattice wavefunctions on the measures of quantum complexity discussed above.
Flowing magic into the infrared using the RG has been implemented in MERA using dis-entanglers~\cite{White_2021} 
(see also, Ref.~\cite{Sarkar_2020,Ellison2021}), 
giving rise to the concept of long-range nonstabilizerness~\cite{Korbany_2025}.
In the context of lattice QCD, quasi-local smearing is a necessary ingredient as it suppresses noise in correlation functions, 
significantly reducing the classical computing resource requirements to achieve target precisions in low-energy observables.  
As an example, HYP-smearing~\cite{Hasenfratz_2001} 
is used extensively in large-scale lattice QCD calculations.
Smearing resembles MERA conceptually, 
but the number of degrees of freedom, and the local gauge invariance makes analysis challenging, except by direct simulation.
For lattice QCD, the smearing flow is typically 
from the scale of the lattice spacing to approximately between the scale of chiral symmetry breaking and the size of a hadron, 
as opposed to the far IR.
The tensor methods described in Ref.~\cite{Meurice_2022} present a well-defined framework in this regard.
\\

Consider an arbitrary wavefunction with $N$ qubits 
from which measures of entanglement and quantum magic are computed.  Representing the wavefunction on a 1D lattice, where each of the basis states is mapped to a lattice site, the wavefunction can be increasingly smeared across the lattice.  
This systematically suppresses the ultra-violet (UV) momentum-space modes of the wavefunction, with minimal modifications to the infra-red (IR) modes.
In the case of Gaussian smearing, the value of the wavefunction at site-$x$ is replaced by the Gaussian-averaged value of the values of the wavefunction at nearby sites, which can be repeated many times.
Each iteration of smearing reduces the information in the wavefunction.
An alternate smearing can be accomplished by applying a Hadamard-Walsh (HW) transform to the wavefunction, and systematically reducing the number of HW vectors, 
starting from the highest sequency,
that are retained to transform back to a reduced-space wavefunction~\footnote{
In exploring the use of sequency truncations in compressing quantum circuits,
Ref.~\cite{Li:2024lrl} studied the flow of magic in a discretized Gaussian wavefunction 
versus sequency cut off.}.
At each level of smearing, the same measures of entanglement and magic can be computed.
With sufficient smearing, the amount of information in the wavefunction will be reduced to that supported by two qubits, and further smearing will reduce it to that of just one qubit.
This corresponds to reducing the low-energy effective model space to support just two states.
Therefore, we expect the measures of entanglement and quantum magic to flow in such a way 
to be compatible with 
the surviving information being compressed into 
a reduced number of qubits.

\begin{figure}[ht!]
    \centering
\includegraphics[width=0.9\linewidth]{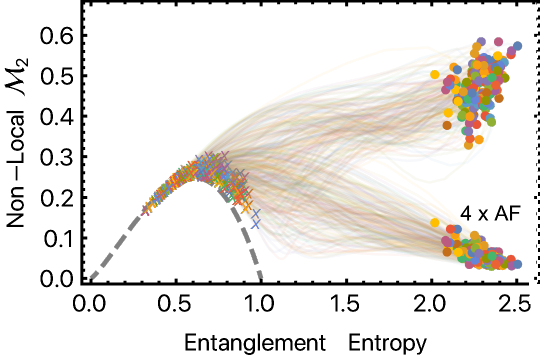} \\
 \vspace{0.5cm}
\includegraphics[width=0.9\linewidth]{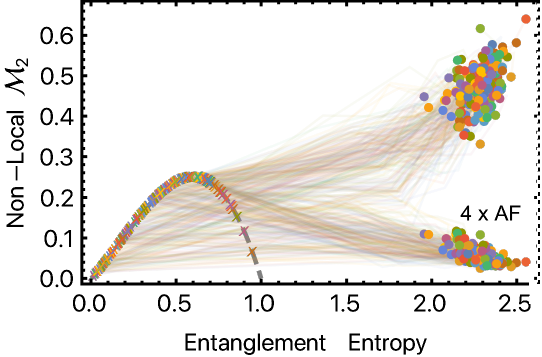}
\caption{
The renormalization flow of non-local linear magic, anti-flatness and entanglement entropy for 
Gaussian (upper panel)  and Hadamard (lower panel)  smearing.
A set of 200 6-qubit Haar-random states was selected and flowed by successive Gaussian smearing 
(from 0 to $2^6$ applications of a width=1 Gaussian profile)
or Hadamard transform truncations over a finite interval (by subtracting $2, 4, 8, 16, 32$ and $55$ states, respectively).
The dashed gray lines correspond to the 2-qubit relation between the entanglement entropy 
and four times the anti-flatness (equal to the non-local magic).
}
    \label{fig:Flow}
\end{figure}
As an example, we start with a set of 6-qubit Haar-random states and flow them into the infrared by a) successive Gaussian smearing and b) successive truncations of the Hadamard transform of the 6-qubit state,
and evaluate the half-chain entanglement entropy, the non-local magic and the anti-flatness.
The results of this evolution are shown in Fig.~\ref{fig:Flow}.
The 
non-local magic, anti-flatness and entanglement entropy
resulting from successive Hadamard truncations are seen to flow to those obtained from a 2-qubit system.
The results from Gaussian smearing also evolve toward those of the 2-qubit system,  but do not exactly recover it after $2^6$ smearing steps.
The total magic has not flowed to that of 2-qubits during the same evolution intervals, 
because of the values at each of the sites having relative phases, see appendix~\ref{sec:M2flow} for more details.
In addition, for a generalized discussion of the quantum complexity structure in two-qutrit systems, see
appendix~\ref{sec:TwoQudits}.

%% file: Section_Motivations.tex
\subsection{Motivations for Physics Applications}
\label{sec:Motivations}

\begin{figure}[ht!]
\centering
\includegraphics[width=.7\columnwidth]{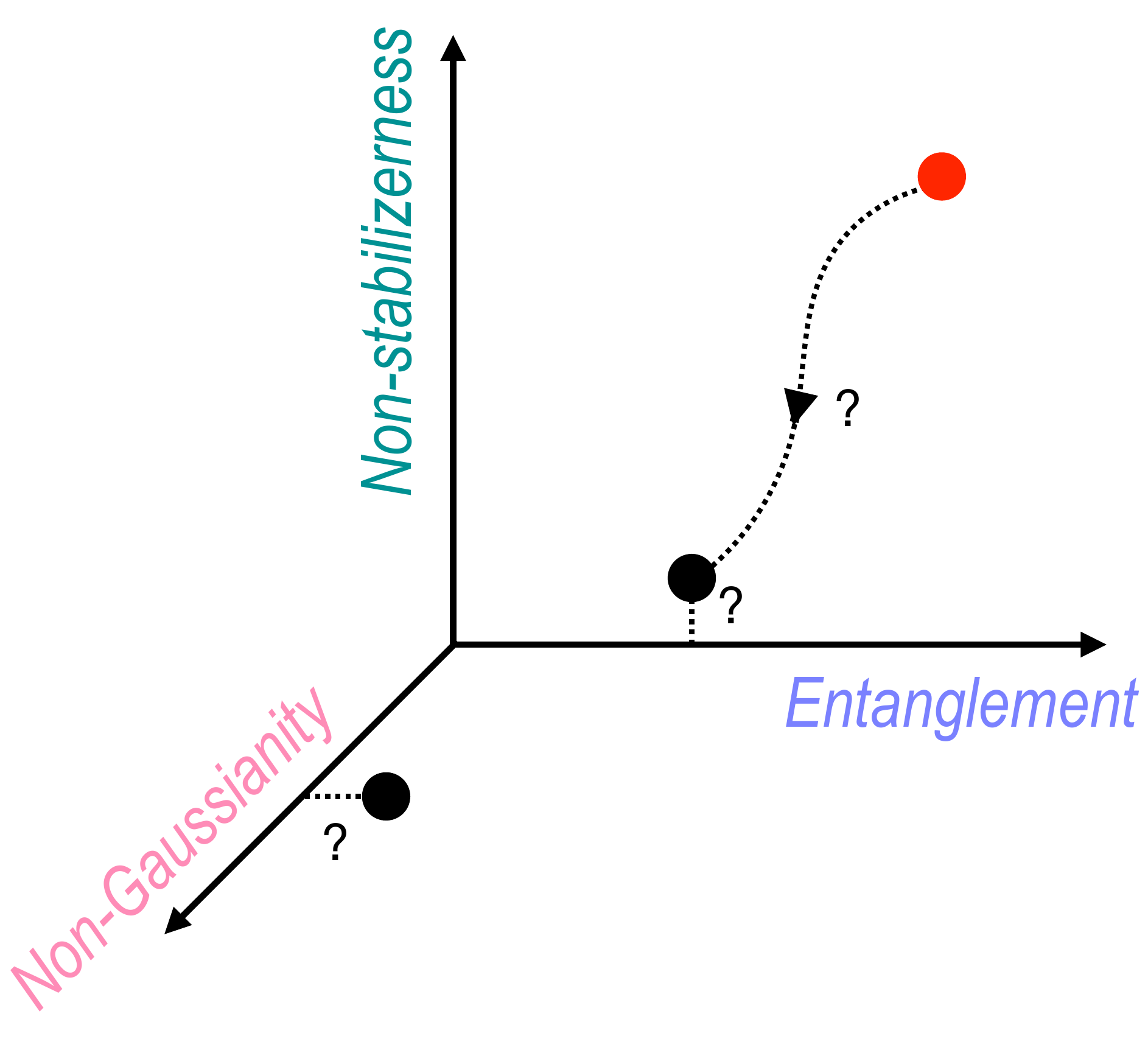}
\caption{Resources characterizing the computational complexity of quantum many-body states:  entanglement, non-stabilizerness (magic), and non-Gaussianity.
Figure inspired by Ref.~\cite{Sierant:2025fax}. }
\label{fig:complexity_triangle}
\end{figure}

As the aspects of quantum complexity reviewed above directly dictate the classical limitations and need for quantum computers, 
a central objective in the context of physics simulations is 
to identify where the quantum states of interest reside within the complexity landscape in Fig.~\ref{fig:complexity_triangle}, and to understand the trajectories that relevant dynamical processes follow through this diagram.

Characterizing the detailed structures of entanglement, magic and/or non-Gaussian features of physical states of interest reveals how far they lie from easy, classically-tractable states, and
allows for locating them 
relative to different resource-theoretic axes, thereby providing guidance on the most effective pathways for their simulations, and potentially informing the choice of suitable quantum platforms.

For example, if the ground states of certain classes of Hamiltonians tend to lie near the zero-magic axis, this indicates that classical stabilizer-based methods can already yield accurate approximations. In such cases, only a modest amount of non-Clifford resources 
(e.g., T-gates) may be needed to reach the target state, for instance through low-rank superpositions of stabilizer states or via T-doped circuits.
Conversely, since measures of complexity generally depend on the representation used to describe a state, configurations that appear highly complex in one basis may become significantly simpler under an appropriate transformation. Similarly, effective Hamiltonian approaches that reorganize information flow can reduce apparent complexity, potentially relocating the state toward a more tractable region of the landscape.
Further, understanding how quantum complexity is generated and spreads during dynamical processes, and whether alternative, less resource-intensive pathways exist, is essential for designing optimal quantum simulation of real-time evolution.

Beyond computational implications, examining physical systems through the lens of quantum information offers a powerful and complementary perspective on the phenomena we seek to understand and predict.
Examining how interactions in nature generate complexity, and how this complexity distributes and evolves in many-body settings, also aims at addressing fundamental questions about the emergence of collective behaviors, effective degrees of freedom and phases of matter. At the small scale of particle physics, quantum information may trace back to the nature of the fundamental interaction themselves and possibly help uncover new physics.

%% file: Section_Nuclear_Forces.tex
\subsection{Nuclear Forces}

The low-energy forces among protons and neutrons in nuclei are dominated by 2-body interactions, 
with much smaller contributions from 3-body interactions, 
and no significant evidence for the presence of 4-body or higher.
This statement is subject to adjustment with changes of renormalization scale 
(away from the scale of chiral symmetry breaking)
when working with chiral interactions, 
see, {\it e.g.}, Refs.~\cite{Platter_2005,Epelbaum_2007,RevModPhys.81.1773,Birse_2011}, 
but is generally valid for all modern analyses of nuclear structure.
Therefore, the quantum complexity of nuclei and its evolution during nuclear reactions is 
expected to have much of its origin in the 2-body nuclear forces.

The S-matrix provides a unique description of the 2-body scattering processes 
below inelastic thresholds.
Neglecting electroweak processes, this threshold is set by the incident energy required to create a pion in the final state (along with two nucleons).
Working directly with the S-matrix provides direct connections to experimental data 
(through fitting measured scattering cross sections, both total and differential),
avoiding the need to fit the multiple components of the nuclear force models, and counterterms in nuclear 
chiral effective field theory ($\chi$EFT) descriptions.
While the S-matrix does not reveal information about the quantum complexity of initial states, it
contains all of the information about fluctuations in complexity.

In 2018, 
Beane, Kaplan, Klco and Savage~\cite{Beane:2018oxh}
demonstrated connections between fluctuations in entanglement induced by scattering processes and global symmetries.  
A scattering process in which the entanglement remains unaltered for all states in the Hilbert space, corresponds to a classical interaction represented by an identity (singlet) 
operator in the Hilbert space(s).
Using low-energy two-baryon scattering systems, 
including nucleon-nucleon, nucleon-hyperon and hyperon-hyperon, 
it was shown~\cite{Beane:2018oxh} that in the limit of vanishing entanglement power
in spin-space 
of the S-matrix, corresponding to vanishing 
changes of entanglement during the scattering process,
these systems display an enhanced emergent global symmetry.
It was also highlighted that the entanglement power vanishes at points of conformal symmetry~\cite{Beane:2018oxh}.
In the case of nucleon-nucleon scattering, the emergent symmetry corresponds to Wigner's SU(4) spin-flavor symmetry, where the spin states of the proton and neutron can be combined into the fundamental representation of SU(4), and the scattering amplitudes display an invariance under SU(4) transformations.
This symmetry also emerges in the  large-N$_c$ limit of QCD~\cite{Kaplan_1996}.
An analogous result was obtained in the limit of vanishing entanglement power 
in hyperon-nucleon and hyperon-hyperon scattering.  However, while the large-N$_c$ limit of QCD predicts an emergent SU(6) symmetry~\cite{Kaplan_1996}, 
the vanishing of fluctuations in entanglement gives rise to a SU(16) emergent symmetry~\cite{Beane:2018oxh}.
This is a larger symmetry group than predicted in the large-N$_c$ limit, and
is consistent with numerical simulation 
results found at unphysical pion masses~\cite{Wagman:2017tmp} by the NPLQCD lattice QCD collaboration~\footnote{ {\tt https://www.ub.edu/nplqcd/} .}.
The connection between symmetry and fluctuations in entanglement 
in baryon-baryon scattering
was returned to in 2023~\cite{Liu_2023,Liu_2024}.
Considerations of entanglement-power minimization 
was extended to low-energy $\pi\pi$ ($\pi$'s are isospin triplets)
and $\pi N$ scattering by Beane and Farrell~\cite{Beane:2021zvo}.  
It was found that minimizing the entanglement power of the S-matrix in these channels is indistinguishable from the behavior in the large-N$_c$ limit.  However, it was noted that the motivation for 
contracted spin-flavor symmetries from the scaling of $g_A$ are absent.
Further, this analysis was extended to include scattering channels beyond the s-wave, e.g., Refs.~\cite{bai2023NN,Miller:2023ujx,Miller:2023snw}.
An analogous analysis of entanglement power has been applied to $\Omega\Omega$ scattering~\cite{Hu:2025lua}, 
where similar results have been obtained, and systems with other spins~\cite{Hu_2025}.
More recently, a comprehensive study of 
spin-entanglement induced by the nuclear forces, 
beyond the s-wave up to N$^2$LO in Weinberg's power counting has been performed~\cite{Cavallin:2025kjn},
building on the results obtained from examining higher-order terms in the EFT expansion~\cite{Beane:2018oxh}.  
Higher-order terms in the expansion that give rise to fluctuations in entanglement are found to have smaller coefficients than those that do not.
This type of analysis has also been extended to 
multi-nucleon systems~\cite{Bai_2022,Kirchner:2023dvg},
including nucleon-deuteron and deuteron-deuteron scattering~\cite{Kirchner:2023dvg} 
for which corresponding emergent symmetries were not identified,
and $p+{}^3{\rm He}$, $n+{}^3{\rm He}$ where the entanglement power was found to be smaller than in nucleon-nucleon scattering.  
Importantly, conceptual designs of experiments 
have been developed 
to reveal the multi-nucleon spin entanglement structures in nucleon-nucleus collisions~\cite{Bai_2024,Bai:2024omg}.

\begin{figure}[!ht]
  \includegraphics[width=0.8\columnwidth]{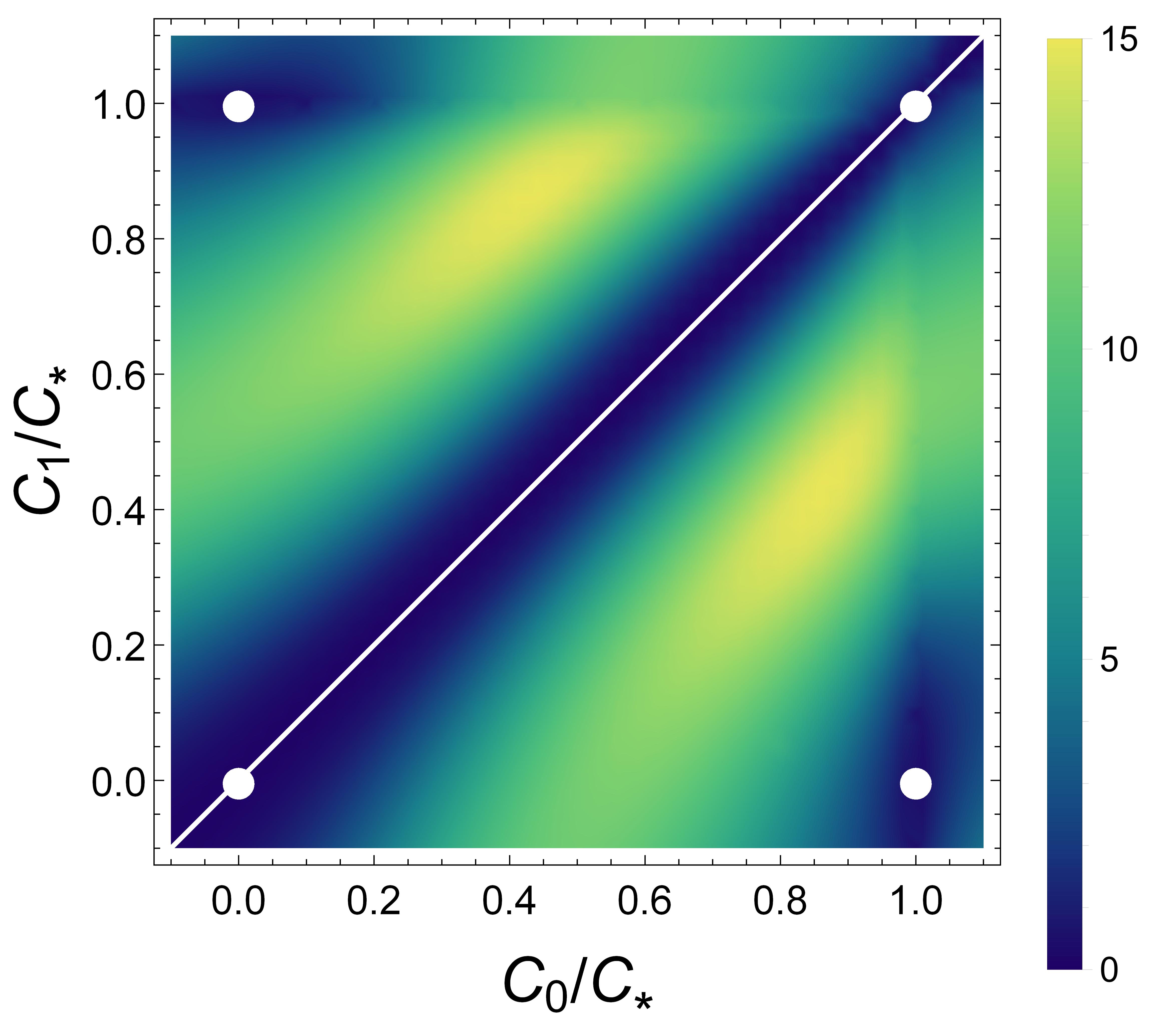}
  \caption{
  The density of entanglement power
    $\overline{\mathcal{E}}(\hat {\bf S})$ of the $S$-matrix integrated
    over center of mass momenta $0\le p\le m_\pi/2$, versus the
    Lagrangian couplings $\overline{C}_0/C_\star$ and $\overline{C}_1/C_\star$ where
    $C_\star$ is the critical coupling for unitary scattering~\cite{Beane:2018oxh}. 
    The entanglement power vanishes
    at the four conformal fixed-points (white points), as
    well as the fixed line corresponding to Wigner $SU(4)$ symmetry
    (white diagonal).
    [Figure reproduced from Ref.~\cite{Beane:2018oxh} with permission from the authors and the American Physical Society.]
    }
  \label{fig:ESBKKS}
\end{figure}

For the s-wave scattering of two spin-$\frac{1}{2}$ particles, 
the scattering operator below inelastic thresholds can be written as
\begin{align}
    \hat {\bf S} & = \frac{1}{4} \left( 3 e^{i 2 \delta_1} + e^{i 2 \delta_0}\right) \ \hat I
    \ +\ 
     \frac{1}{4} \left(  e^{i 2 \delta_1} 
     - e^{i 2 \delta_0}\right) \sum_a \hat \sigma^a\otimes \hat\sigma^a
     \ ,
\end{align}
where $\delta_{0,1}$ are the energy-dependent scattering phase shifts in the spin-0,1 channels.
Mapping the s-wave nucleons to two qubits, 
the entanglement power of the S-matrix~\cite{Beane:2018oxh} 
is defined as the average entanglement induced by $\hat {\bf S}$ 
over the tensor-product 
stabilizer states $\ket{\Psi_i}$~\cite{Robin:2024oqc}
(which form a complex projective 3-design, see section \ref{subsubsec:power}),
\begin{align}
    \overline{\mathcal{E}}(\hat {\bf S}) \equiv \frac{1}{\mathcal{N}_{stab}^{TP}} \sum_{i=1}^{\mathcal{N}_{stab}^{TP}}  \mathcal{E} \left( \rho_i^{(1)}(\hat {\bf S}) \right) 
    \; ,
\label{eq:Entang_Power}
\end{align}
where $\mathcal{E} \equiv \mathcal{E}_{\mathrm{lin}}$ denotes the linear entanglement entropy~\footnote{In this section we use the letter $\mathcal{E}$ to denote quantities related to entanglement entropy (instead of the standard notation $S$ used previously), to avoid confusion with the scattering matrix.},
$\mathcal{N}_{stab}^{TP}$ is the number of tensor-product stabilizer states, and 
$\rho_i^{(1)}(\hat {\bf S}) = \mbox{Tr}_2 \left[ \rho_i^{(12)}(\hat {\bf S}) \right]$  
is the outgoing reduced density matrix for particle-1, obtained by tracing the full outgoing density matrix 
$\rho_i^{(12)}(\hat {\bf S}) = \hat {\bf S} \ket{\Psi_i}\bra{\Psi_i} \hat {\bf S}^\dagger$ over particle-2.
As nucleon-nucleon scattering phase-shifts are precisely known from experimental measurements, and subsequent phase-shift analysis, the entanglement power can be well determined, using the exact solution 
\begin{align}
\overline{\mathcal{E}}(\hat {\bf S}) & = 
\frac{1}{6}\ \sin^2 2\Delta \delta
\ ,
\label{eq:S4NN}
\end{align}
where  $\Delta \delta \equiv \delta_1-\delta_0$,
and
which vanishes when $\delta_1 = \delta_0 + n\frac{\pi}{2}$,
which includes the SU(4) symmetric point where $\delta_1 = \delta_0$,
as seen in the density plot in Fig.~\ref{fig:ESBKKS}.
Interestingly, 
the S matrices at the four conformal fixed points 
($\delta_0=\delta_1=0$; 
$\delta_0=\frac{\pi}{2}, \delta_1=0$; 
$\delta_0=0, \delta_1=\frac{\pi}{2}$; and
$\delta_0=\delta_1=\frac{\pi}{2}$)
are a representation of the Klein four-group,
$Z_2\otimes Z_2$, where the theory has a conformal symmetry~\cite{Beane:2018oxh}.

\begin{figure}[!ht]
    \centering
    \includegraphics[width=0.45\textwidth]{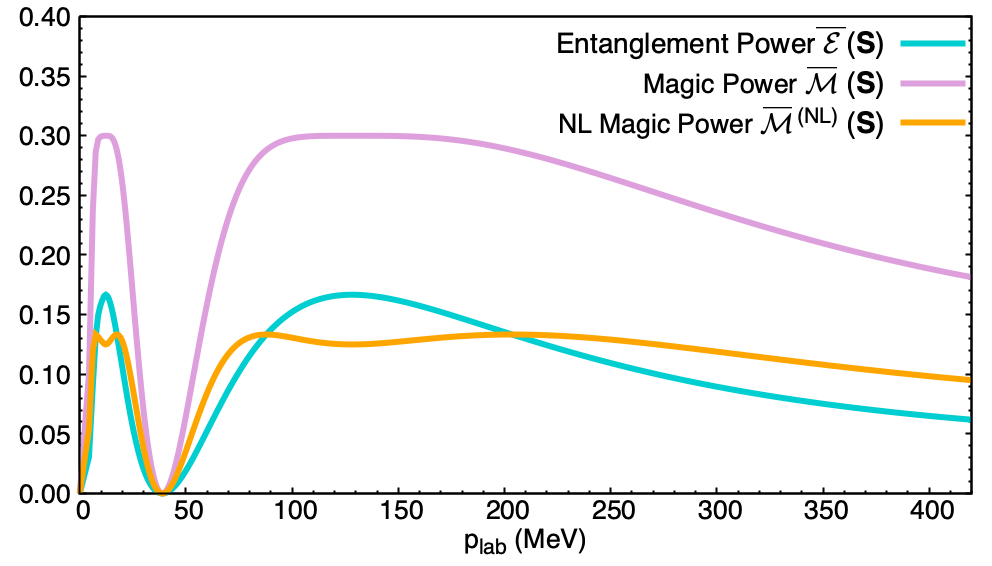}
    \caption{Entanglement power $\overline{{\cal E}}(\hat S)$ (turquoise), 
    total magic power $\overline{{\cal M}}(\hat S)$ (pink),
    and  non-local magic power $\overline{{\cal M}^{(NL)}}(\hat S)$ (orange),
    for low-energy s-wave nucleon-nucleon scattering
    as a function of momentum in the laboratory~\cite{Robin:2024oqc,Robin:2025ymq}, obtained from Nijmegen phase shifts~\cite{PhysRevC.49.2950,NNonline}.
    }
    \label{fig:NNnlm}
\end{figure}

\begin{figure}[!ht]
    \centering
    \includegraphics[width=\columnwidth]{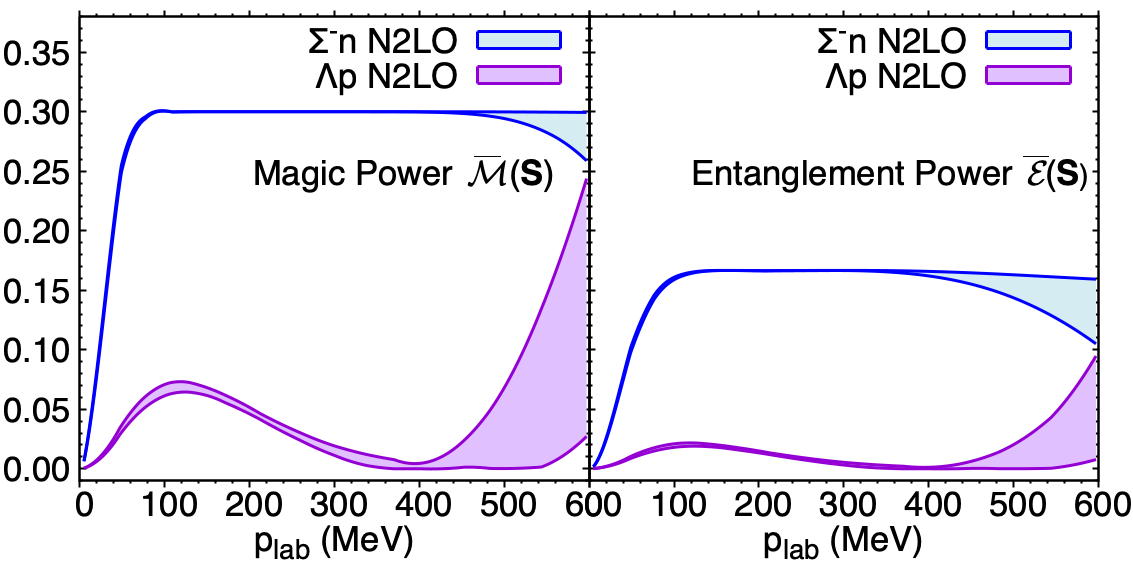}
    \caption{Magic power $\overline{\mathcal{M}}(\hat {\bf S})$ (left panel) and entanglement power $\overline{\mathcal{E}}(\hat {\bf S})$ (right panel) in $\Sigma^-n$ and $\Lambda$p scattering, obtained using N$^2$LO-$\chi$EFT phase shifts from Ref.~\cite{Haidenbauer:2023qhf}.
    Isospin symmetry between $\Sigma^+p$ and $\Sigma^-n$ has been assumed, and Coulomb interactions have been neglected.
    The uncertainty bands represent the maximum and minimum values in magic and entanglement derived from the N$^2$LO phase-shift uncertainty bands~\cite{Haidenbauer:2023qhf}.
    [Figure from Ref.~\cite{Robin:2024oqc} used with permission from the authors under 
{\it Creative Commons Attribution 4.0 International license}~\cite{cc_by_4.0}.]
}
    \label{fig:YN_magic_entang}
\end{figure}

The first studies of fluctuations in magic in nucleon-nucleon and hyperon-nucleon 
scattering were performed in Ref.~\cite{Robin:2024oqc}, 
which calculated 
the linear magic power of the S-matrix, $\overline{\mathcal{M}}(\hat {\bf S}) \equiv \overline{\mathcal{M}}_{\mathrm{lin}}(\hat {\bf S})$, defined according to Eq.~\eqref{eq:magic_power_def}
as the average linear magic induced by scattering operator 
$\hat {\bf S}$ on all $2$-qubit stabilizer states $\ket{\Psi_i}$:
\begin{align}
    \overline{\mathcal{M}}(\hat {\bf S}) \equiv \frac{1}{\mathcal{N}_{stab}} \sum_{i=1}^{\mathcal{N}_{stab}}  \mathcal{M} \left( \hat {\bf S} \ket{\Psi_i} \right) \; ,
\label{eq:Magic_Power}
\end{align}
where $\mathcal{N}_{stab}=60$ is the number of $2$-qubit stabilizer states,
and $\mathcal{M} \left( \hat {\bf S} \ket{\Psi_i} \right)$ is the linear magic in the scattered state defined in Eqs.~\eqref{eq:2SRE} and \eqref{eq:linSRE}.
In terms of the s-wave scattering phase shifts,
the linear magic power is found to be
\begin{align}
\overline{{\cal M}} (\hat S)  &= 
\frac{3}{20} \Bigl( 3 + \cos 4\Delta \delta \Bigr) \sin^2 2\Delta \delta
\ .
\label{eq:Magic_Entang_Power_NN_2q}
\end{align}
The entanglement and magic power in nucleon-nucleon (neutron-proton) scattering are displayed in Fig.~\ref{fig:NNnlm} (turquoise and pink curves) as a function of laboratory momentum, p$_{\rm lab}$,
and those for hyperon-nucleon scattering are shown in Fig.~\ref{fig:YN_magic_entang}.
They have been obtained from experimentally determined phase shifts~\cite{PhysRevC.49.2950,NNonline}, and phase shifts derived from $\chi$EFT~\cite{Haidenbauer:2023qhf}, respectively.
It is seen that the entangling and magic powers in the nucleon-nucleon and nucleon-hyperon sectors are significantly different.
The behavior of the magic power in low-energy $n\Sigma^-$ scattering, in particular, is striking.   
The phase shifts are such that the magic power rapidly rises to its maximum value 
and remains there over a large range of energies.   
The same behavior is seen in the entanglement power.
This invites speculations that, using a simple model without decoherence, 
$\Sigma n$ scattering may provide a  potential catalyst for spreading quantum complexity 
in dense matter via successive neutron scatterings~\cite{Robin:2024oqc}. 
\\

The non-local magic power and anti-flatness power 
of the S-matrix for 2-body scattering~\cite{Robin:2025ymq} 
can be defined in a way analogous to the magic power 
(Eq.~(\ref{eq:Magic_Power})).
To further isolate the induced changes of entanglement, 
the Hilbert-space averaging is restricted to the 36 tensor-product stabilizer states. 
That is,
\begin{align}
   \overline{\mathcal{M}^{(NL)}}(\hat {\bf S}) &\equiv \frac{1}{36} \sum_{i=1}^{36}  \mathcal{M}^{(NL)} \left( \hat {\bf S} \ket{\psi_i} \right) \; , 
   \label{eq:NL_Magic_Power}\\
    \overline{\mathcal{F}_A}(\hat {\bf S}) &\equiv \frac{1}{36} \sum_{i=1}^{36}  \mathcal{F}_A \left( \hat {\bf S} \ket{\psi_i} \right) \; .
\label{eq:AF_Power}
\end{align}
For s-wave nucleon-nucleon scattering,
direct optimization over the six Euler angles defining the two Bloch spheres 
of nucleon spins
yields a non-local magic power of
\begin{eqnarray}
4\ \overline{{\cal F}_A} (\hat S)   & = &
 \overline{{\cal M}^{(NL)}} (\hat S)  \nonumber \\
& = &
\frac{1}{48}\left( 11 + 5 \cos 4 \Delta\delta\right) \sin^2 2\Delta\delta
\ ,
\label{eq:NLMagNN}
\end{eqnarray}
which is equal to four times the anti-flatness power~\footnote{
This is found to be true for any two-qubit state $\ket{\psi}$~\cite{Robin:2025ymq}, 
{\it i.e.} $4\ \mathcal{F}_A (\ket{\psi}) = \mathcal{M}_{\rm lin}^{(NL)}(\ket{\psi})$. 
}.
This is to be compared with the total magic power of s-wave 
nucleon-nucleon scattering~\cite{Robin:2024oqc}, given in 
Eq.~(\ref{eq:Magic_Entang_Power_NN_2q}).
The non-local magic power for neutron-proton scattering
is shown in Fig.~\ref{fig:NNnlm}. 
It is seen that in the regime $p_{lab} \lesssim 150 $ MeV, 
only about one
third of the magic generated is non-local, while the rest can be eliminated via local basis transformations.
In hyperon-nucleon scattering, non-local magic is found to capture less than
half of the total magic, and this fraction remains constant even at large momenta~\cite{Robin:2025ymq}.
\\

For 2-qubit scattering the stabilizer states, displayed in  
Table~\ref{tab:TwoQstabs} in App.~\ref{app:stabs},
can be divided into 3 groups, with all states in each group yielding the same 
entanglement, total magic, and non-local magic after scattering~\cite{Robin:2024oqc}.
Further, it was found that the total linear magic is proportional to the non-local magic (and thus anti-flatness), with a different proportionality factor for each group~\cite{Robin:2024oqc}.

Each group includes tensor-product initial states,
which is useful
from the standpoint of conceptual designs for experiments, as they can be easily prepared.
Tensor-product stabilizer states representative of each group, 
denoted by $|\psi_{1,2,3}\rangle$, 
that are candidates for initial-state preparation are
\footnote{
Experimentally, the wavefunction associated with the selected Group 3 state can be prepared by a single-spin $\pi/2$ rotation about the y-axis.
},
\begin{eqnarray}
&& |\psi_1\rangle \ =\ |\uparrow\rangle \otimes |\uparrow\rangle 
\ \ ,\ \ 
 |\psi_2\rangle \ =\ |\uparrow\rangle \otimes |\downarrow\rangle  
 \ \ , \ \ \nonumber \\
 && |\psi_3\rangle \ =\ 
\frac{1}{\sqrt{2}} \left[\ |\uparrow\rangle + |\downarrow\rangle \ \right]\otimes |\uparrow\rangle 
 \ .
 \label{eq:NNstabs}
\end{eqnarray}

Returning to the connection between symmetries and fluctuations in entanglement.
It is interesting to note that Nature's proximity to points of emergent symmetries, 
Wigner's SU(4) and conformal symmetry for two flavors of light quarks, 
as well as 
SU(16) symmetry for three flavors,
which have been shown to be the result of minimizing the entanglement power in the spin sector 
of the low-energy S-matrix~\cite{Beane:2018oxh},
could also be attributed to the minimization of 
non-local magic power. \\

As $s$-wave scattering is valid at low energies, works to go beyond this approximation and include higher-partial waves have been accomplished in the context of entanglement within phenomenological or $\chi$EFT frameworks, see {\it e.g.}, Refs.~\cite{Beane:2018oxh,bai2023NN,Miller:2023ujx,Cavallin:2025kjn}.

\begin{figure}[ht!]
\centering
\includegraphics[width=\columnwidth]{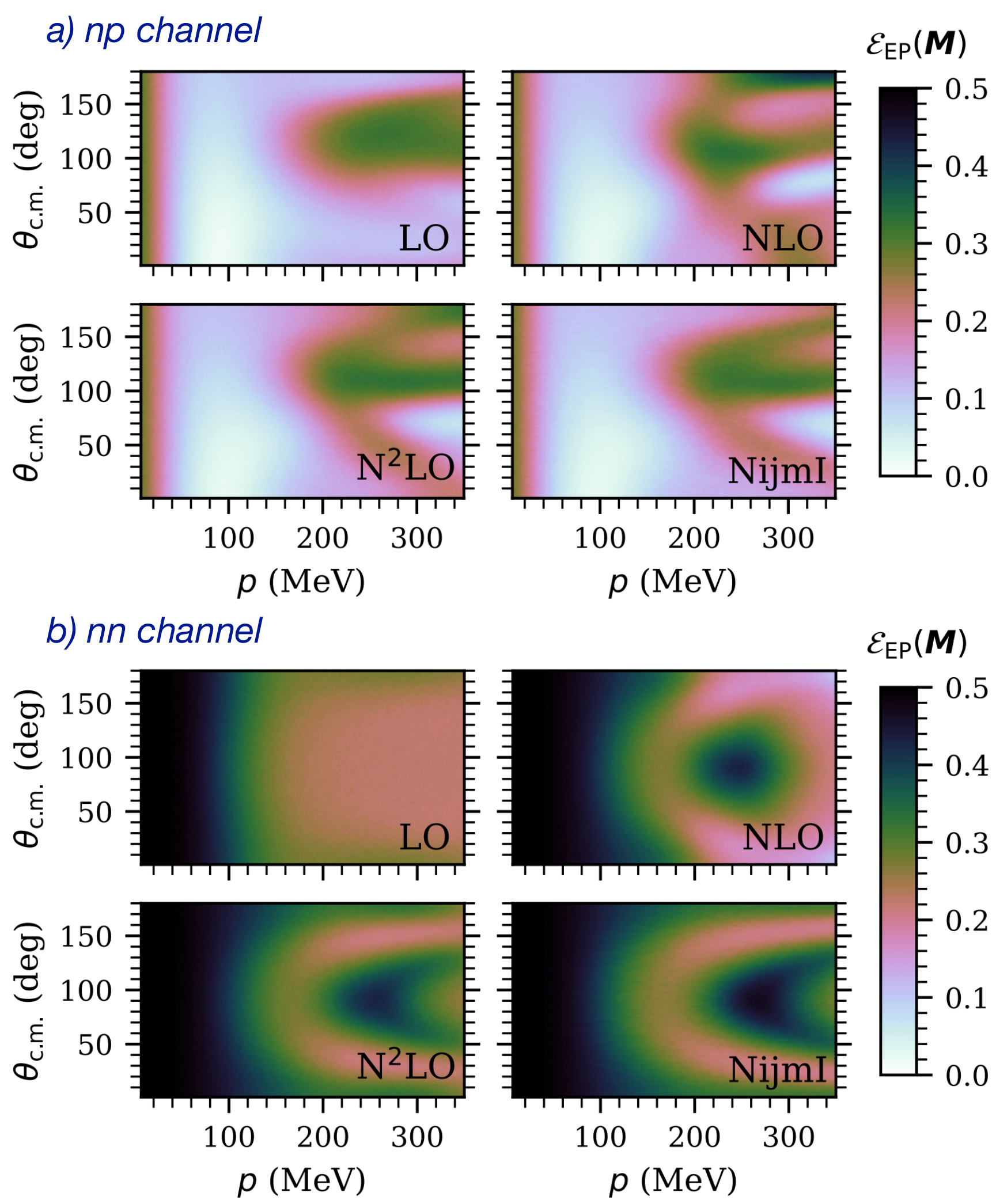} 
\caption{
Entanglement powers for the $np$ (upper panel a)) and $nn$ (lower panel b)) systems as a function of the scattering angle $\theta_{\rm{c.m.}}$ and momentum $p$ in the center-of-mass frame.
The displayed results are obtained using the WPC potential from Ref.~\cite{PhysRevX.6.011019} at different  orders in the chiral expansion, and using the phenomenological NijmI potential from Ref.~\cite{PhysRevC.49.2950}.
[Figure adapted from Ref.~\cite{Cavallin:2025kjn} with permission from the authors under {\it Creative Commons Attribution 4.0 International license}~\cite{cc_by_4.0}]
}
\label{fig:Cavallin}
\end{figure}

Investigations into connections among accidental symmetries and spin-entanglement fluctuations 
within NN scattering in the context of $\chi$EFT have been performed~\cite{Cavallin:2025kjn},
building upon the earlier consideration of higher-dimensional operators~\cite{Beane:2018oxh}.
Using the full S-matrix, the authors performed a detailed analysis of the effect of Wigner-$SU(4)$ and Serber symmetry-breaking terms in the chiral potential at leading order (LO) in Weinberg power counting (WPC)~\cite{Weinberg:1990rz,vanKolck:1994yi,Ordonez:1995rz}.
\footnote{For reviews of chiral nuclear forces, see,
{\it e.g.},
Refs.~\cite{Bedaque:2002mn,Epelbaum:2008ga,Evgeny2020a,Tews:2022yfb,machleidt2024},
and for alternate power-counting schemes, see, {\it e.g.},
Refs.~\cite{Kaplan:1998tg,Kaplan:1998we,Beane:2001bc,Gantenberg:2026xzy}. }
Order-by-order up to N$^2$LO,
spin-entanglement generation 
was systematically studied in both neutron-proton ($np$) and neutron-neutron ($nn$) channels. 
The entanglement powers obtained from chiral potentials at different orders~\cite{PhysRevX.6.011019} 
in the expansion are shown in Fig.~\ref{fig:Cavallin}. 
The results are compared to those obtained from the NijmI phenomenological potential~\cite{PhysRevC.49.2950}.
The $np$ and $nn$ channels exhibit substantially different profiles of entanglement power,
which are seen to significantly evolve from LO to NLO and N$^2$LO in the chiral expansion, especially in the $nn$ channel. At N$^2$LO the profiles largely resemble those obtained from the phenomenological potential.
In the $np$ channel, consistent with previous studies, the enhanced entanglement power at low-energy was understood to be related to Wigner-symmetry-breaking parts of the contact term, 
while the tensor term was found to be the main driver for pronounced entanglement power at higher momenta $p \gtrsim 150$ MeV. The important role of the tensor term on entanglement power has also been noted~\cite{Beane:2018oxh,Miller:2023ujx}.
The suppression of induced entanglement fluctuations in the region $ 50 \lesssim p \lesssim 150$ MeV is seen to persist for all scattering angles, indicating approximate Wigner symmetry, while it is not found in the $nn$ channel due to the Pauli principle which significantly entangles the neutrons (as was also noted in Ref.~\cite{Beane:2018oxh} and elaborated upon in Ref.~\cite{Hu:2025lua}).
Natural extensions of this work would be to investigate the (non-local) magic powers of chiral potentials.

Overall, the aforementioned studies point towards possible power counting schemes, based upon entanglement and magic generations.
\\

Going from in-vacuum nucleon-nucleon interactions to low-energy many-body environments, nuclear forces can be flowed via SRG transformations that decouple high- and low-momentum modes.
In nuclear many-body calculations, rather than working in spin, isospin and momentum (or position) representation, one typically uses bases of Slater determinants (occupation number representation), which are naturally anti-symmetrized under permutation.
Due to this change of representation, the information encoded into the initial operators is then translated into a drastically larger set of interaction matrix elements which connect initial and final nucleon states~\cite{Zhu:2021pis}.
This indicates that irrelevant information is introduced, that could be subsequently compressed.

\begin{figure}[ht!]
\centering
\includegraphics[width=\columnwidth]{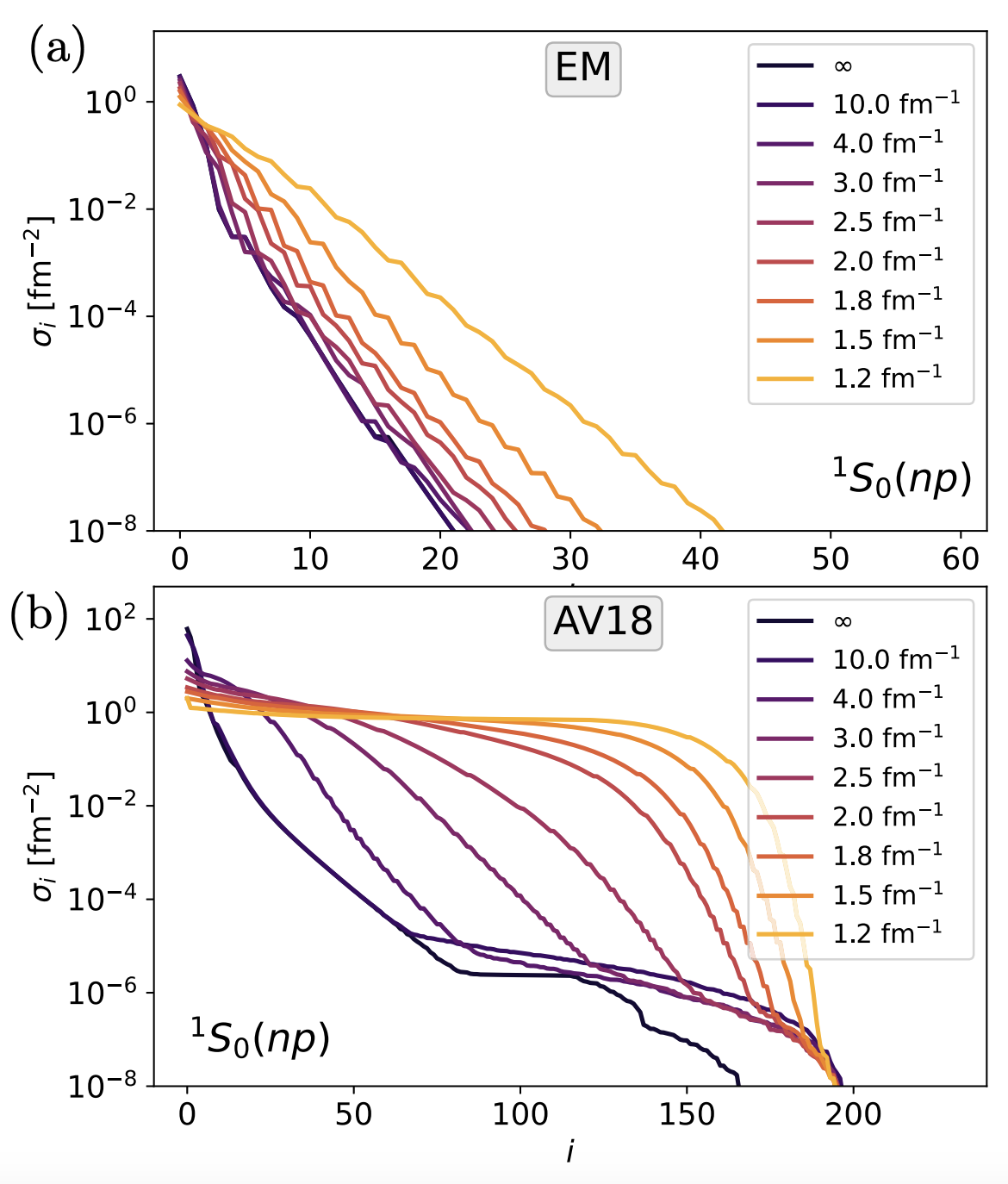} 
\caption{Singular value spectra of the neutron-proton $^1 S_0$ partial wave
for the SRG-evolved 
(a)
Entem-Machleidt (EM) interaction derived from $\chi$EFT~\cite{PhysRevC.68.041001}
and (b)
phenomenological AV18 interaction~\cite{PhysRevC.51.38} 
at different resolution scales $\lambda$.
[Figure reproduced from Ref.~\cite{Zhu:2021pis} with permission from the authors and the American Physical Society.]
}
\label{fig:SVD_NN_int}
\end{figure}

In an effort to reduce computational storage requirements in nuclear structure calculations, Refs.~\cite{Tichai:2021rtv,Zhu:2021pis} performed SVDs of nucleon–nucleon interactions in a partial-wave representation and analyzed how the singular-value spectrum evolves under SRG, thereby tracking the evolution of information along the RG flow.
They showed that interactions derived from $\chi$EFT present a rapidly decaying singular value spectrum, allowing for low-rank truncations while preserving two-nucleon observables to good accuracy. 
The behavior of the singular value spectrum was found to be largely unchanged under SRG evolution~\cite{Glazek:1993rc,Wegner:1994,Bogner:2006pc}, 
up to $\sim 1.8 \; \mathrm{fm}^{-1}$, before entering the domain of pion dynamics. 
This is shown in the top panel of Fig.~\ref{fig:SVD_NN_int}.
This confirms that the information contained in the original $\chi$EFT is already restricted to the relevant low-energy physics, making it sparse and largely compressible. As long as a clear separation of scales persists, only a limited amount of relevant information is flowed through SRG. 
Once the pion dynamics regime is approached the integrated information becomes larger, and thus the rank increases.

The fall-off of singular values was found to be much slower for interactions that are local in space, see lower panel of Fig.~\ref{fig:SVD_NN_int}, 
making low-rank truncations ineffective in that framework~\cite{Tichai:2021rtv,Zhu:2021pis}. 
As these interactions have greater 
extension in momentum space (hard core), strategies for SVD in coordinate space would be more appropriate~\cite{Tichai:2021rtv}.
Furthermore, these works also found that the required rank appears to be largely uniform across all partial waves, suggesting that the relevant singular components could, in fact, be projections of a small set of underlying operators into these different channels~\cite{Zhu:2021pis}. 
The nature of these operators remains an open question.

%% file: Section_Nuclei.tex
\subsection{Many-Body Nuclear Structure}
\label{sec:Qcomplex_nuclei}

As prime example of mesoscopic systems, atomic nuclei exhibit structure and properties that emerge from a delicate interplay between quantum and classical effects. They display features common to other many-body systems, ranging from single-particle shell structure, to collective phenomena as superfluid pairing, deformation and quantum shape fluctuations.
In recent years, significant effort has been devoted to understanding how various correlations and phenomena are associated with different forms of quantum complexity. A range of studies has explored these aspects, from bipartite entanglement in both pure and mixed states along isotopic chains towards instability and drip-lines, to recent analyses of multipartite entanglement and non-stabilizerness, to investigate the build up of collectivity.

\paragraph{Proton-neutron bipartite entanglement}

As nuclei are two-species systems, 
their wave functions feature a natural bi-partitioning in terms of proton and neutron subsystems,
\begin{equation}
\ket{\Psi} = \sum_{\alpha_\pi=1}^{d_\pi} \sum_{\alpha_\nu=1}^{d_\nu} C_{\alpha_\pi \alpha_\nu} \ket{\Phi_{\alpha_\pi}} \otimes \ket{\Phi_{\alpha_{\nu}}} \; ,
\label{eq:nuclear_wf_bipartite}
\end{equation}
where $\ket{\Phi_{\alpha_\pi}} $ and $ \ket{\Phi_{\alpha_{\nu}}}$ denote proton and neutron many-body configurations, respectively.
Tracing the density matrix of the full state $\rho=\ket{\Psi} \bra{\Psi} $ over, 
{\it e.g.}, the proton sector, 
one obtains the neutron reduced density matrix (RDM) $\rho_\nu = \text{Tr}_\pi \rho$, which
provides information on how proton and neutron subsystems are entangled with each other in a collective global manner. 

\begin{figure}[ht]
\centering
\includegraphics[width=\columnwidth]{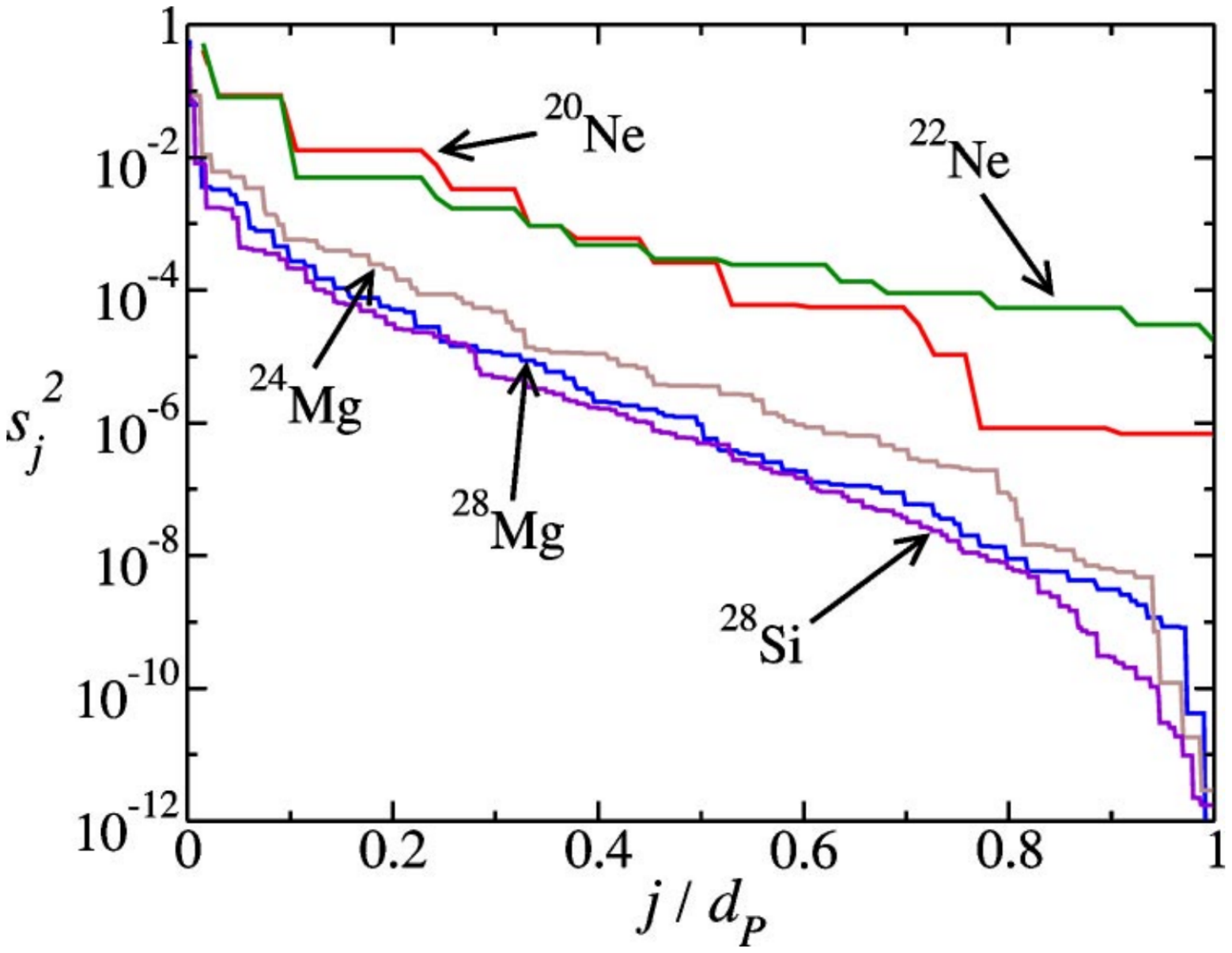}
\caption{Singular values $s_j$ (eigenvalues of $\rho_\nu$) for ground states of a few nuclei in the $sd$ and $pf$ shells. The dimension of the proton many-body space is denoted $d_{\mathrm{p}}$, which corresponds to $d_\pi$ in the main text. Note that $^{20}$Ne, $^{24}$Mg and $^{28}$Si have $N=Z$.
[Figure reproduced from Ref.~\cite{Papenbrock:2003az} with permission from the authors and the American Physical Society.]
}
\label{fig:SVD_nuclei}
\end{figure}

In the early 2000's, 
Papenbrock and Dean computed the proton-neutron entanglement spectrum (eigenvalue spectrum of $\rho_\nu$) of nuclear ground states in the 
$sd$ and $pf$ shells~\cite{Papenbrock:2003bj,Papenbrock:2003az}. 
The wave functions were obtained within the interacting shell model (ISM) framework~\cite{BROWN2001517,Caurier:2004gf}, 
{\it i.e.},
via exact diagonalization in these active valence spaces, 
using high-accuracy empirical interactions.
Focusing mainly on $N=Z$ nuclei, 
a rapid fall-off of the singular value spectrum was found, 
hinting at the fact that the RDM is dominated by a few contributions and thus, 
bipartite proton-neutron entanglement is relatively low in these nuclei. 
Examples are shown in Fig.~\ref{fig:SVD_nuclei}.
This result may appear, at first, as somewhat surprising, as nuclei with similar numbers of protons and protons, are expected to have increased contributions of the proton-neutron interaction.

\begin{figure}[ht]
\centering
\includegraphics[width=\columnwidth]{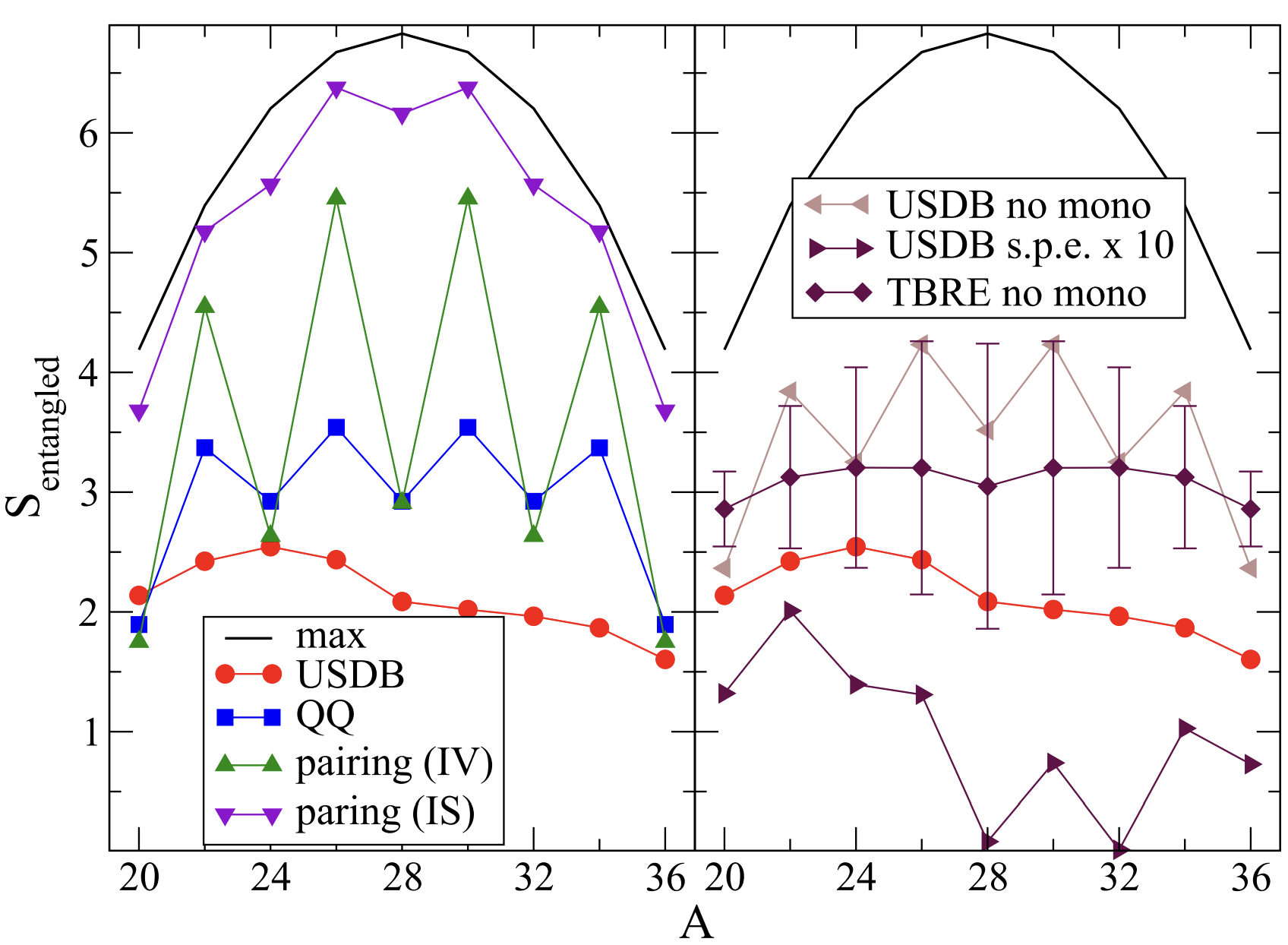}
\caption{Proton-neutron entanglement entropy in $N=Z$ $sd$-shell nuclei as a function of total nucleon number A$=N+Z$. The entropies are calculated with the natural logarithm $S_{pn} \equiv S_{\mathrm{entangled}}= - \text{Tr} \left( \rho_\nu \, \text{ln} \rho_\nu \right)$. 
The black curve shows the maximum possible entanglement entropy $S_{pn}^{\mathrm{max}}$ based on dimensionalities.
The left panel shows the entanglement $S_{pn}$ for the high-quality empirical interaction USDB~\cite{Brown:2006gx}, as well as for schematic attractive isoscalar quadrupole–quadrupole ("QQ") interaction, isovector ("IV") and isoscalar ("IS") pairing interactions.
The right panel displays results for USDB with single-particle energies and
monopole interactions set to zero, which eliminates shell structure ("no
mono"), and with single-particle energies multiplied by a factor of ten, which amplifies shell
structure ("s.p.e. x 10"). 
The right panel also gives the average and standard deviation for calculations drawn from a two-body random ensemble~\cite{Johnson:1998uj}, with shell structure eliminated ("TBRE no mono").
[Figure from Ref.~\cite{Johnson:2022mzk} used with permission from the authors under {\it Creative Commons Attribution 4.0 International license}~\cite{cc_by_4.0}.]
}
\label{fig:nuclei_pn_entropy}
\end{figure}

Proton-neutron entanglement in shell model wave functions was recently further investigated by Gorton and Johnson~\cite{Johnson:2022mzk}, 
who computed proton-neutron von Neumann entanglement entropies and found values 
significantly below the maxima allowed by the dimensionalities of the system. In $N=Z$ $sd$-shell nuclei, for example, the ratio $S_{pn}/S_{max}$ ranges between $\simeq 0.3 - 0.5$, and decreases with neutron excess. 
This can be seen in the left panel of Fig.~\ref{fig:nuclei_pn_entropy}.

These findings were directly tied to the nature of the mean-field and shell-model effective interactions. 
In particular, the single-particle energies were found to play a driving role in lowering the entanglement entropies (as they weaken the effect of the interaction). This is illustrated in the right panel of Fig.~\ref{fig:nuclei_pn_entropy}. 
This observation is in accordance with Ref.~\cite{Robin:2020aeh}, 
who recently studied the impact of single-particle basis optimization on minimizing orbital entanglement in a no-core {\it ab-initio} context (see section~\ref{sec:QI_driven_nuclear_methods} below). 
The role of the underlying mean field in reducing complexity was also pointed out in early studies of the Shannon information entropy~\cite{ZELEVINSKY1995141,PhysRevLett.74.5194}.

On the other hand, the interaction, which induces correlations between nucleons, competes with the simple mean field picture and tends to increase the entanglement entropy. 
With the goal of understanding what drives the entanglement, the entropy generated by schematic isoscalar (IS) and isovector (IV) pairing interactions, 
as well as a quadrupole-quadrupole (QQ) force, 
was studied~\cite{Johnson:2022mzk}.
\footnote{These are the most important terms in a multipole decomposition of the nuclear force~\cite{Dufour:1995em}, 
and drive proton-neutron pairing in the $T=0$ channel, like-particle and proton-neutron pairing in the $T=1$ channel, and collective deformation. }
The results are shown in the left panel of Fig.~\ref{fig:nuclei_pn_entropy}. 
Clearly, the isoscalar pairing alone generates near-maximal proton-neutron entanglement, 
while other channels give rise to smaller values and large odd-even staggering effects. 
Yet, the physical USDB interaction yields the smallest values, hinting at the fact that the $T=0$ pairing channel acts with limited strength in nuclei, 
but competition between different channels may lead to relative cancellations.
Ref.~\cite{Shinde:2025xud} later performed a decomposition of the same USDB interaction into central, spin-orbit and tensor terms. 
By investigating $N=Z$ nuclei  the authors found that the central term generates the largest amount of proton-neutron entanglement, and that, while the tensor term acts constructively, the spin-orbit term tends to act destructively by lowering the entropy, with stronger effects in systems with large nucleon numbers.
Further, 
proton-neutron bi-partitioning typically leads to smaller entanglement ratios over the maximal values, compared to other 
equal-sized bi-partitions, such as the one separating time-reversed single-particle states, 
{\it i.e.}, breaking BCS pairs~\cite{Perez-Obiol:2023wdz}. 
This again highlights the prominent role of the like-particle $T=1$ pairing interaction over the proton-neutron channel. 

More broadly, understanding which channels of the underlying interaction generate different aspects of complexity, in both phenomenological and {\it ab-initio} frameworks, 
is expected to be valuable both for tackling the problem and deriving nuclear interactions.
\\

\begin{figure}[h]
\centering
\includegraphics[width=.8\columnwidth]{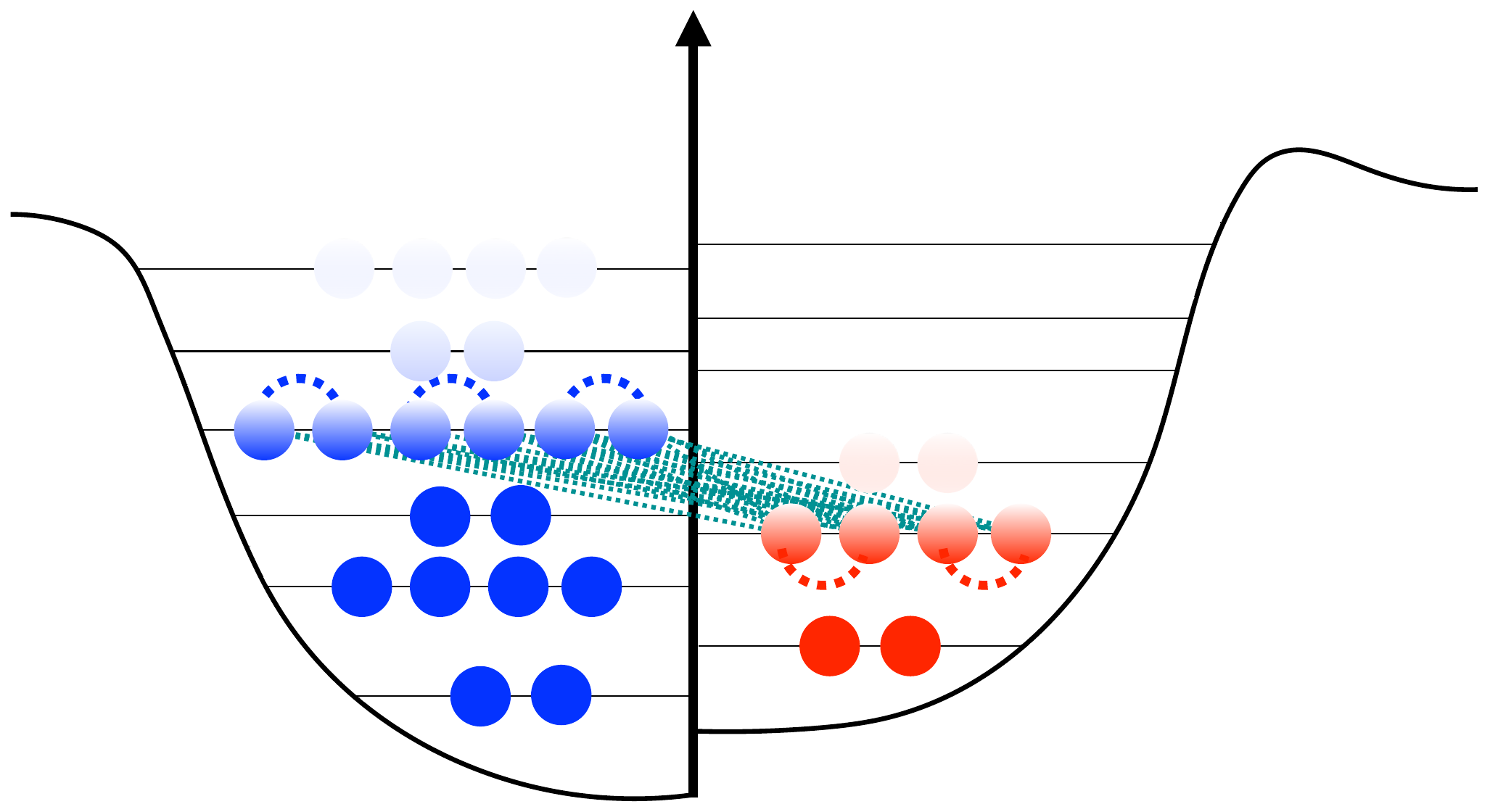}
\caption{ The proton-neutron interaction unlocks more communication channels than in the proton-proton and neutron-neutron sector. } 
\label{fig:info_pn_shellmodel}
\end{figure}

Away from $N=Z$ nuclei, 
for ISM calculations in a single active shell, the proton-neutron entanglement further decreases with neutron excess, as the neutron active space becomes filled, and that this behavior is characteristic of the effective nuclear force, as it is not observed for a random interaction~\cite{Johnson:2022mzk}. 
This  line of work was recently extended to larger cross-shell model spaces ($sd-pf$) with IMSRG interactions to study the island of inversion towards the neutron drip-line~\cite{Shinde:2026hus}. 
Interestingly, the proton-neutron entanglement increases as neutron excess becomes large, 
indicating that neutrons become further entangled with the protons for $N \gg Z$. 
\begin{figure}[ht]
\centering
\includegraphics[width=\columnwidth]{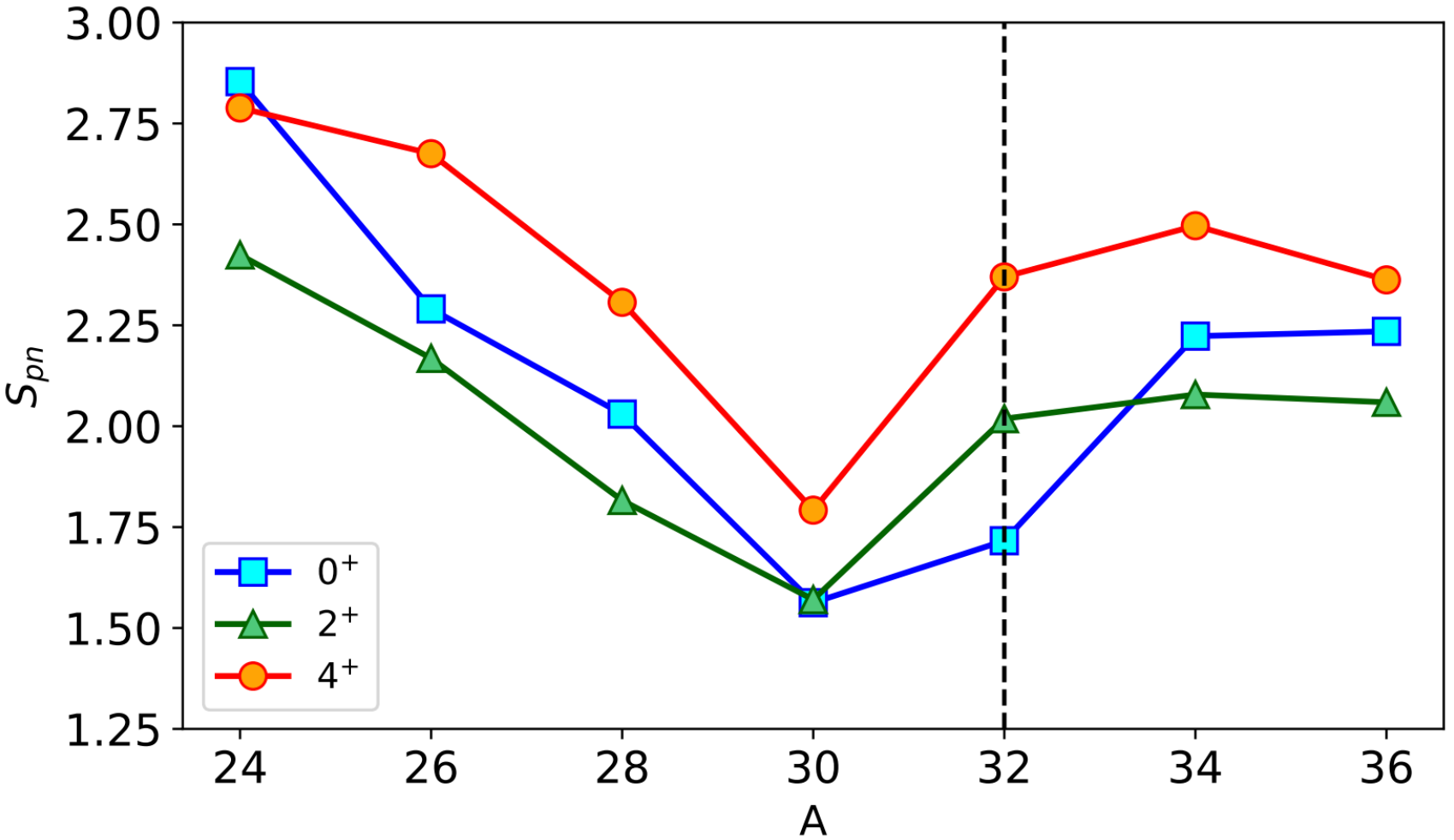}
\caption{Proton-neutron entanglement entropy $S_{pn} = - \text{Tr} \left( \rho_\nu \, \log_2 \rho_\nu \right)$ in the Mg isotopic chain ($Z=12$), as a function of total nucleon number A. The vertical line indicates the neutron shell closure at $N=20$. Results are shown for the ground state with angular momentum and parity 
$J^{\text{p}} = 0^+$, as well as for the first $2^+$ and $4^+$ excited states.
The non-zero entropy observed at $N=20$ signals the disappearance of this shell closure, characteristic of the island of inversion.
[Figure from Ref.~\cite{Shinde:2026hus} used with permission from the authors under {\it Creative Commons Attribution 4.0 International license}~\cite{cc_by_4.0}.]
}
\label{fig:nuclei_pn_entropy_inversion}
\end{figure}
This is shown in Fig.~\ref{fig:nuclei_pn_entropy_inversion} for the Mg isotopic chain,
and
attributed to the spreading of neutron occupation number, 
thus fragmenting the many-body wave function and increasing entanglement entropies.
The correlations reflected in the fractional occupations number indeed signal information sharing with other components of the system, which appears to be facilitated in the proton-neutron channel. 
Intuitively, this can be understood as follows: when neutrons predominantly occupy a single shell, they can, to a good approximation, only entangle with other neutrons within the same shell, since the remaining orbitals are effectively either fully occupied or empty. A similar “information blocking” mechanism applies to protons within their own sector. In contrast, the proton–neutron channel provides a wider range of possibilities for information exchange through the proton–neutron interaction. 
A schematic illustration is shown in Fig.~\ref{fig:info_pn_shellmodel}.
As discussed further below, this results in multipartite proton–neutron entanglement dominating over multi-proton and multi-neutron entanglement due to enhanced collectivity.

\paragraph{Orbital entanglement: from few to many, towards the emergence of collectivity}
The bipartite proton-neutron entanglement studied above has allowed for connections between traditional concepts in nuclear structure and entanglement properties of shell-model nuclei, which in turn has motivated the development of improved shell model approaches based on proton-neutron factorization (see section~\ref{sec:QI_driven_nuclear_methods} below). 
Bi-partitioning between global proton and neutron sectors however somewhat hides the information about entanglement between the various smaller degrees of freedom, such as individual nucleons or orbitals. 
Understanding how collectivity emerges from individual constituents, or, more precisely, how these constituents organize themselves and share information in a collective manner, requires partitioning the system into smaller sub-parts.
To that aim, one can write the nuclear state in terms of occupation numbers of the nucleon orbitals (modes), as
\begin{align}
& \ket{\Psi}  = \sum_{n_1...n_M}  C_{n_1...n_M} \ \ket{n_1...n_M}   
\; , 
\nonumber \\
& = \sum_{n^\pi_1...n^\pi_{M_\pi}}  \sum_{ n^\nu_1...n^\nu_{M_\nu}} C_{ n^\pi_1...n^\pi_{M_\pi} n^\nu_1...n^\nu_{M_\nu}  } \ \ket{n^\pi_1...n^\pi_{M_\pi} n^\nu_1...n^\nu_{M_\nu}   } \; ,
\label{eq:Nuclear_wf_mode}
\end{align}
where $n_i \in \{0,1\}$ denotes the occupation number of orbital $i$. The superscripts $\pi$ and $\nu$ are used to differentiate between proton and neutron orbitals, respectively. 
The number of proton (resp. neutron) orbitals is given by $M_\pi$ (resp. $M_\nu$), and $M_\pi + M_\nu = M$.
Partitioning the wave function in Eq.~\eqref{eq:Nuclear_wf_mode} in different ways reveals how various orbitals are correlated with one another within the nuclear state.
This type of orbital entanglement, formulated in Fock space, has been shown to be relevant for systems of identical fermions, as it circumvents partitioning ambiguities arising from the anti-symmetrized tensor-product structure of the underlying Hilbert space~\cite{Benatti:2014gaa}. Moreover, it is the natural notion of entanglement in the context of classical configuration-interaction methods, such as the ISM or no-core shell-model in nuclear physics, as well as quantum algorithms, that often map orbital occupation numbers to single bits or qubits.

Up to now, orbital entanglement (or mode entanglement) has been mostly studied in a bipartite manner, in pure states via single-orbital entanglement entropy, 
to quantify 
the entanglement of each orbital with the rest of the system,
or in mixed states via two-orbital mutual information, negativity or concurrence, to quantify how two orbitals are correlated within the full nuclear system. 
It is found that such measures are able to identify the emergence of shell closures, and to be connected to detailed aspects of correlations related to particular channels of the nuclear force.

\begin{figure}[ht]
\centering
\includegraphics[width=\columnwidth]{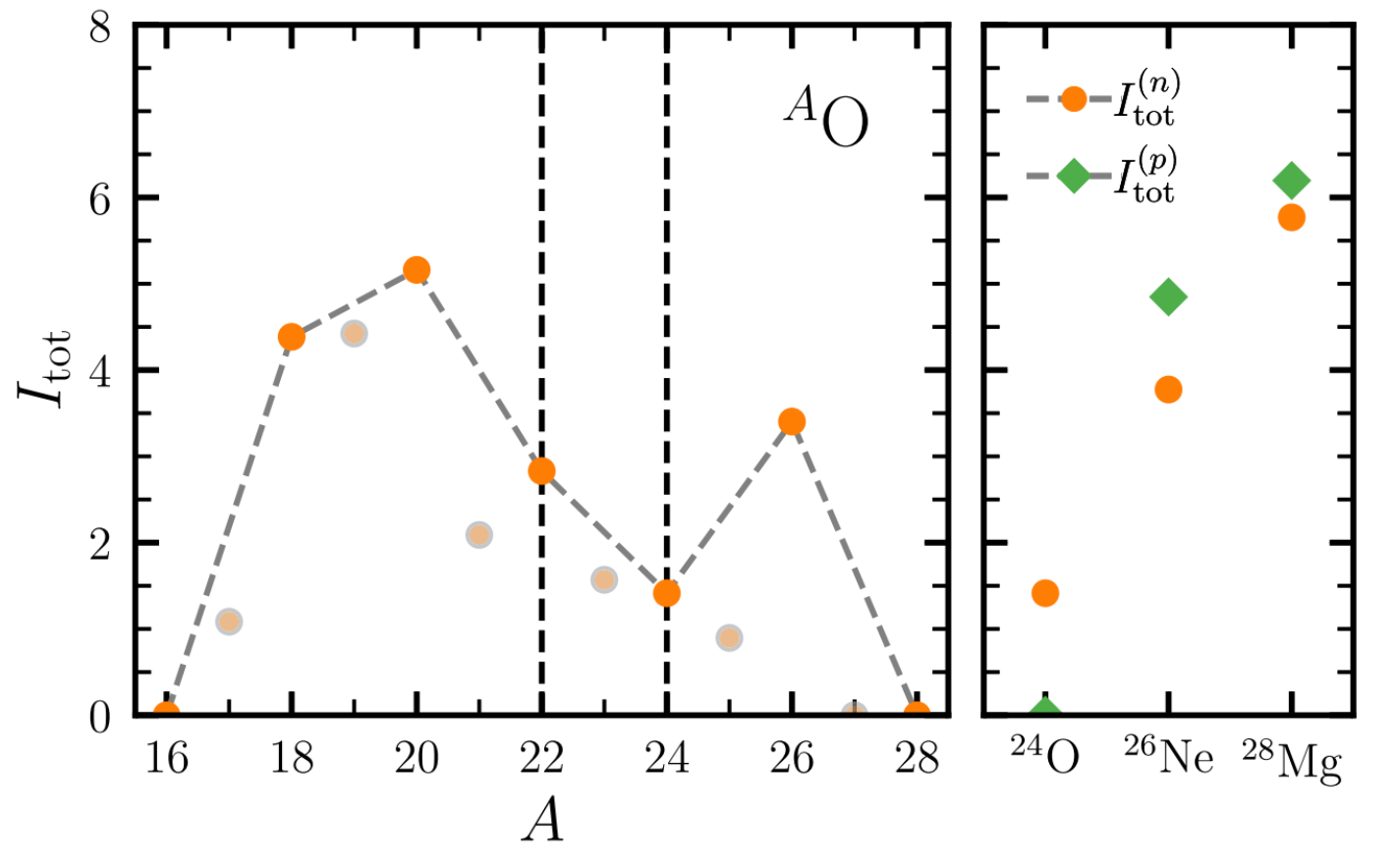}
\caption{Left: Total one-orbital neutron entropies $I_{tot}^\nu = \sum_{i=1}^{M_\nu} S_{i}$ where $S_i = - \mathrm{Tr} \left( \rho_i \, \mathrm{ln} \rho_i \right)$, from valence-space DMRG calculations for the oxygen chain ($Z=8$). The low entropy reveals the emergence of the shell closure at $N=16$, indicated by the dashed vertical line. To a lesser extent, the sub-shell closure at $N=14$ also appears. The lighter symbols show the results for odd-mass nuclei.
Right: Total proton and neutron one-orbital entropies for nuclei with $N=16$, and increasing proton number $Z$. 
[Figure from Ref.~\cite{Tichai:2022bxr} used with permission from the authors under {\it Creative Commons Attribution 4.0 International license}~\cite{cc_by_4.0}.]
}
\label{fig:Oshell_Itot}
\end{figure}

For example, Ref.~\cite{Robin:2020aeh} performed {\it ab-initio} no-core calculations of $^6$He and used mutual information to reveal the emergence of a core+valence structure, with a $^4$He core and two neutrons in the $1p$-shell, when using an optimized single-particle basis that encompasses two-body correlations effects of the system. 
The emergence of shell closures as a signature of orbital entanglement entropies was also found 
in {\it ab-initio} valence-space computations of medium-mass nuclei with the  DMRG~\cite{Tichai:2022bxr}, 
see Fig.~\ref{fig:Oshell_Itot}.

Measures of two-orbital mutual information have been found 
to be useful in pinning down 
detailed correlations corresponding to various channels of the underlying interaction, such as quadrupole or pairing interaction, between like-particles or in the proton-neutron sector,
see {\it e.g.}, Refs.~\cite{Legeza:2015fja,Kruppa:2020rfa,Robin:2020aeh,Stetcu:2021cbj,Tichai:2022bxr,Perez-Obiol:2023wdz,Kovacs:2024zoz,Tichai:2024cyd,Shinde:2026hus}. 
As an example, Ref.~\cite{Kovacs:2024zoz} recently performed a detailed analysis of entanglement and correlations in two-nucleon systems, in angular-momentum (AM) and isospin-coupled formulation, in the presence of schematic isovector and isoscalar pairing interactions, with both $jj$ and $LS$ coupling schemes for the AM. 
This complements and extends earlier works on two-nucleon entanglement~\cite{Kwasniewicz:2013cqa,Kwasniewicz:2017dbc}. 
The one-body entropy is found to be typically larger for proton-neutron pairs 
than like-particle pairs for $T=1$ pairing,
and proton-neutron pairs typically present more complex and fragmented patterns of mutual information when subjected to the $T=0$ pairing interaction than the $T=1$ channel in the $jj$ coupling.
On the other hand, 
these patterns were simplified when using the $LS$ coupling scheme, which highlights the importance of symmetries and representations in constraining entanglement and correlation features.

Mutual information, however, not only accounts for entanglement, but for all classical and quantum correlations. 
Measures isolating two-orbital entanglement, such as negativity and concurrence (equivalent to 2-tangles), 
have been computed in light and mid-mass nuclei~\cite{Robin:2020aeh,Brokemeier:2024lhq} and were found to be generally weak compared to mutual information as they are strongly constrained by symmetry, such as conservation of proton and neutron numbers, 
and are also sensitive to the nature of the underlying single-particle basis~\cite{Robin:2020aeh,Faba:2021dki,Hengstenberg:2023ryt,Brokemeier:2024lhq}. 

The natural orbital basis, which diagonalizes the one-body density matrix, in particular, is known to minimize the total one-orbital entropy~\cite{Gigena:2015wso}
and to yield reduced two-orbital entanglement. 
In nuclear physics, this was observed~\cite{Robin:2020aeh,Faba:2021dki,Bulgac:2022cjg}, 
and was also noted in the context of quantum chemistry, see {\it e.g.}, Refs.~\cite{Ding:2021kbs,Materia:2024ijj}. 
As a result, two-orbital correlations in systems 
with large degree of symmetry 
or expressed with physically-motivated bases are typically mostly classical in nature.
In particular, the symmetry constraint related to proton and neutron numbers explains the vanishing of two-orbital entanglement and large reduction of mutual information
in the proton-neutron sector, as observed in {\it e.g.} Refs.~\cite{Brokemeier:2024lhq,Perez-Obiol:2023wdz,Shinde:2026hus}.
\\

\begin{figure}
\centering
\includegraphics[width=\columnwidth]{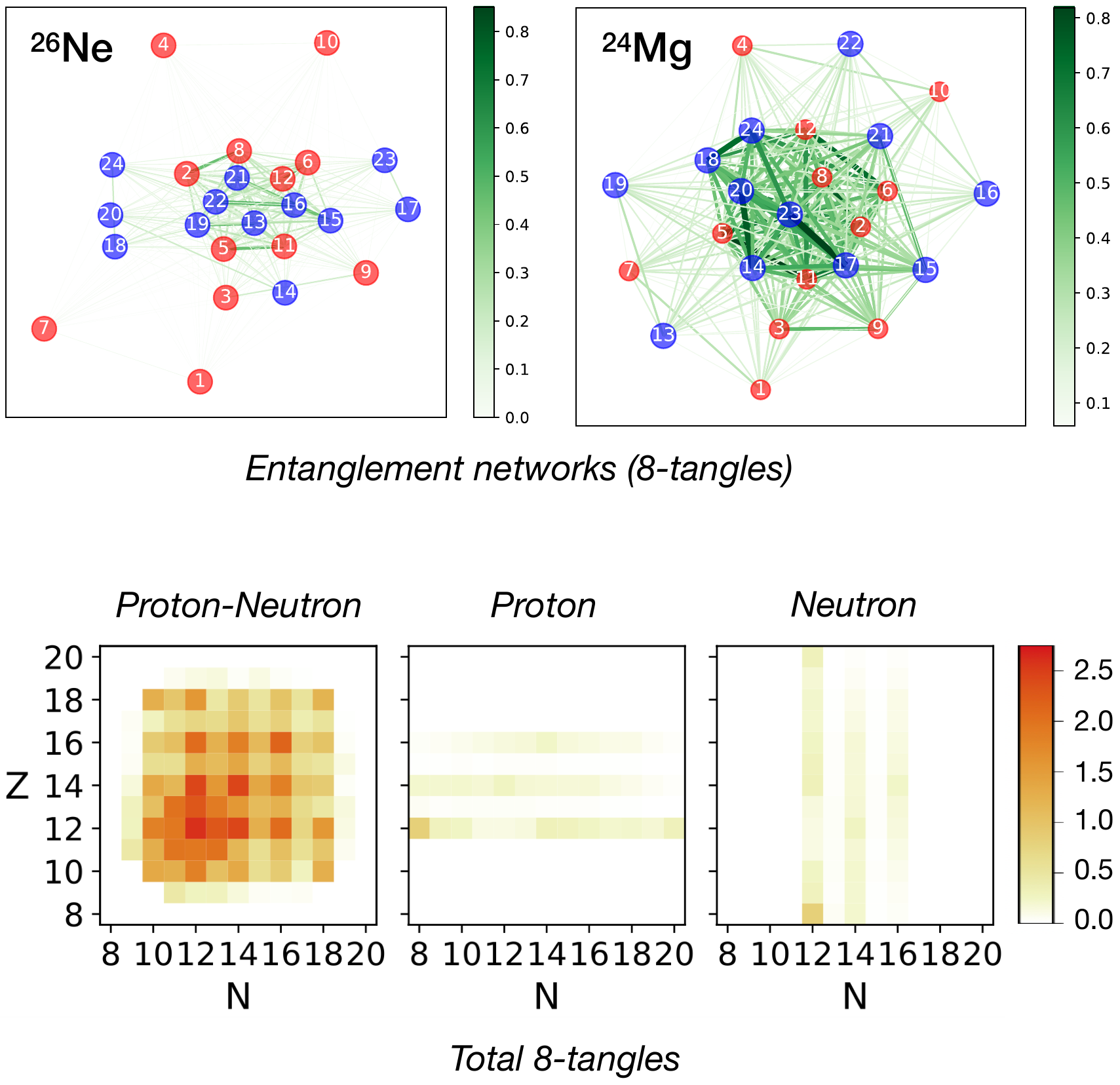}
\caption{ 
Upper: Network representations of 8-orbital entanglement in the ground state of $^{26}$Ne ($Z=10$, $N=16$)
and $^{24}$Mg ($Z=12$, $N=12$), known to be spherical and prolate, respectively. 
The nodes show the proton (red) and neutron (blue) orbitals.  The values of the edges, representing the entanglement, are given by $e^{(8)}_{i_1 i_2} = \sum_{i_3 < i_4 < i_5 < i_6 < i_7 < i_8}   \tau^{(8)}_{(i_1, i_2, i_3, i_4, i_5, i_6, i_7, i_8)}$
and are indicated by both the darkness and the thickness of the lines. Orbitals that are shown closer together are more entangled with each other. 
Lower: Total 8-tangles in the proton-neutron (left), proton (middle) and neutron (right) sectors, $\overline{\tau}^{(8)}_{\pi\nu}$, $\overline{\tau}^{(8)}_{\pi}$, and $\overline{\tau}^{(8)}_{\nu}$, defined in Eq.~\eqref{eq:ntangles_sum}.
[Figures adapted from Ref.~\cite{Brokemeier:2024lhq} used with permission from the authors under 
{\it Creative Commons Attribution 4.0 International license}~\cite{cc_by_4.0}.]
}
\label{fig:ntangles_sd_shell}
\end{figure}

\begin{figure}[!ht]
\centering
\includegraphics[width=.85\columnwidth]{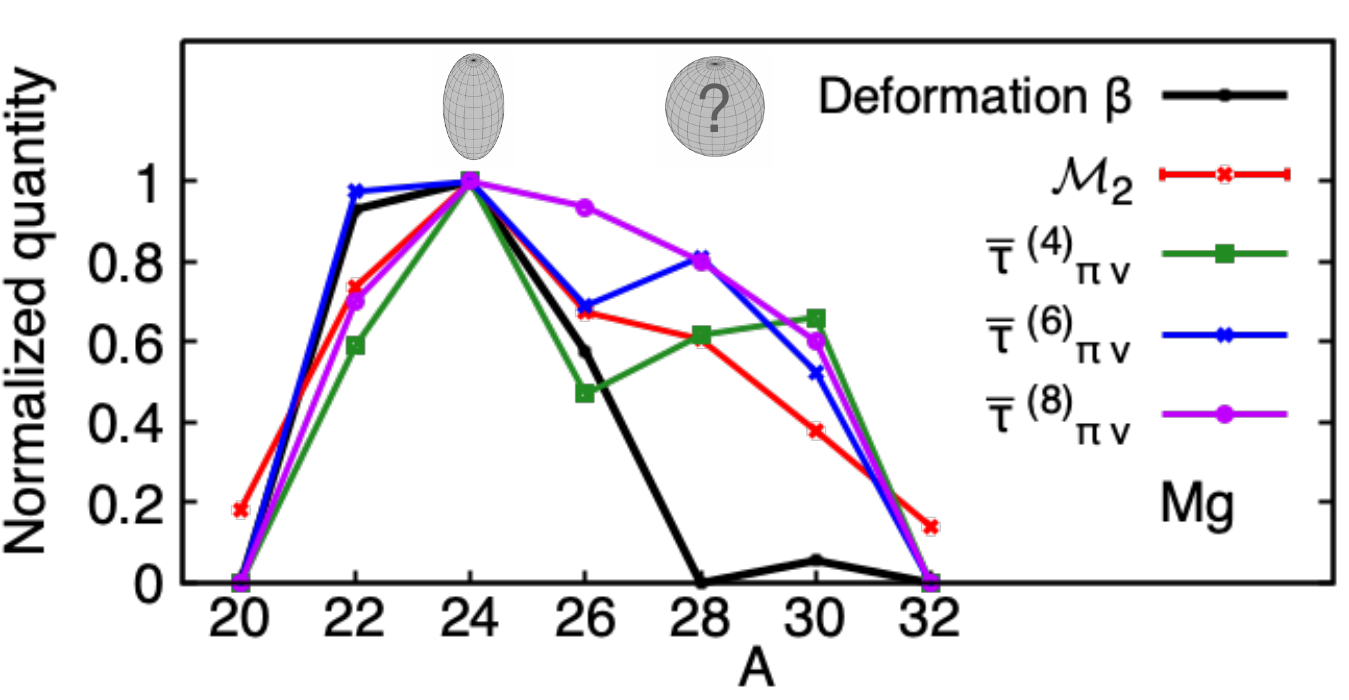}
\caption{
Non-stabilizerness (magic) ${\cal M}_2$, $n=4,6,8$-tangles $\bar{\tau}_{\pi\nu}^{(n)}$ in the proton-neutron sector, and axial deformation parameter $\beta$, in even-even Mg nuclear ground states, as a function of total nucleon number A. The values of $\beta$ were taken from \href{https://www-phynu.cea.fr/science_en_ligne/carte_potentiels_microscopiques/tables/HFB-5DCH-table_eng.htm}{Summary Tables}~\cite{PhysRevC.81.014303} reproduced at the website~\cite{phynu-cea}. Each quantity has been normalized to its maximum value in the chain.
[Figure adapted from Ref.~\cite{Brokemeier:2024lhq} used with permission from the authors under 
{\it Creative Commons Attribution 4.0 International license}~\cite{cc_by_4.0}.]
}
\label{fig:Mg_magic_deformation}
\end{figure}

Understanding how few-body or collective phenomena, such as clustering or deformation, emerge from fundamental constituents from an information-theoretic perspective requires access to multipartite entanglement, to characterize how few or many orbitals entangle with each other. 
Explorations of multi-orbital entanglement in nuclear systems is only beginning, with initial studies in selected models via few-orbital quantum discord~\cite{Faba:2022qop} and $n$-tangles~\cite{Hengstenberg:2023ryt,Robin:2025wip}, and in light and mid-mass nuclei within shell-model calculations through the calculation of up to $8$-tangles~\cite{Brokemeier:2024lhq}.
Multi-orbital entanglement in $p$- and $sd$-shell nuclei are typically most collective in the mixed proton-neutron sector, through the presence of many small components, while pure proton and pure neutron sectors exhibit few but larger $n$-tangles~\cite{Brokemeier:2024lhq}. 
This gives  total $n$-tangles that are typically an order of magnitude larger in the proton-neutron sector than in the like-particle sector, for a given value of $n$.
For example, the top panel of Fig.~\ref{fig:ntangles_sd_shell} displays collective entanglement networks in the strongly-deformed nuclei $^{24}$Mg, in comparison with the spherical ground state of $^{26}$Ne. A few thick edges between orbitals of the same specie are seen, while many thin edges appear across proton and neutrons sectors. 
The bottom panel of Fig.~\ref{fig:ntangles_sd_shell} shows the total summed 8-tangles in each sector, defined as 
\begin{align}
 \overline{\tau}^{(8)}_{\pi}  \equiv \sum_{\substack{ i_1, i_2, .. i_8 \\ \text{all protons} } } & \tau^{(8)}_{(i_1 i_2 .. i_8)}  \; , \ \ \
\overline{\tau}^{(8)}_{\nu}  \equiv \sum_{\substack{ i_1, i_2, .. i_8 \\ \text{all neutrons} } } \tau^{(8)}_{(i_1 i_2 .. i_8)} \; , \nonumber \\
& \overline{\tau}^{(8)}_{\pi\nu}  \equiv \sum_{\substack{ i_1, i_2, .. i_8 \\ \text{mixed } } } \tau^{(8)}_{(i_1 i_2 .. i_8)} 
\; .
\label{eq:ntangles_sum}
\end{align}
Additionally, these $n$-tangles were found to be larger in nuclei near $N=Z$, 
which exhibit large axial deformation.
For example, $^{24}$Mg, for which $\overline{\tau}^{(8)}_{\pi\nu}$ is the highest, is known to be one of the most axially deformed nucleus of the $sd$-shell. 
Fig.~\ref{fig:Mg_magic_deformation} also displays the total proton-neutron 4-, 6- and 8-tangles as a function of predicted axial deformation in the Mg chain. 
Interestingly, the 4- and 6-tangles, related to two-body and 3-body correlations 
again 
increase in the region that precedes the island of inversion, above $^{26}$Mg, which, in nature are characterized by configuration mixing and shape co-existence. While the shell-model calculations, which are restricted to a single oscillator shell, cannot realistically capture these effects, the wave function contains correlations that signal the onset of these phenomena via the effective empirical interaction. Multipartite proton-neutron entanglement appears to be sensitive to these effects, and thus may constitute a good indicator of their presence
\footnote{We note that the Ref.~\cite{Brokemeier:2024lhq} study used an orbital-to-qubit Jordan-Wigner mapping to 
compute the $n$-tangles using Eq.~\eqref{eq:ntangles}, 
which is defined for qubits and thus does not explicitly account for fermionic anti-symmetrization. The resulting $n$-tangles are then dependent on the ordering of the orbitals. Empirically, we find that although the values vary, the overall trends and conclusions discussed above remain robust under different orderings.}.

These results now provide additional insights into the global bipartite proton-neutron entanglement studied above, by revealing substructure of the latter, which can be seen as a "residue" of the multi-proton-neutron entanglement.
Again the symmetries and model space play a major role in the characterization of multipartite entanglement, and it would be interesting to assess whether this conclusion holds in crossed-shell or no-core calculations.
\\

\begin{figure}[!ht]
\centering
\includegraphics[width=\columnwidth]{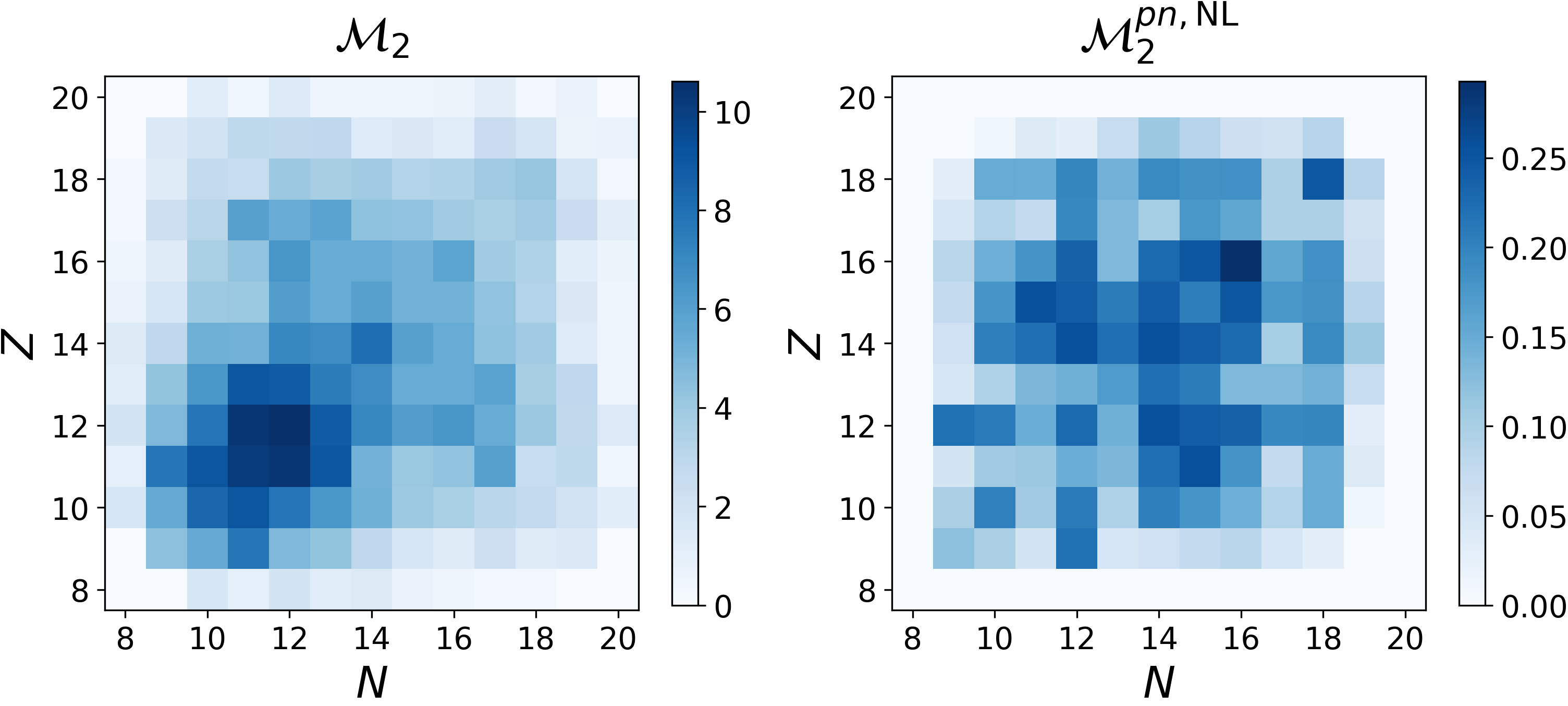}
\caption{ Full-state magic $\M_2$ and bi-partite proton-neutron non-local magic $\M_2^{pn,NL}$ of $sd$-shell nuclear ground states. 
The calculations are performed using the minimum value of the angular momentum projection $J_z = J_{z,min}$ for a given ground state~\cite{Keeble:2026}.
}
\label{fig:NL_magic_sd}
\end{figure}

\paragraph{Non-stabilizerness}
As practically-computable measures of non-stabilizerness have only been introduced recently, investigations of non-stabilizerness in nuclear systems, similarly to other many-body systems, are only in their infancy. A first study of $p$- and $sd$-shell nuclei, in a traditional shell model framework, has been performed in Ref.~\cite{Brokemeier:2024lhq} using exact and Markov-Chain Monte-Carlo computations of the stabilizer R\'enyi entropies.
Non-stabilizerness was found to follow a similar trend as the multi-proton-neutron entanglement ($n$-tangles), although being more closely correlated with the axial deformation patterns around $N=Z$, with a sharp peak at the maximum deformation, and decreasing with neutron excess, yet sustaining non-zero values in $N>Z$ nuclei known to possibly exhibit shape co-existence in nature. This can be seen in Fig.~\ref{fig:Mg_magic_deformation} for the Mg chain. The values of the stabilizer 2-R\'enyi entropy $\mathcal{M}_2$ is displayed for the ground states of all $sd$-shell nuclei in the left panel of Fig.~\ref{fig:NL_magic_sd}. Note that the detailed patterns of multipartite entanglement have been found to be highly sensitive to the underlying non-stabilizerness content, in the context of holography~\cite{Cao:2024nrx,Nezami:2016zni}.

As mentioned in section~\ref{sec:measures_complexity}, the concept of non-local (NL) magic has been recently introduced in Ref.~\cite{Cao:2024nrx} in order to capture the interplay between entanglement and non-stabilizerness (magic). 
Given a partitioning of the quantum state (in a bipartite or multipartite manner), 
NL magic is obtained by minimizing the magic over the set of unitary transformations acting on the local sub-systems,
as in Eq.~\eqref{eq:NLMdef}. 
As such, non-local magic is a quantity which is independent on the local bases, and quantifies the magic that resides in the correlations between the subsystems.
First calculations of NL magic in nuclei are now being performed in the same shell-model framework as discussed previously, considering, as a first step, a proton-neutron bi-partitioning of the nuclear ground state.
The values of the NL magic have been computed using the upper bound given in Eq.~\eqref{eq:NL_M2_upperbound},
and
the results for $sd$-shell nuclei are displayed in the right panel of Fig.~\ref{fig:NL_magic_sd}. 
It is seen that the patterns of proton-neutron NL magic differ from the ones of the full-state magic, with a maximum value obtained for $^{32}$S. The origins of these differences are yet to be determined.

\paragraph{Excited States}

How entanglement, and more generally quantum complexity, evolves from ground states to excited states 
requires further exploration.
Several studies have investigated how entanglement varies with total angular momentum $J$ in low-lying states~\cite{Kruppa:2021yqs,Sarma:2024vqx,Gorton:2024hbb,Shinde:2026hus}, but no clear trend has emerged.
A more systematic analysis has been carried out, 
in which bipartite proton-neutron entanglement entropy was studied for excitations up to $\sim 20$ MeV~\cite{Johnson:2022mzk}. It was found that, on average, the entropy increases and then decreases as a function of excitation energy within the shell model. In particular, nuclei near $N = Z$ exhibit a semicircular, symmetric pattern, which was largely attributed to single-particle energies.

\paragraph{Entanglement and short-range correlations}

As is customary in finite fermionic systems, the bi-partite entanglement related to a division of the single-nucleon basis in term of a particle and hole sector (the sectors above and below the Fermi level, respectively) 
has been investigated in nuclear systems. In that case the wave function is partitioned into 
\begin{equation}
\ket{\Psi} = \sum_{ph} C_{ph} \ket{\phi_p} \otimes \ket{\phi_h} \; ,
\end{equation}
where the indices $p$ and $h$ refer to the particle and hole sectors, respectively.
One traces, for instance, over the hole sector to compute the corresponding entanglement entropy. 
Such entanglement has been studied
in closed-shell nuclei with $N=Z$ using {\it ab initio} coupled-cluster framework~\cite{Gu:2023aoc}.
Due to the large number of single-particle states involved in the calculations, 
the depletion number as approximate entanglement measure, 
finding that nuclei are expected to follow a volume-law scaling with nucleon number $A=N+Z$. 
This behavior was attributed to the short range nature of the nuclear interaction (long range in momentum space), and the entanglement was found to increase with interaction cut-off.
A volume-law scaling of entanglement was also 
found in the context of nuclear short-range correlations~\cite{Pazy:2022mmg},
the effects of which 
have been investigated~\cite{Bulgac:2022ygo}.

\paragraph{Quantum Complexity in Nuclear Phase Transitions}

As finite-size mesoscopic systems, atomic nuclei lie at a particularly interesting regime: they are too small to exhibit true thermodynamic phase transitions, yet sufficiently large to display pronounced collective behavior. Accordingly, while they do not display genuine non-analyticities and discontinuities in observables, they nevertheless show clear signatures of phase transitions that manifest through rapid structural changes  of the quantum state as control parameters, such as neutron or proton numbers, are varied. Most notable examples include transitions from spherical to deformed shapes and the emergence of pairing correlations analogous to superfluidity, which are two dominant and competing collective effects in nuclei.
To gain a better understanding of which aspects of quantum complexity are associated with these phenomena,
one can consider schematic Hamiltonians, or models that share similar properties to nuclei, and study how complexity evolves with interaction strength, or scales with system size.
\\

Early work in this area employed information measures of the full many-body state to study pairing phase transitions. For example, the invariant correlation entropy \cite{PhysRevE.58.56}, defined as the entropy of a density matrix averaged over a small parameter region to probe correlations between many-body components in the transition, was applied to the BCS Hamiltonian and selected shell-model nuclei~\cite{Volya:2003su}
By varying pairing interaction strengths parameters, this measure allowed for identification and characterization of pairing phases. For physical parameters, typical nuclei were found to lie near the transition to a $T=1$ paired phase, whereas $T=0$ pairing was typically too weak to induce coherence. 
Later works also used classical Shannon information entropy to specify pairing transitions 
in nuclei, from light to heavy,  within several models~\cite{Guan:2013isa,Guan:2016nxg,Liang:2024kna}.

Measures of entanglement and non-stabilizerness
have been calculated for the Richardson pairing Hamiltonian~\cite{PhysRev.141.949,Dukelsky:2004re} with transition to a superfluid phase, 
the Lipkin-Meshkov-Glick (LMG) model~\cite{LIPKIN1965188} that presents a phase transition analogous to a deformed phase in nuclei, 
in the Agassi model~\cite{Agassi:1968zsu} that combines both of these effects, 
as well as in the interacting boson and boson-fermion models with quantum shape phase transitions~\cite{Jafarizadeh:2023lvg,Jafarizadeh:2022kcq}.

For example, bipartite entanglement between particle and hole spaces in the pairing model 
has been found to satisfy an area law scaling~\cite{Gu:2023aoc}.
This explains the good performance of the DMRG for this model in, {\it e.g.}, Refs.~\cite{Dukelsky:1999rj}, or more recently, Ref.~\cite{Rausch:2021ifu}. 

\begin{figure}[!ht]
\centering
\includegraphics[width=\columnwidth]{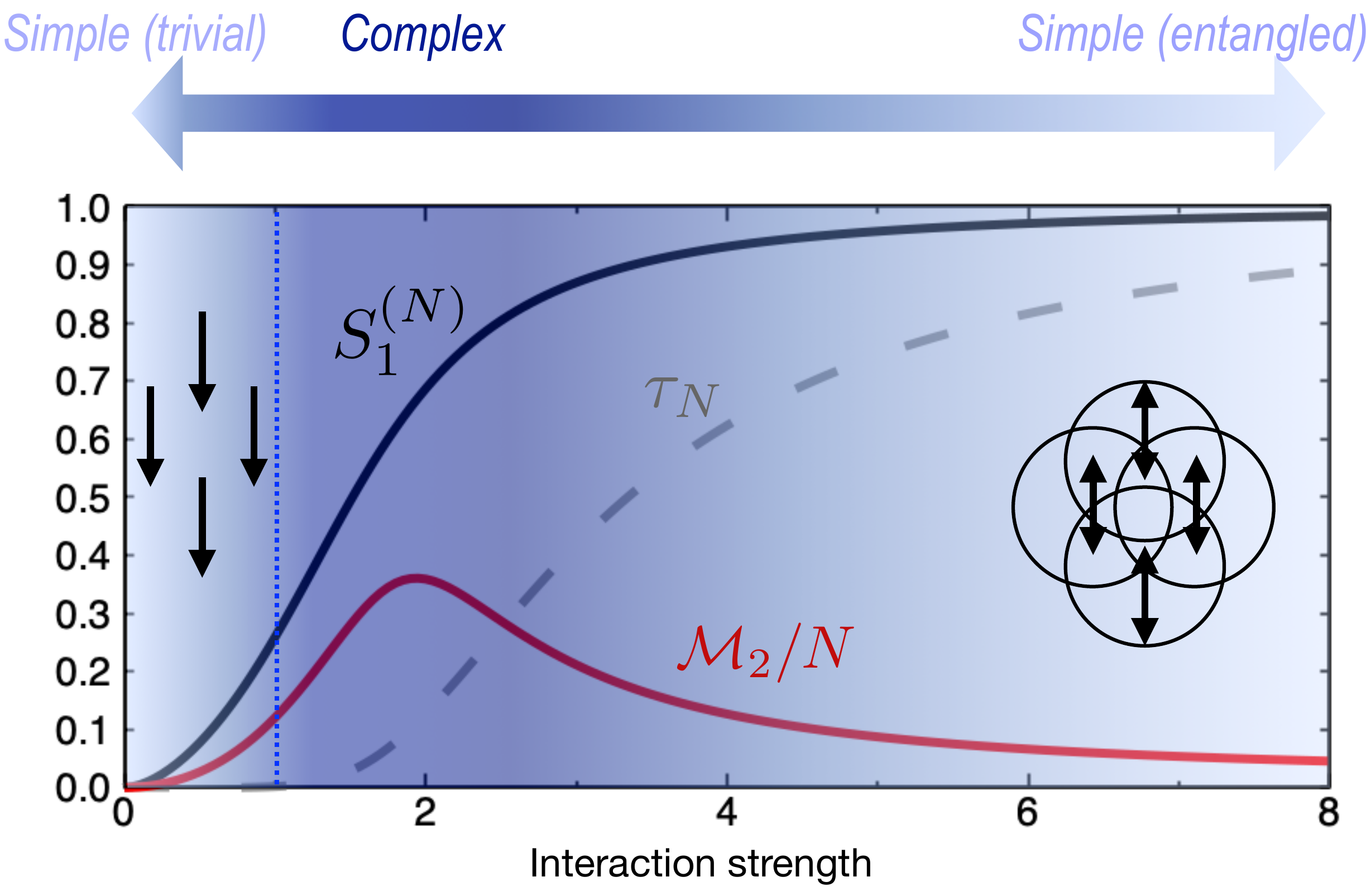}
\caption{Complexity diagram of the LMG model. The red curve shows the magic quantified by 2-SRE density, the plain black and dashed gray curves display the single spin (or one-orbital) entropy and $N$-tangle, respectively. 
The darker shaded area corresponds the the region of high complexity, with both high-entanglement and extensive magic, while the lighter regions correspond to low-complexity ground states: unentangled zero-magic non-interacting ground state (left), genuinely $N$-partite entangled low-magic state (right).
The values of the curves correspond to a system with $N=8$ spins (16 orbitals) obtained in Ref.~\cite{Robin:2025wip}.  } 
\label{fig:LMG_phase_diag}
\end{figure}

Aspects of quantum complexity in the LMG model have been extensively studied due to the relevance of this model in many areas of physics and quantum information. Besides presenting a spontaneous parity-symmetry breaking similar to deformation in nuclei, 
it can also model (and be realized in) two-mode Bose-Einstein condensates~\cite{PhysRevA.55.4318,PhysRevA.67.013607,Fuentes-Schuller:2006xby}, trapped ion systems~\cite{Molmer:1998mq,PhysRevA.66.042101,PhysRevLett.90.133601,Lee:2014ftq}, and cavity QED~\cite{PhysRevA.77.043810}. The LMG Hamiltonian is also able to produce spin squeezed states for quantum sensing and metrology~\cite{PhysRevA.68.012101,PhysRevA.80.012318,PhysRevA.90.022111,Lee:2014ftq,Calixto_2021}. 
Studied measures span from entanglement entropies for various bi-partitionings, to measures of few- and multi-partite entanglement and correlations in both low-lying states and dynamics of the system, 
see {\it e.g.}, Refs.~\cite{PhysRevA.68.012101,PhysRevA.69.022107,PhysRevA.69.054101,PhysRevA.70.062304,PhysRevA.71.064101,Morrison:2008exn,PhysRevA.77.052105,PhysRevLett.101.025701,PhysRevA.80.012318,Lee:2014ftq,PhysRevA.100.062104,PhysRevB.101.054431,Calixto_2021,Faba:2021dki,Faba:2021kop,Faba:2022qop,Hengstenberg:2023ryt,Lacroix:2024drc,Robin:2025wip}, and more recently, measures of non-stabilizerness~\cite{Passarelli:2024tyi,Robin:2025wip}.
Multipartite entanglement has been studied in various ways via, for example, global entanglement~\cite{PhysRevA.77.052105}, quantum relative entropy~\cite{PhysRevB.101.054431}, Quantum Fisher Information (QFI)~\cite{PhysRevA.80.012318,Lee:2014ftq}, quantum discord~\cite{Faba:2022qop}, and $n$-tangles~\cite{Hengstenberg:2023ryt,Robin:2025wip}. 
It was found from QFI and $N$-tangles that genuine $N$-partite entanglement becomes near-maximal and plateaus after the phase transition~\cite{PhysRevA.80.012318,PhysRevA.90.022111,Lee:2014ftq,Robin:2025wip}, 
while other measures of $k$-partite entanglement with $k<N$, typically peak nearby, 
and disappear in the collective parity-broken regime.
Non-stabilizerness behaves similarly~\cite{Passarelli:2024tyi,Robin:2025wip}, 
reaching its maximum value in the region of interplay between single-particle and collective degrees of freedom. 
These aspects are summarized in Fig.~\ref{fig:LMG_phase_diag}.

It is seen that the system transitions from one simple state (a tensor product) to another simple state, which features near-maximal entanglement and low magic, and thus resembles a stabilizer state.
In between, the system goes though a region of high complexity that is characterized by extensive magic, and significant (but not maximal) entanglement.

Although the LMG Hamiltonian constitutes an idealized model, this study already illustrates how quantum information, by expanding our notion of what is simple and complex, can offer fresh insights into physical phenomena while also enabling more efficient ways to describe them.
Indeed, in traditional (nuclear) many-body theory, the regime of strong collectivity beyond the phase transition is typically described in terms of a (deformed) mean-field state, {\it i.e.}, a Gaussian state, which has long been regarded as the canonical notion of simplicity in this context. However, mean-field approaches generally rely on symmetry breaking to incorporate the relevant correlations, necessitating a subsequent symmetry restoration step that can be computationally demanding.
In contrast, this study suggests that the collective region is in fact well described by a single stabilizer state, 
{\it i.e.}, another type of simple state without need for symmetry breaking (for details, see section~\ref{sec:QI_driven_nuclear_methods}).

Combining pairing and LMG Hamiltonians yields the Agassi model, which was studied via two-orbital quantum discord~\cite{Faba:2021dki}, and recently through measures of multipartite entanglement and non-stabilizerness~\cite{Hengstenberg:2026}. 
These were found to clearly reveal the phase diagram of this model, 
while yielding information about the specific types of quantum complexity related to the various phases. 
In particular the deformed region, similarly to the LMG model, displays high multipartite entanglement but low magic, while the pairing region is characterized by higher magic but lower entanglement, with maxima of complexity reached at the phase transitions and regions of interplay between the different regimes, where magic becomes extensive.
Measures of fermionic non-Gaussianity, such as FAF in Eq.~\eqref{eq:FAF}, 
however appear to be largely insensitive to the nature of correlations in this model~\cite{Hengstenberg:2026}. 
This is because for systems with good particle number, 
FAF can be reduced to occupation numbers, which was also noticed in shell model nuclei~\cite{UB:private}.

\subsection{Dynamical Excitations and Reactions}

Quantum complexity in dynamical nuclear processes has so far received limited attention and has primarily been explored through entanglement. 
Studies include dynamical vibrational and rotational modes
and time-dependent processes such as multi-nucleon transfer reactions and fission.

Studies of  entanglement in nuclear wobbling and chiral modes have been performed~\cite{Chen:2024vvd,Chen:2025xbj,Dai:2025izp},
which have been previously interpreted as spin-coherent and spin-squeezed states~\cite{Chen2022,Chen:2024iki}.
A phenomenological model was used, 
where the core of the nucleus is described as a rotor with tri-axial deformation which is coupled 
to one or two nucleons with high angular momenta,
to establish examples of bi- and tripartite quantum systems.
In the wobbling motion, the angular momentum of the odd particle aligns either longitudinally (along the medium axis of the rotor nucleus) or transversally (along the short or long axis of the rotor). While the entanglement entropy between rotor and the odd particle is relatively insensitive 
to the spin, it increases with the wobbling number, which characterizes the amplitude of the motion~\cite{Chen:2024vvd}.
The coupling of the rotor to a particle-hole proton-neutron pair can exhibit chiral modes. 
In that case, a pronounced asymmetry in the distribution of entanglement among the rotor, proton and neutron is observed~\cite{Chen:2025xbj}. 
Using the concurrence triangle to quantify the amount of total entanglement, 
it is found to increase substantially with increasing spin.
\\

The entanglement between rotation and shape degrees of freedom 
has been studied~\cite{Wang:2025acg} 
using a phenomenological framework in which the nuclear state is defined over five collective coordinates (two shape variables and three Euler angles describing rotations) thereby exhibiting bipartite entanglement between shape and rotational degrees of freedom. 
Significant differences in the patterns of entanglement entropy are found for different types of deformations.\\

First investigations of entanglement in multi-nucleon transfer reactions have been performed~\cite{Li:2024jdb}. 
Specifically, a collision of two closed-shell nuclei 
(a projectile, $^{40}$Ca, and a target nucleus, $^{208}$Pb) 
was considered,
where an initially un-entangled state gives rise to two entangled fragments in the final state.
The reaction was described within a time-dependent relativistic density functional theory.
As the projectile comes close to the target, the entanglement was found to rise sharply, reaching its peak at the point of closest approach or maximal overlap. As the fragments subsequently separate, the entanglement diminishes slowly, until reaching its asymptotic value.
The study further identified a positive correlation between entanglement and the incident energy, indicating that stronger projectile–target interactions yields more pronounced entanglement. In addition, a linear relationship was observed between the nucleon-number fluctuations of the fragments and the entanglement entropy.
\\

\begin{figure}[h]
\centering
\includegraphics[width=\columnwidth]{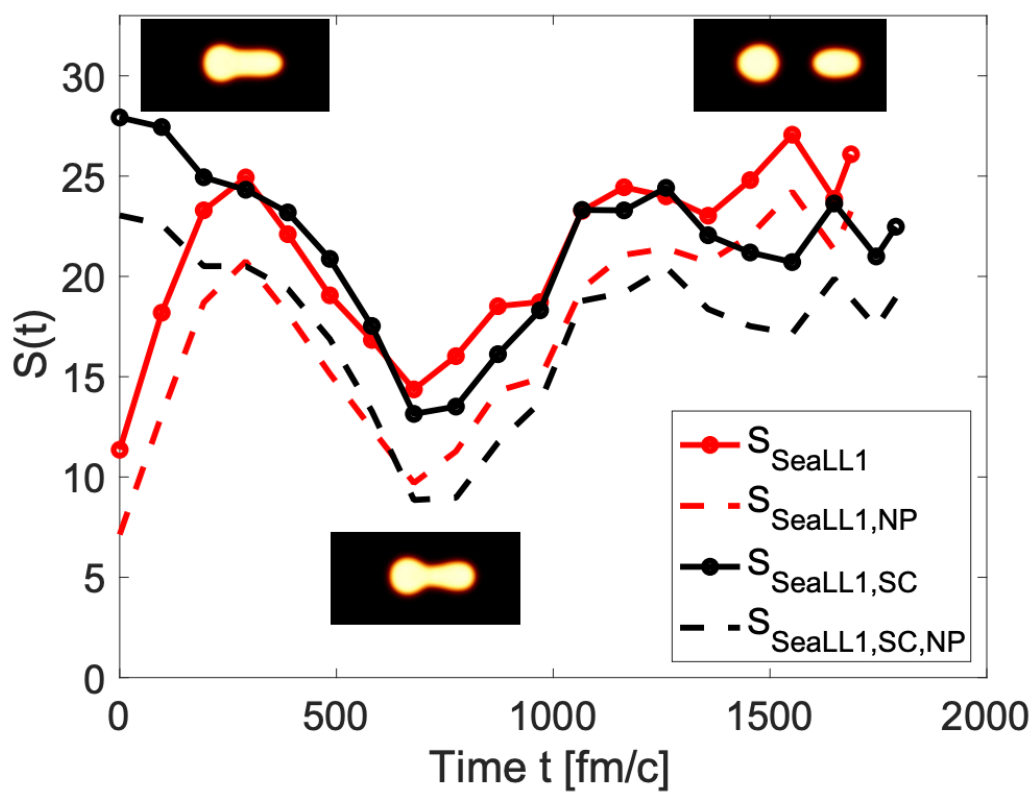}
\caption{One-body entanglement entropy in the fission process of $^{235}$U$(n,f)$ induced by a low-energy neutron, as a function of time. The solid (resp. dashed) curves correspond to entropies evaluated without (resp. with) projection onto particle number of the total many-body wave function. The black and red curves differ in their treatment of the initial state of the compound nucleus. See Ref.~\cite{Bulgac:2022cjg} for more details about the framework. 
[Figure reproduced from Ref.~\cite{Bulgac:2022cjg} with permission from the authors and the American Physical Society.]
} 
\label{fig:entang_fission}
\end{figure}
Entanglement in the dynamical fission of a heavy nucleus has been investigated~\cite{Bulgac:2022cjg}. 
Using natural orbitals, the evolution of the one-body entanglement entropy 
was examined during the induced fission of $^{235}$U with a time-dependent mean-field framework including superfluid pairing. 
Fig.~\ref{fig:entang_fission} displays the results obtained for two different initial states. 
Interestingly, the one-body entanglement decreases until the point of scission, 
before again increasing, once the fragments separate. 
This decrease is attributed to the formation of a neck during nuclear elongation,
which inhibits nucleon exchange between the fragments.
The evolution of entanglement between the two fragments during the dynamical fission of $^{240}$Pu
was subsequently studied~\cite{Qiang:2024syr}.
The fragments are found to become increasingly entangled as the system approaches scission, 
and that this entanglement persists after separation due to highly delocalized wave functions. 
These results indicate that entanglement is crucial for reproducing the characteristic “saw tooth” behavior in fragment excitation energies, 
while it has little impact on the neutron excess of the fragments.
\\

Given the complexity of nuclei, nuclear reactions promise to reveal
even more aspects of the complexity in many-body nuclear systems.
To the best of our knowledge, there are only a couple of on-going studies investigating multipartite entanglement and magic in nuclear decays and reactions~\cite{BroekemeierMaster,WendtAPS2026}.

%% file: Section_NBody_methods.tex
\section{Many-Body Techniques Emerging from the Interface}
\label{sec:QI_driven_nuclear_methods}

The steadily growing understanding of nuclear structure and phenomena from the perspective of (quantum) information, has allowed for improvements of existing many-body methods, through their re-interpretation in terms of entanglement and other aspects of complexity; and is now also enabling the developments of new approaches to nuclear many body problems in directions that had not been anticipated before.
In this section, we review advances in this area and outline future directions for addressing current challenges in developing robust, predictive simulations of nuclear systems across classical and hybrid quantum approaches.

\subsection{Rearranging Complexity}

\paragraph{Single-particle basis ordering and optimization}

Because the apparent complexity of a nuclear state often depends on the basis used to 
represent it, it can be can be useful to apply basis transformations to decrease that complexity or re-organize entanglement or magic structures into localized sectors of the Hilbert space. This can enable more controlled truncations that preserve essential information while improving the efficiency of classical computations. 

For example, the nature and ordering of the single-particle basis play a crucial role in the DMRG. 
In this method, the fermionic orbitals are mapped onto a one-dimensional chain, starting from a small set, which is iteratively enlarged by adding additional orbitals. 
At each step, the new Hamiltonian in the enlarged space is solved for the target state, 
and a SVD is performed to retain only the largest singular values, thereby truncating the state while retaining the important entangled pieces~\cite{PhysRevLett.69.2863,PhysRevB.48.10345}. 
If long-range correlations (correlations between distant orbitals in terms of their ordering in the basis) are present, the convergence of the DMRG can be significantly hindered~\cite{PhysRevB.68.195116,Rissler_2006}. 
Thus, grouping strongly-correlated orbitals together in the mapping is crucial for performance of the method. 
To address this point, basis orderings based on orbital entanglement entropies or mutual information have been introduced in the context of quantum chemistry~\cite{PhysRevB.68.195116,Rissler_2006,Barcza:2010xtn}, interacting orbitals near each other, typically toward the center of the chain. 
Specifically, the ordering can be determined so as to minimize a "correlation distance" given by
\begin{equation}
I_{dist} = \sum_{ij} I_{ij} |i-j|^\eta \; ,
\label{eq:corr_distance}
\end{equation}
where $I_{ij}$ is the MI between orbitals number $i$ and $j$, and $\eta$ typically chosen to be $\eta=2$.
Such quantum-information-based ordering strategies have been subsequently adopted in nuclear physics applications, 
{\it e.g.}, 
Refs.~\cite{Papenbrock:2004jd,Legeza:2015fja} which allowed for significant improvements of previous particle-hole DMRG (ph-DMRG) calculations of $sd$- and $fp$-shell nuclei with empirical effective interactions, and more recently with {\it ab-initio} interactions~\cite{Tichai:2022bxr,Tichai:2024cyd}. 
This ordering technique was also employed in
{\it ab-initio} no-core Gamow-DMRG (G-DMRG) computations of resonances in light Helium nuclei~\cite{Sehovic:2026hni}. 
In this method, the bi-partitioning is defined through a separation of the discrete and continuum states. In the case of narrow resonances the entanglement remains sufficiently low to allow for efficient G-DMRG computations~\cite{Papadimitriou:2013ix,Shin:2016poa,Fossez:2016dch}.
In contrast, for broad resonances, the entanglement can become significantly larger due to stronger coupling to the continuum. 
It has been shown that  optimizing the basis ordering by minimizing the correlation distance in Eq.~\eqref{eq:corr_distance}, combined with a new truncation scheme, enables an accurate description of such broad resonances and extends the reach of G-DMRG beyond previous limitations~\cite{Sehovic:2026hni}. 
\\

Beyond ordering, transforming the orbital basis itself, to reduce the intrinsic complexity of the many-body state, can provide further, more substantial improvements. The use of the natural orbital basis~\cite{PhysRev.97.1474} which diagonalizes the one-body density matrix, provides the most physically meaningful representation of the wave function since it corresponds to the basis in which particle occupations are well defined. As mentioned above, the natural basis minimizes the total one-orbital entropy~\cite{Gigena:2015wso}, thus providing a measure of entanglement that is intrinsic to the wave nuclear state (basis-independent). 
Although originally introduced in quantum chemistry, natural orbitals have also been employed in nuclear physics at various levels of approximation, in which context they have been shown to accelerate convergence with respect to the size of the model space, see {\it e.g.}, Refs.~\cite{Robin:2015lba,Robin:2016wsx,Constantinou:2016urz,Tichai:2018qge,Novario:2020kuf,Robin:2020aeh,Hoppe:2020elo,Fasano:2021ahd,Scalesi:2024pfm,Knoll:2025vzp,Sehovic:2026hni}. Entanglement properties of natural orbitals have been investigated in {\it ab-initio} studies of light nuclei using one- and two-orbital measures~\cite{Robin:2020aeh}, as well as in dynamical fission processes via orbital entanglement entropies and spectra~\cite{Bulgac:2022cjg}.
\\

\begin{figure}[!ht]
\centering
\includegraphics[width=\columnwidth]{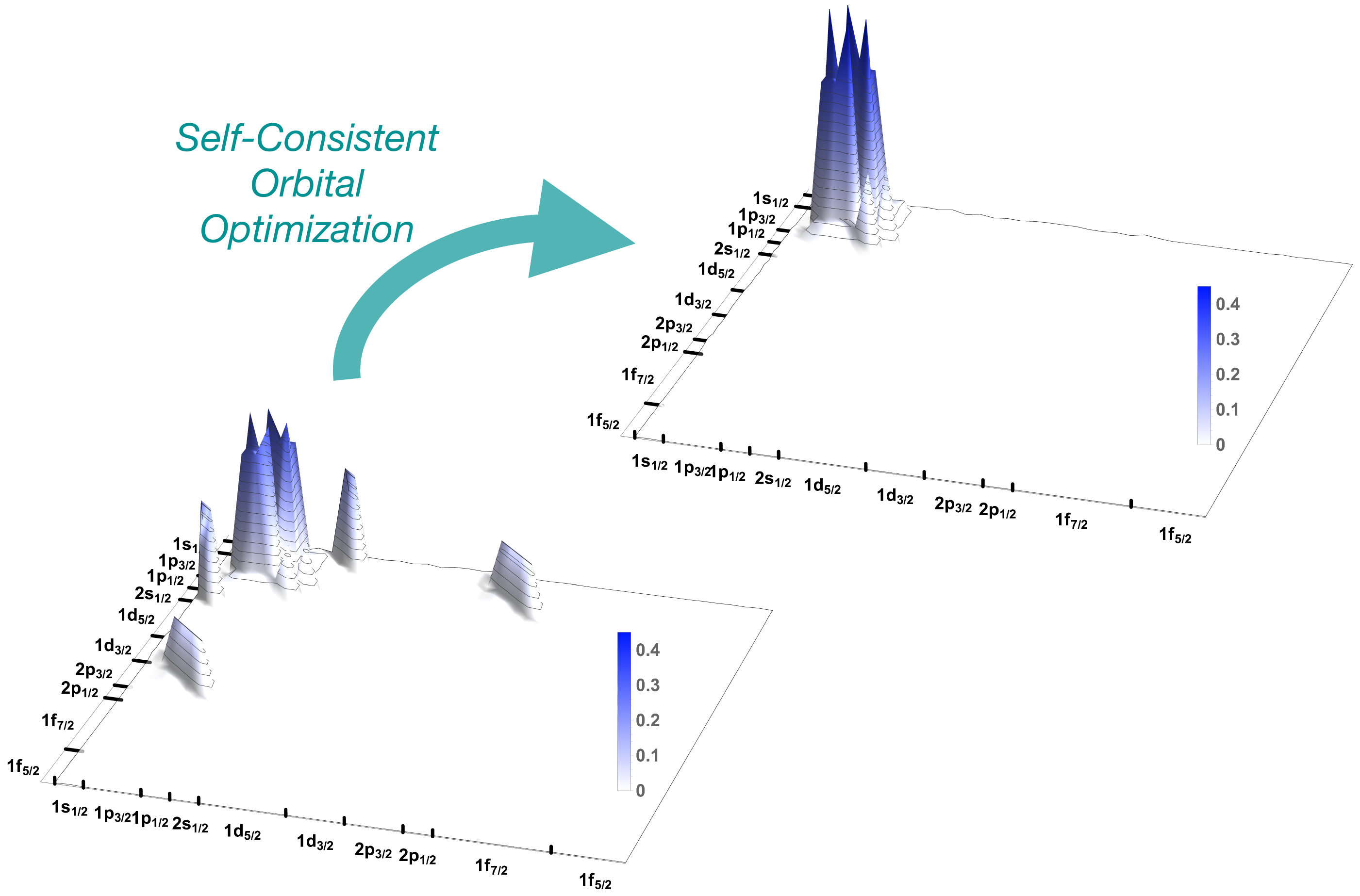}
\caption{The two-neutron-orbital mutual information in $^{6}$He using a harmonic oscillator basis (bottom) and natural orbitals self-consistently optimized with two-body correlations (top). The calculations are performed in an {\it ab-initio} framework, in an active space with all particles active (no frozen core). As the basis is optimized, information is rearranged into a small number of orbitals.
The figure is adapted from Ref.~\cite{Robin:2020aeh}. }
\label{fig:6He_CASSCF_NO}
\end{figure}

While natural orbitals provide a first level of improvement, their effect can become limited when working within truncated active model spaces. In such cases, diagonalizing the one-body density matrix only mixes orbitals within the active space and cannot capture correlations involving orbitals outside this space. To go beyond this limitation, one can incorporate the influence of external orbitals through an effective one-body density operator.
For example, it is possible to account for two-body correlations connecting active and frozen spaces in a variational, self-consistent manner. In quantum chemistry, this approach is known as the multiconfigurational self-consistent field (MCSCF) or complete active space self-consistent field (CASSCF) method~\cite{schmidt1998construction,roos1987complete}, and has also been applied in nuclear physics, see, e.g., Refs.~\cite{PhysRevLett.25.123,FAESSLER197131,PhysRevC.4.1077,Schmid1974,Krewald1974,Robin:2015lba,Robin:2016wsx,Robin:2020aeh}.
In Ref.~\cite{Robin:2020aeh}, an analysis based on mutual information showed that this approach leads to a decoupling of active and frozen spaces in terms of two-orbital correlations, and, at the same time, naturally orders orbitals according to decreasing entanglement entropy, while also minimizing the correlation distance defined in Eq.~\eqref{eq:corr_distance} (see Fig.~\ref{fig:6He_CASSCF_NO}).

Such approaches have also been explored in combination with ground-state quantum algorithms~\cite{Robin:2023pgi}, where the Variational Quantum Eigensolver (VQE)~\cite{Peruzzo:2013bzg} was extended to simultaneously 
optimize both the ground-state wave function and an effective Hamiltonian (or, equivalently in that case, an optimized single-particle basis). This scheme, referred to as Hamiltonian-Learning-VQE (HL-VQE), was applied to the LMG model, where it demonstrated an exponential improvement in convergence with respect to the size of the model space compared to standard VQE.
For that model, the inclusion of the Hamiltonian-learning component enabled a reorganization of the wave function into a small number of dominant components, naturally ordered by decreasing weight. This effectively compresses the state in Hilbert space and allows for more efficient truncation schemes. Furthermore, Ref.~\cite{Hengstenberg:2023ryt} showed that this procedure reduces orbital entanglement and leads to rapid convergence of multi-orbital entanglement measures.
These features are particularly relevant in the context of quantum simulations on NISQ-era devices, as faster convergence and reduced model spaces can significantly lower the number of required qubits and the depth of quantum circuits, thereby mitigating the impact of noise.
Related ideas leveraging basis optimization and entanglement reduction to improve the efficiency of ground state preparation have also been explored in quantum chemistry and condensed matter, see {\it e.g.}, Refs.~\cite{Sokolov:2020fbf,Mizukami:2020icm,Besserve:2021jyx,Ratini:2023fdz,Materia:2023sfq,Ollitrault:2024kne}.

\paragraph{Many-body basis optimization}

In nuclear physics, transformations of the many-body basis to exploit and increase the disentangling of the proton and neutron sectors has been explored. 
As an example, 
the rapid drop of the Schmidt coefficients associated with the proton-neutron bi-partition of the nuclear state in Eq.~\eqref{eq:nuclear_wf_bipartite} was utilized to develop the {\it wave function factorization} (WFF) approach~\cite{Papenbrock:2003bj,Papenbrock:2003az,Papenbrock:2004jd}. 
This method seeks an optimal representation in which the many-body state can be approximated to high accuracy by a truncated Schmidt decomposition or SVD
\begin{equation}
\ket{\Psi} \simeq \sum_{j=1}^{\Omega} s_j \ket{\pi_j} \otimes \ket{\nu_j} \; ,
\end{equation}
where $s_j$ are the singular values, eigenvalues of, for example, the neutron reduced density matrix.
At a conceptual level, this idea is related to the DMRG, 
which also relies on successive 
SVDs to compress the state. 
However, WFF differs significantly in its implementation. 
Rather than mapping single-particle orbitals onto a one-dimensional chain, WFF operates directly in the spaces of proton and neutron many-body states, and involves solving two coupled variational equations (one for each sector) which are linked through the proton–neutron interaction. 
Starting from a random truncated set of $\Omega$ states, for example in the neutron sector, the basis is iteratively refined by solving the equation in the complementary sector.

In contrast to full diagonalization, which scales as $d_\pi \times d_\nu$, the WFF scales as $\Omega \times \text{min}(d_\pi,d_\nu)$~\cite{Papenbrock:2003bj}.
The method was applied to ground and excited states of $sd$- and $fp$-shell nuclei, for which an exponential convergence with respect to $\Omega$ was found, along with increased efficiency for large nuclei due to higher neutron excess. About three orders of magnitude were gained for the largest problems studied~\cite{Papenbrock:2003az}.
The WFF method was developed and applied in the early 2000's, and at the time the main limitation originated from the dimensionalities in each proton or neutron sector. As the computational capabilities have greatly increased since, it would be interesting to re-evaluate how this method compares to current state-of-the-art methods.
\\

More recently, an alternative framework based on a weak proton-neutron entanglement approximation, was developed, which avoids the need for an iterative procedure~\cite{Gorton:2024hbb}. This approach, referred to as {\it proton and neutron approximate shell-model} (PANASh), was motivated by the observation that proton-neutron entanglement is typically lower than the maximal possible value~\cite{Johnson:2022mzk}, as discussed previously in section~\ref{sec:Qcomplex_nuclei}. 
In the PANASh framework, the proton and neutron Hamiltonians are first solved separately and in an exact way, a task made feasible by the significantly reduced dimensionality of the separate sectors compared to the full proton–neutron problem. Forming a few tensor products further coupled to a good angular momentum $J$ and parity p then provides what the authors call a "good enough basis" for representing and solving the proton-neutron Hamiltonian~\cite{Gorton:2024hbb}. 
Precisely,
\begin{equation}
\ket{\Psi} \simeq \sum_{\pi=1}^{d_\pi} \sum_{\nu=1}^{d_\nu} \psi_{\pi\nu} \ket{\pi\nu,J^{\text{p}} } \; .
\end{equation}
PANASh was applied to $fp$- and $fpg_{9/2}$-shell nuclei~\cite{Gorton:2024hbb}, 
for which ground state energies were typically reproduced to higher accuracy than standard shell-model calculations. 
Excited energies, however, were generally described to a lesser extent, in particular in odd-odd nuclei due to the larger proton-neutron entanglement from interaction between the odd proton and odd neutron. 
\begin{figure}[ht]
\centering
\includegraphics[width=.9\columnwidth]{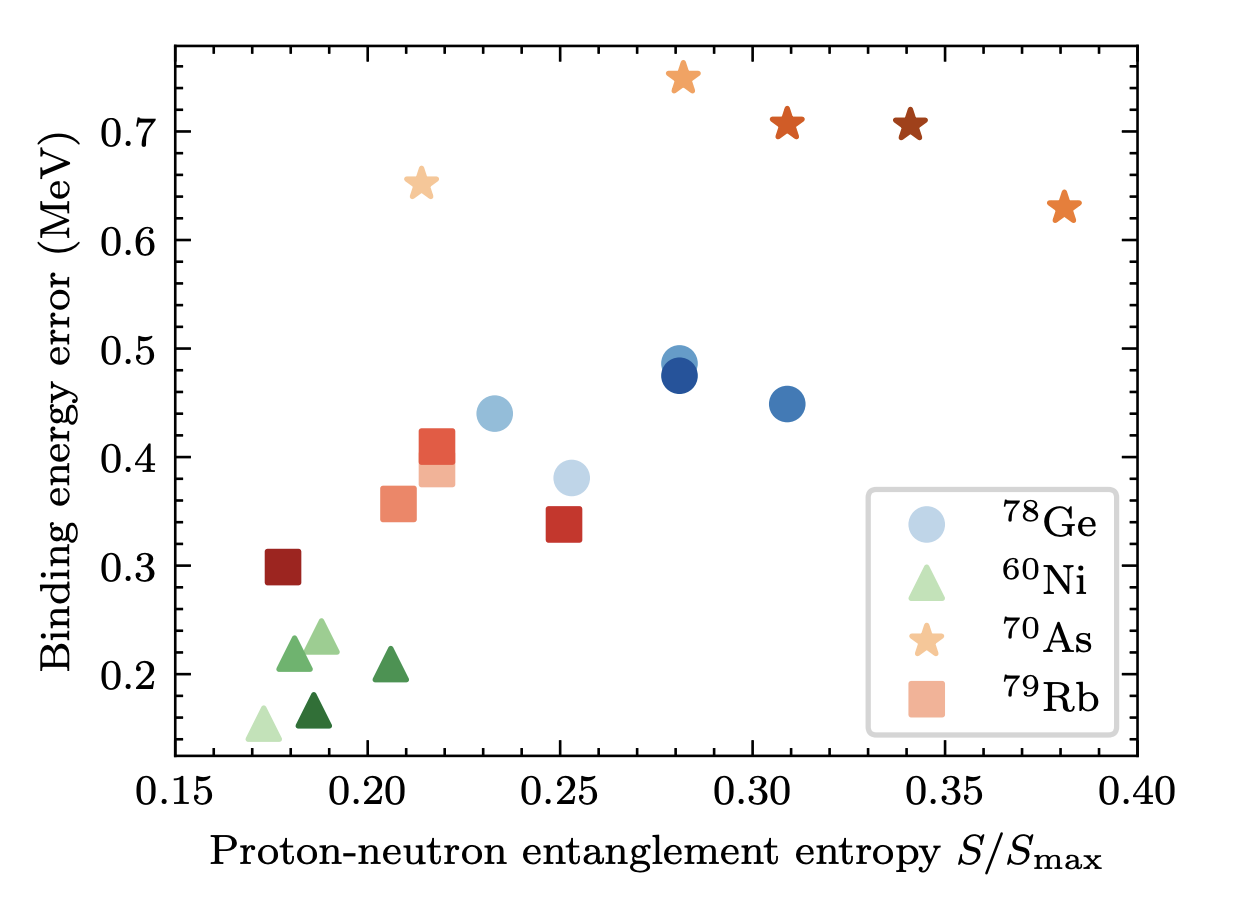}
\caption{Absolute error in the binding energy from the PANASh algorithm, as a function of the relative proton-neutron entanglement entropy (ratio of the calculated entropy over the maximal value allowed by dimensional considerations). States with higher excitation energy are shaded darker.
The even-even nuclei $^{78}$Ge and $^{60}$Ni are deformed and spherical, respectively, while $^{70}$As is odd-odd and deformed, and $^{79}$Rb is odd-A spherical.
[Figure reproduced from Ref.~\cite{Gorton:2024hbb} with permission from the authors and the American Physical Society.] }
\label{fig:PANASh}
\end{figure}
The authors suggest that while pairing correlations are well captured within PANASh, quadrupole proton–neutron collectivity tends to be underestimated.
This interpretation is supported by subsequent applications of the PANASh approach to generator coordinate method (GCM) wave functions with explicit deformation, where improved performance relative to standard GCM calculations was observed.
The recent findings on multi-proton-neutron entanglement also indicate that the truncations applied in PANASh may partly suppress the proton-neutron collectivity~\cite{Brokemeier:2024lhq}. 
A systematic investigation of bi-partite and multi-partite entanglement within the PANASh framework would be of particular interest, especially to clarify how the independent transformations in the proton and neutron sectors impact these correlations.
An appealing feature of PANASh is its ability to access systems beyond the reach of current state-of-the-art interacting shell-model calculations, while also providing a large number of excited states at significantly reduced computational cost. As noted by the authors, the method is particularly well suited for applications targeting averaged properties of highly excited states, where a moderate loss of accuracy is acceptable.
The deviation in the binding energy as a function of entanglement entropy is shown in Fig.~\ref{fig:PANASh}.
\\

Finally, the weak proton–neutron entanglement assumption has also been explored in the context of quantum circuit cutting for quantum simulations of shell-model nuclei~\cite{Perez-Obiol:2024vjo}, as part of continuing efforts to reduce quantum resource requirements. For $^{28}$Ne, a nucleus of the $sd$-shell with large neutron excess, this method was found to accelerate the performance of ADAPT-VQE~\cite{Grimsley:2018wnd}.

\subsection{Stabilizer-State Methods}
While entanglement has been utilized to improve the efficiency of classical many-body methods, how to capture collective behaviors as emergent properties in {\it ab-initio} frameworks remains an open question. 
Nuclear deformation, for example, is still most often explicitly introduced through symmetry-breaking techniques. 
The well-known drawback is that the required symmetry restorations are computationally demanding.

The recent advances on understanding muti-partite entanglement and non-stabilizerness in relation with collective phenomena, and the ability of stabilizer states to describe large collective entanglement in classically efficient ways, have the potential to bring new breakthroughs in that area. 

In this context, Ref.~\cite{Robin:2025wip} developed a method based on stabilizer ground states (SGSs), to explore whether the large-scale entanglement of stabilizer states could provide a way to capture the emergence of collectivity from individual constituents, while preserving symmetries and maintaining computational efficiency.
Note that related stabilizer-based methods and variants have been recently introduced in quantum chemistry, see {\it e.g.} Refs.~\cite{Kirby2021contextualsubspace,Gu:2023ylw,jkcp-6km5} and local 1D systems \cite{Bhattacharyya:2023xad,Sun:2024bvn}. 
The approach developed in Ref.~\cite{Robin:2025wip} consists of 
finding an optimal separation of the Hamiltonian $\hat H$ into a stabilizer part $\hat{H}_{stab}$ and a residual term $\hat W$ inducing magic. Precisely, after mapping $\hat H$ onto tensor products of Pauli operators $\hat P$, one can write
\begin{eqnarray}
        \hat{H }= \sum_{P \in \mathcal{S}} a_P \hat P +  \sum_{P \notin \mathcal{S}} a_P \hat P \equiv \hat{H}_{stab} + \hat W \; ,
    \label{eq:Hami_stab}
\end{eqnarray}
where $\mathcal{S}$ is a stabilizer group.
The ground state of $H_{stab}$, which, by definition, is a stabilizer state, 
provides a first approximation to the ground state which can be obtainable classically efficiently.
Magic can be incorporated as a second step using classical many-body techniques or, 
at scale, quantum algorithms implemented on quantum platforms.

The decomposition in Eq.~\eqref{eq:Hami_stab} is reminiscent of mean-field-based methods in many-body physics, where the Hamiltonian is divided into a mean field ({\it e.g.}, Hartree-Fock) and a residual interaction. In that case, the mean-field ground state is an un-entangled Slater determinant (or some symmetry-breaking generalization). 
Here in contrast the SGS can encompass large-scale multi-orbital entanglement, and thus is expected to provide a closer approximation to the true ground state (especially in systems with all-to-all interactions, such as atomic nuclei). 
This method was applied to the LMG model, based on the observation that magic is extensive only around the phase transition, while it decreases in the deformed phase which is characterized by near-maximal entanglement, see Fig.~\ref{fig:LMG_phase_diag}. The SGS was efficiently determined using techniques of graph states and stabilizer tableaus, and was found to capture to a large extent both bi-partite and collective multi-partite entanglement features of the exact solution in the region of large deformation, without resorting to symmetry breaking and restoration techniques. 
Non-stabilizerness was subsequently incorporated via an emulated quantum imaginary-time evolution, whose convergence was found to be significantly faster than when starting from a simple uncorrelated ground state, due to the larger initial overlap with the exact ground state.
Further applications of this approach to more complex systems and nuclei are needed to assess whether it can provide an efficient and accurate description of the emergence of deformation in such systems.

\subsection{Neural Quantum States}

Neural quantum states (NQS)~\cite{Carleo:2016svm,Lange:2024nsr} have emerged as promising tools for describing strongly-correlated systems. One attractive aspect of these networks is their versatility, and the way they can adapt to the targeted system.
For example, it is known that NQS can efficiently encompass MPS, as well as volume-law entangled stabilizer states, even in their simplest restricted Boltzmann machines (RBMs) architectures with a single hidden layer~\cite{Deng:2017uik,Gao:2017xhj}, which is a highly desirable feature for the description of nuclear systems.
Previous works using NQS in the context of nuclear physics include~\cite{Adams:2020aax,Rigo:2022ces,Lovato:2022tjh,Yang:2022esu,Yang:2023rlw,Gnech:2023prs,Yang:2025mhg,Keeble:2026rto}, that employed various classes of networks and mappings
(for a recent review, see Ref.~\cite{Lovato:2026erx}).

From a broad many-body perspective, it is desirable to employ network architectures that mirror the underlying physical system, particularly in how information is distributed among components.
In this context, it is beneficial to 1) understand the quantum complexity structure of the targeted systems, {\it i.e.}, how information is shared and spreads over the degrees of freedom in dynamical processes, and 2) explore the expressive power and limitations of various networks in reproducing different aspects of complexity. Such insights can then inform the selection or design of models that provide an efficient and faithful representation of the system.

While the ability of NQS to capture entanglement has been extensively studied in general settings (see {\it e.g.} Refs.~\cite{Levine:2019gtf,Passetti:2022ilw,Denis:2023dww,Jreissaty:2025qip,Paul:2025uew}), their connection to non-stabilizerness remains largely unexplored.
In the context of random networks,
it has been shown that a class of random RBMs with volume-law entanglement reach a maximum value of non-stabilizerness density $m_2 = \mathcal{M}_2 /N \simeq 0.24$ in the large $N$ limit~\cite{Sinibaldi:2025cst}. 
This is to be compared to Haar random states (states of maximal complexity) which are characterized by $m_2 \rightarrow 1$~\cite{Leone:2021rzd,Szombathy:2025euv}, 
while tensor-product states can attain a maximum $m_2 \simeq 0.585$, as discussed in section~\ref{sec:QC}. 

\begin{figure}[ht]
\centering
\includegraphics[width=\columnwidth]{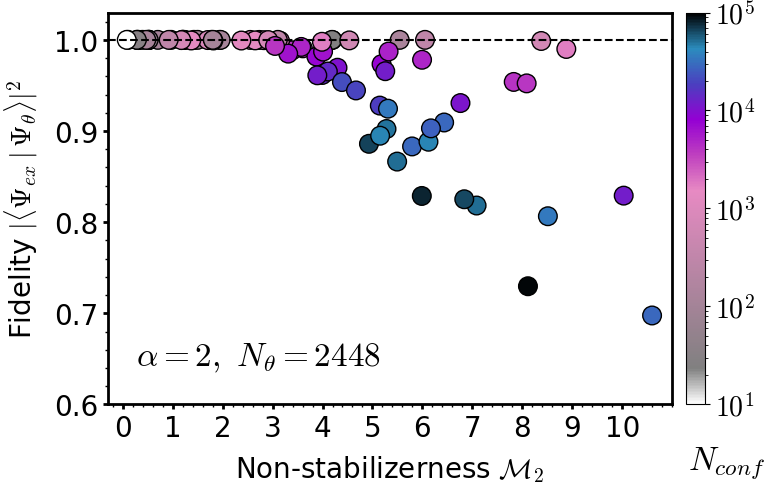}
\caption{Fidelity of NQS nuclear ground states $\ket{\Psi_\theta}$ with respect to the exact ground state, as a function of the non-stabilizerness $\M_2(\ket{\Psi_{ex}})$, for nuclei of the $sd$ shell~\cite{Keeble:2026rto}. 
The color represents the total number of many-body configurations $N_{conf}$ (size of the many-body Hilbert space) in the chosen symmetry sector.
The NQS wavefunction is a RBM with hidden-node density $\alpha=2$, which corresponds to $N_\theta = 2448$ networks parameters.
For a given $N_{conf} > N_\theta$, the fidelity decreases as a function of $\M_2(\ket{\Psi_{ex}})$.
}
\label{fig:RBM_sdshell_M2}
\end{figure}

Within nuclear physics, a first study connecting network efficiency and quantum complexity has been performed~\cite{Keeble:2026rto}.
A RBM representation of medium-mass nuclei in the shell-model valence space 
was constructed using a second-quantization framework, where orbital occupation numbers are mapped to visible nodes, and the performance of the network as a function of the complexity of the target state was carried out.
For a given number of configurations exceeding the number of parameters in the network, 
states with higher non-stabilizerness were found to be consistently more difficult to represent with the network, as presented in Fig.~\ref{fig:RBM_sdshell_M2}. 
This trend was present independently of the width of the hidden layer, suggesting that non-stabilizerness is a key factor influencing the compression and representational efficiency of RBMs for entangled systems.
Further studies with more expressive architectures are needed to assess whether this behavior is a general feature of neural quantum states or specific to RBMs.

As discussed in section~\ref{sec:Qcomplex_nuclei}, 
nuclei are known to exhibit various types of correlations, 
such as pairing and deformation, 
that contribute to differing degrees, according to their number of protons and neutrons. 
Since these correlations are associated with different regimes in terms of entanglement and magic, 
the flexibility of neural networks in representing matrix-product states or stabilizer states efficiently 
may make them a well-suited tool for capturing the relevant aspects of nuclei in a natural way.
Overall, employing complexity-guided network architectures is expected to ensure maximal representational power for targeted physical systems, that may be further enhanced by quantum resources.
\\

Finally, let us mention another group of emerging approaches aimed at combining tensor network techniques and stabilizer techniques, not yet explored in nuclear physics. 
These include stabilizer tensor networks, 
{\it e.g.}, Refs.~\cite{Nezami:2016zni,Masot-Llima:2024doz,Nakhl:2024gfr} and Clifford-augmented matrix product states, {\it e.g.}, Refs.~\cite{PRXQuantum.6.010345,PhysRevLett.133.150604,Qian:2024vea,Huang:2024ron}. 
Such methods also have the potential of describing interplays between stabilizer and magic phases, although they have been shown to capture multipartite entanglement in a limited way~\cite{Nezami:2016zni}.

%% file: Section_Fundamental_Particles.tex
\subsection{Fundamental Particles}
\label{sec:SM}
\noindent
Beyond tests of Bell inequalities and hidden variables, and the role of decoherence, 
pursued in high-energy experiment and theory
(see Ref.~\cite{Bertlmann2004} for enlightening review),
new techniques from QIS are diversifying the study of quantum complexity in high-energy physics.
In the last decade, the role of entanglement as a potential guiding principle for the Standard Model interactions has been considered.  
By considering tree-level Standard Model scattering amplitudes, 
the pioneering work of Cervera-Lierta, Latorre, Rojo and Rottoli~\cite{10.21468/SciPostPhys.3.5.036,cerveralierta2019thesis}
pointed out that such processes typically maximize final state entanglement.
Two-body scattering processes in quantum electrodynamics (QED), 
including $e^+e^-\rightarrow \mu^+\mu^-$, Mott, M{\o}ller and Bhabha scattering, as well as pair-annihilation, show that the couplings of QED permit maximum entanglement to be produced in final states.
Further, if the QED couplings are treated as free parameters, 
the principle of maximum entanglement (MaxEnt) has
QED as one possible solution.  
Applying MaxEnt to the electroweak sector constrains the weak-mixing angle to be 
$\theta_W=\pi/6$, which is close to the experimental value.
Explicitly,
defining the concurrence, $\Delta$,
of the normalized wavefunction $|\psi\rangle$~\cite{10.21468/SciPostPhys.3.5.036,cerveralierta2019thesis},
\begin{eqnarray}
    |\psi\rangle & = & 
    \alpha |00\rangle+
    \beta |01\rangle+
    \gamma |10\rangle+
    \delta |11\rangle
    \ ,\nonumber\\
    \Delta & = & 2|\alpha\delta-\beta\gamma|
    \ ,\ 0\le\Delta\le 1
    \ ,
\end{eqnarray}
evaluation of the tree-level helicity amplitudes for high-energy $e^+e^-\rightarrow\mu^+\mu^-$,
with contributions from photons and $Z^0$ (and neglecting the $Z^0$ mass), 
gives
\begin{eqnarray}
\Delta_{RL} & = & \frac{4\sin^2\theta}{6\cos\theta+5(1+\cos^2\theta)}
\ ,
\nonumber\\
\Delta_{LR} & = & \frac{\sin^2\theta\sin^2\theta_W}{\cos^4\frac{\theta}{2} 
+ 4 \sin^4 \frac{\theta}{2} \sin^4\theta_W}
    \ ,
\end{eqnarray}
for scattering angle $\theta$ and weak mixing angle $\theta_W$.
Maximizing these concurrences gives
\begin{eqnarray}
\theta =\arccos \left(-\frac{1}{3}\right) \ \forall\  \theta_W 
\ ,\ \ \ 
\theta_W\ =\ \arcsin \left( \frac{\cot\frac{\theta}{2}}{\sqrt{2}} \right)
    \ ,
\end{eqnarray}
which are satisfied for $\theta_W=\pi/6$.

\begin{figure}[!ht]
    \centering
    \includegraphics[width=.9\columnwidth]{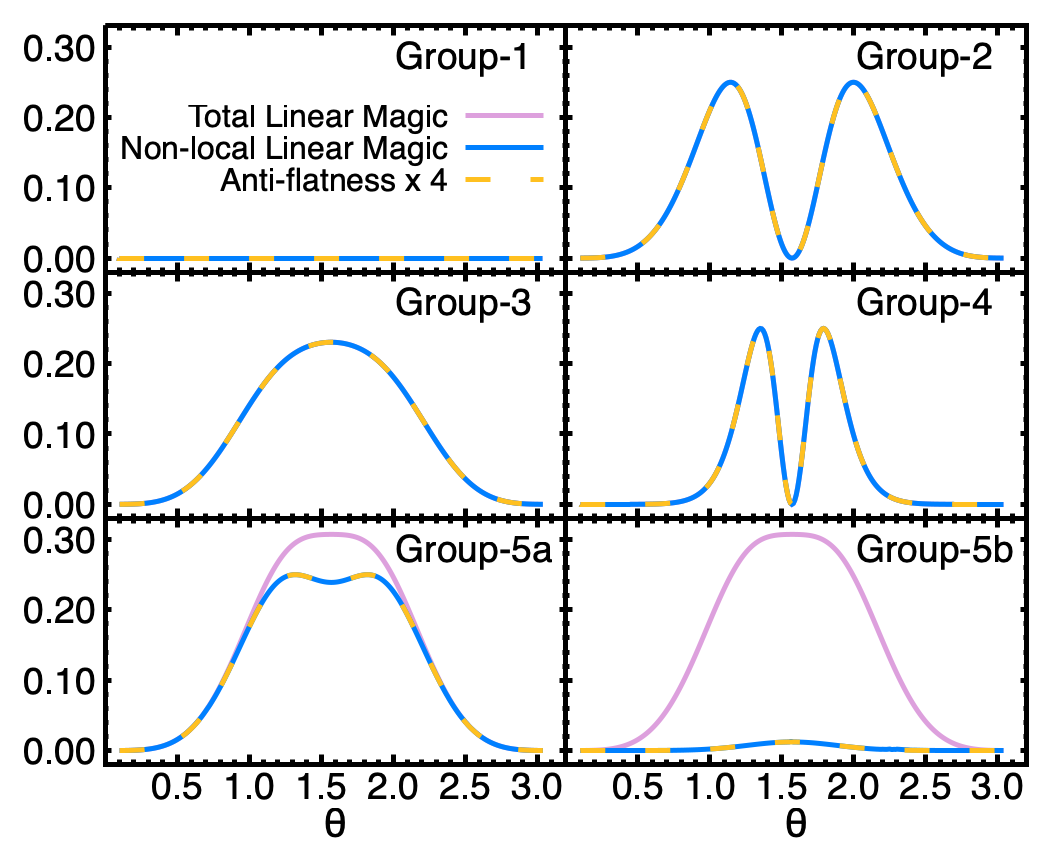}
    \caption{
    The linear magic, ${\cal M}_{\rm lin}(\ket{\chi_i})$, the non-local linear magic, 
    ${\cal M}_{\rm lin}^{(NL)}(\ket{\chi_i})$ and the anti-flatness, ${\cal F}_A(\ket{\chi_i})$ (multiplied by a factor 4), for outgoing states 
    $\ket{\chi_i}$ associated with the distinct groups of initial-state wavefunctions $\ket{\psi_i}$ (defined in Eq.~(\ref{eq:Mollerinitialstates}))
    for high-energy \Moller scattering~\cite{Robin:2025ymq}.
        }
    \label{fig:MOLLHECOMPgr1to5b}
\end{figure}

High-energy \Moller scattering, $e^-e^-\rightarrow e^-e^-$,  provides a sensitive probe for new physics 
and places precise constraints on sectors of the Standard Model.  
Following the work of Ref.~\cite{Robin:2024oqc} 
in defining the magic-power of the S-matrix  in the NN and YN sectors
as a measure of total magic fluctuations in two-particle scattering and hence quantum complexity, 
fluctuations in magic in quantum electrodynamics processes, 
including \Moller scattering
has been studied~\cite{Liu:2025qfl}.
It was determined that quantum electrodynamics appears to be inefficient in 
generating total magic in two-body scattering processes.
In subsequent work, it was found that on the $Z^0$-pole, the total magic production is minimized
when the weak-mixing angle takes the value $\sin^2\theta_W = 0.2317$, 
which 
is remarkably close to the experimental value of $0.23122(3)$
\footnote{The obtained value of $\sin^2\theta_W$ differs slightly from that obtained from minimizing final state entanglement, with gives $\sin^2\theta_W=0.25$~\cite{cerveralierta2019thesis,10.21468/SciPostPhys.3.5.036}.  }.
It has been recently shown~\cite{Robin:2025ymq} that \Moller scattering can  be used to probe the 
production of non-local magic and anti-flatness.
While this does not represent a computation challenge, 
it does provide a glimpse at an underlying process contributing to the 
computational complexity in a larger multi-particle systems.
Representative initial-state  tensor-product wavefunction are selected from each of six groups of stabilizer states, 
\begin{eqnarray}
&& |\psi_1\rangle \ =\ |\uparrow\rangle \otimes |\uparrow\rangle 
\ \ ,\ \ \nonumber \\
&& |\psi_2\rangle \ =\ 
 \frac{1}{\sqrt{2}} \left[\ |\uparrow\rangle + |\downarrow\rangle \ \right]\otimes 
 \frac{1}{\sqrt{2}} \left[\ |\uparrow\rangle + |\downarrow\rangle \ \right]
 \ \ , \ \ \nonumber \\
&& |\psi_3\rangle \ =\ 
 \frac{1}{\sqrt{2}} \left[\ |\uparrow\rangle + |\downarrow\rangle \ \right]\otimes 
 \frac{1}{\sqrt{2}} \left[\ |\uparrow\rangle - |\downarrow\rangle \ \right]
 \ \ , 
 \nonumber\\
&&  |\psi_4\rangle \ =\ 
|\uparrow\rangle \otimes |\downarrow\rangle 
 \ \ ,\ \ \nonumber \\
&& |\psi_{5a}\rangle \ =\ 
 \frac{1}{\sqrt{2}} \left[\ |\uparrow\rangle + i |\downarrow\rangle \ \right]\otimes 
 \frac{1}{\sqrt{2}} \left[\ |\uparrow\rangle + |\downarrow\rangle \ \right]
 \ \ ,\ \ \nonumber \\
&& |\psi_{5b}\rangle \ =\ 
|\uparrow\rangle \otimes 
 \frac{1}{\sqrt{2}} \left[\ |\uparrow\rangle + |\downarrow\rangle \ \right]
 \ ,
 \label{eq:Mollerinitialstates}
\end{eqnarray}
which are acted on by a leading-order insertion of the \Moller scattering amplitude, to generate an
exclusive $e^-e^-$ final state, 
$\hat S \ket{\psi_i} \rightarrow \mathcal{N}\ \hat{\cal A}|\psi_i\rangle=\ket{\chi_i}$.
The non-local magic and anti-flatness can be computed from these six distinct final states.
While experimentally reconstructing the non-local magic of the process from each initial state would require measuring the spins of both particles and reconstructing the complete density matrix, 
the anti-flatness (which bounds non-local magic from below) requires measuring only one of the spins.
Figure~\ref{fig:MOLLHECOMPgr1to5b} shows the 
linear magic, ${\cal M}_{\rm lin}(\ket{\chi_i})$, the non-local linear magic, 
${\cal M}_{\rm lin}^{(NL)}(\ket{\chi_i})$ and the anti-flatness, ${\cal F}_A(\ket{\chi_i})$, defined in Eqs.~\eqref{eq:2SRE}-\eqref{eq:linSRE}, \eqref{eq:NLMdef} and \eqref{eq:Antiflat},
as a function of center-of-mass scattering angle for the six different groups of initial stabilizer states in Eq.~(\ref{eq:Mollerinitialstates}).

\begin{figure}[ht!]
    \centering
\includegraphics[width=0.9\linewidth]{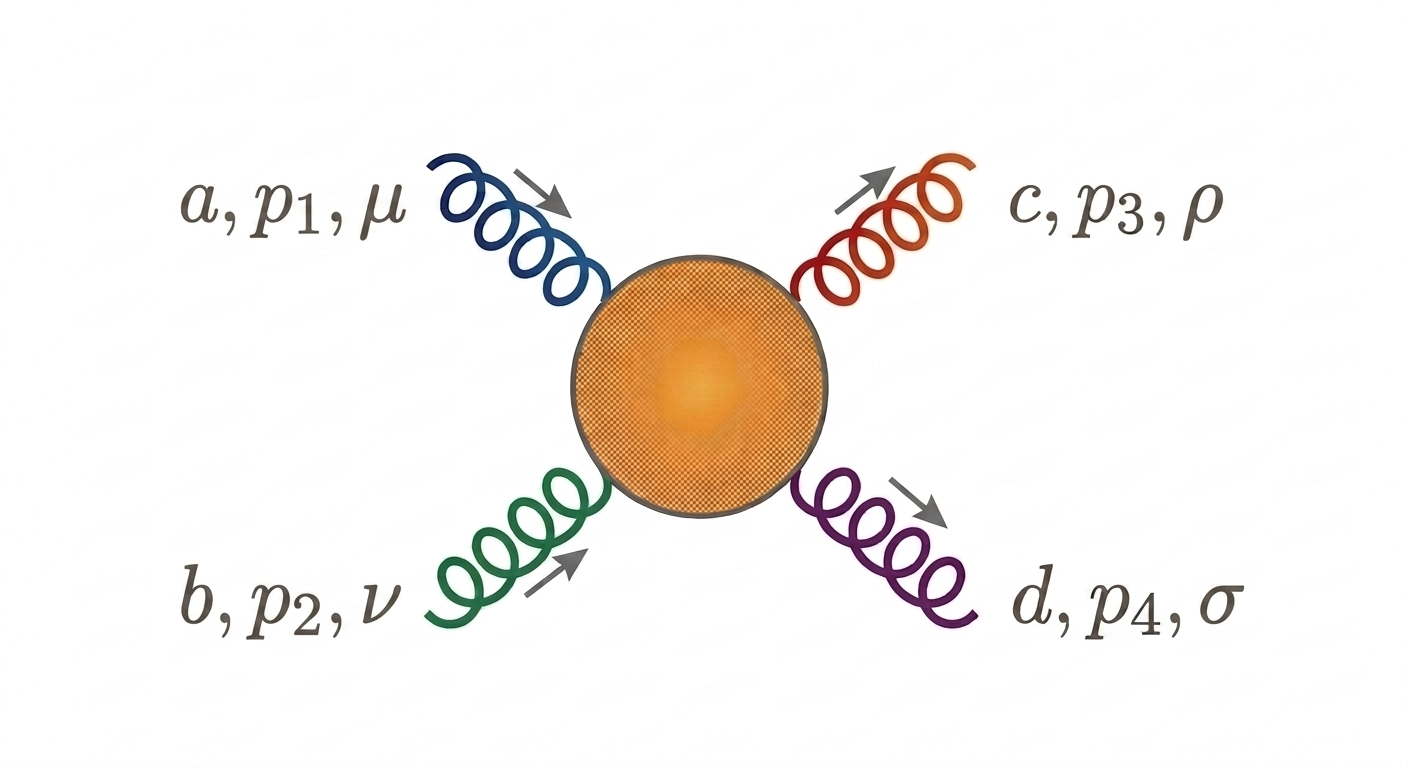}
\caption{
A  depiction of high-energy gluon-gluon elastic scattering. 
The indices are color, momentum and Lorentz, respectively.
}
    \label{fig:gggART}
\end{figure}

\begin{figure}[ht!]
    \centering
\includegraphics[width=0.95\linewidth]{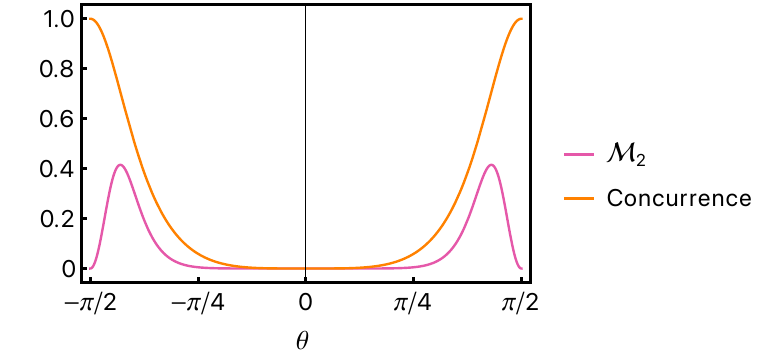}
\caption{
The  concurrence and ${\cal M}_2$ magic 
of gluon-gluon scattering from an initial helicity state $|+,-\rangle$
as a function of center-of-momentum scattering angle, 
$\theta$~\cite{gargalionis2025spinversusmagiclessons},
from Eq.~(\ref{eq:M2ggRL}).
}
    \label{fig:glgr}
\end{figure}

The entangling power, magic power and non-local magic power 
of high-energy gluon-gluon scattering,  $gg\rightarrow gg$,
as depicted in Fig.~\ref{fig:gggART},
have been computed at tree-level from the Yang-Mills Lagrange density,
\begin{align}
{\cal L}_{\rm YM} & =  -\frac{1}{4} \sum_{a=1}^8 G_{\mu\nu}^a G^{a, \mu\nu} 
\ ,\nonumber\\
G_{\mu\nu}^a &=  \partial_\mu A_\nu^a - \partial_\nu A_\mu^a  + g f^{abc}A_\mu^b A_\nu^c
\ ,
\label{eq:YMgggg}
\end{align}
where $f^{abc}$ are SU(3) structure constants, 
$g$ is the QCD coupling constant and $A_\alpha^d$ is the gluon field.
The two helicity states of a gluon, being massless vector particles, 
can be embedded into a qubit, from which 
the concurrence and ${\cal M}_2$ magic generated in
$gg\rightarrow gg$,
from specific initial states, have been computed~\cite{gargalionis2025spinversusmagiclessons}.  
At tree level, in terms of Mandelstam variables,
closed-form expressions for 
the magic and concurrence 
from a $|+,-\rangle$ initial state are given by~\cite{gargalionis2025spinversusmagiclessons}
\begin{align}
    {\cal M}_2 & = 
    -\log 
    \begin{pmatrix}
        \frac{t^{16}+14 t^8 u^8 + u^{16}}{(t^4+u^4)^4}
    \end{pmatrix}
    \ ,\ 
    \Delta\ =\ \frac{2 t^2u^2}{t^4+u^4}
     \ ,
\label{eq:M2ggRL}
\end{align}
which are shown in  Fig.~\ref{fig:glgr}.

Non-local magic generation in these processes has been recently considered~\cite{Gargalionis:2026onv}, motivated by the desire for basis independent measures of quantum complexity.   
It was found that the helicity basis is optimal for considering the quantum complexity, 
however, this is modified by the presence of higher-dimension operators.
These calculations were extended to include gravitons, gravitinos, and gluinos, for which analogous results were obtained~\cite{Gargalionis:2026onv}.
Questions about deviations from gauge invariance have also been asked~\cite{Nunez:2025xds}.
Permitting gauge invariance to be violated by tree-level operators contributing to gluon-gluon and graviton-graviton scattering amplitudes, it was found that the point of gauge invariance coincides with maximal entanglement and minimal magic generation.

Top-quark events reconstructed at the LHC are providing a 
surprising
new window into entanglement and magic in high-energy collisions.
During 2023 and 2024, ATLAS and CMS  reported observations of top-quark-anti-top-quark spin-entanglement~\cite{ATLAS2024,CMS2024,PhysRevD.110.112016}, produced in 
$q\overline{q}/gg\rightarrow t\overline{t}\rightarrow (l^-\nu b)(l^+\overline{\nu}\overline{b})$,
by measuring the relative directions between the charged leptons.
The most general form for the helicity-state density matrix of the $t\overline{t}$-pair
is 
\begin{align}
\hat \rho & =  \frac{1}{4} \left[
\hat I\otimes  \hat I +  \left(B_i^+\hat \sigma^i\otimes \hat I + B_i^- \hat I\otimes\hat \sigma^i \right)
+ C_{ij}\  \hat \sigma^i\otimes\hat \sigma^j
\right]
    \ ,
    \label{eq:ttrho}
\end{align}
where repeated indices are summed over, and 
the $B_i^\pm$ and $C_{ij}$ are coefficients determined from the $t\overline{t}$ ensemble.
The differential cross section for such events is
\begin{align}
    \frac{1}{\sigma} \frac{d\sigma}{d\cos\phi} & =  \frac{1}{2}\left( 1 - D \cos\phi \right)
    \ ,
\end{align}
where $\phi$ is the angle between the two charged leptons in the 
$t\overline{t}$ center-of-momentum frame.
$D$ is an entanglement proxy, such that 
$D < -\frac{1}{3}$ indicates that the $t\overline{t}$-pair are not separable~\footnote{For a more general discussion of separability in mixed (Werner) states, 
see Ref.~\cite{PhysRevA.40.4277}.}.
Experimentally, ATLAS measures $D=-0.537(02)(19)$ for 
$\sqrt{s_{t\overline{t}}}$ between 340 and 380 GeV, 
while CMS measures $D=-0.480^{(26)}_{(29)}$
between 345 and 400 GeV.  A discussion of detailed aspects of the measurements can be found in Ref.~\cite{demina2024localitycollidertestsquantum}.
It has also been pointed out,
because the density matrix is that of a mixed-state,
that other measures of quantum correlations may be observable in $t\overline{t}$ data, including quantum discord~\cite{Han:2024ugl}.

The measurement of non-separability in $t\overline{t}$ production has spurred analogous investigations 
as to the viability of observing entanglement in other analogous processes.  
A 2025 simulation study of $pp\rightarrow\tau^+\tau^- X$, 
with the same structure of mixed-state density matrix as  Eq.~(\ref{eq:ttrho}), 
shows a clear measurement potential for quantum entanglement in the $\tau^+\tau^-$-pair
spin sector~\cite{zhang2025tau}.   
\footnote{Machine learning played an important role in neutrino reconstruction in the numerical simulations.}
\begin{figure}[ht!]
    \centering
\includegraphics[width=\columnwidth]{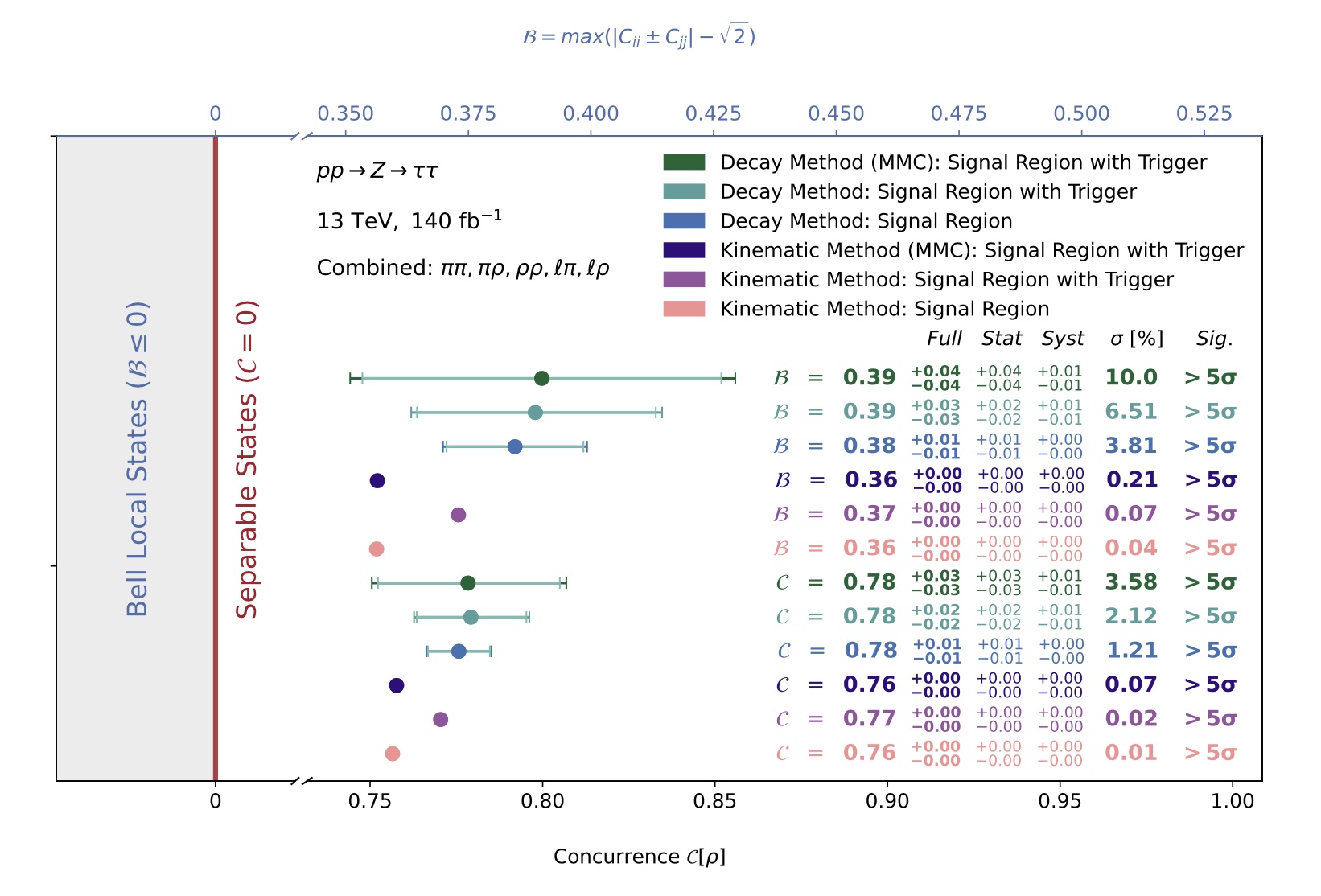}
\caption{
A summary of 
the results of simulations of
non-separability in the spins of $\tau^+\tau^-$ produced in 
$pp\rightarrow\tau^+\tau^- X$~\cite{zhang2025tau} for  the LHC.
[Figure from Ref.~\cite{zhang2025tau} used with permission from the authors under {\it Creative Commons Attribution 4.0 International license}~\cite{cc_by_4.0}.]
}
    \label{fig:LHCtautau}
\end{figure}
Figure~\ref{fig:LHCtautau} provides a summary of the simulation analysis results 
for the non-separability of the 
$\tau^+\tau^-$ pair.

This analysis has been extended to measures of magic~\cite{White:2024mag}, 
appropriately using the mixed-state relations for ${\cal M}_2$~\cite{Leone:2021rzd}.
From their reconstructed 
$t\overline{t}$ spin-density matrix~\cite{PhysRevD.110.112016},
the CMS collaboration has reported results for experimental determinations of 
${\cal M}_2$~\cite{CMS:2025cim,Yazgan:2025pah}, as a function of $\sqrt{s_{t\overline{t}}}$, 
that are consistent with expectations.
\begin{figure}[ht!]
    \centering
\includegraphics[width=0.9\columnwidth]{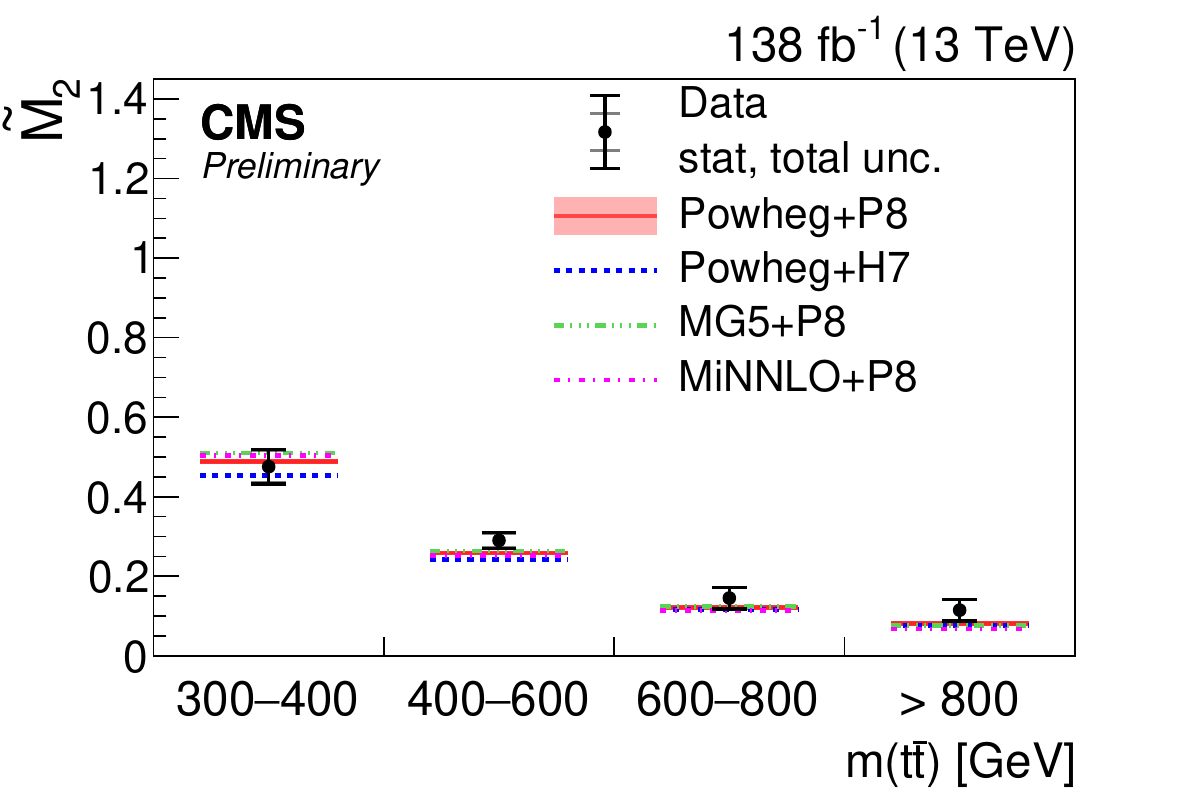}
\caption{
The ${\cal M}_2$ magic in the helicity states of  $t\overline{t}$ produced at LHC measured by the CMS collaboration~\cite{CMS:2025cim}.  A cut on scattering angle has been applied.
[Figure from Ref.~\cite{CMS:2025cim} by the CMS Collaboration, used under 
{\it Creative Commons Attribution 4.0 International license}~\cite{cc_by_4.0}.]
}
    \label{fig:CMSmagic}
\end{figure}
Figure~\ref{fig:CMSmagic} shows the experimentally determined  mixed-state 
${\cal M}_2$ as a function of 
$t\overline{t}$ invariant mass~\cite{CMS:2025cim}, with comparisons to predictions from event generators,
as determined from their reconstructed spin-density matrix.

It has been suggested that 
the structure of the
Higgs sector, specifically the form of 2-Higgs doublet model extensions,
could be constrained by the entanglement power of the Higgs-Higgs scattering
($\phi_i\phi_j\rightarrow \phi_k\phi_l$),
S-matrix~\cite{Carena:2023vjc,Carena:2025wyh}. 
Requiring maximal entanglement fluctuations in the Higgs sector S-matrix
leads to the emergence of symmetries among the quartic couplings, $U(2)\otimes U(2)$.
In contrast, requiring the S-matrix 
to minimize fluctuations in entanglement 
gives rise to a  global $SO(8)$ symmetry among the quartic couplings,
analogous to the results obtained in baryon scattering.
The $SO(8)$ symmetry gives the "natural alignment" in the Higgs sector, 
where one of the CP-even scalars 
can be identified as the Standard Model Higgs boson.
These observations hint that 
traditional "naturalness" arguments could be revised to include quantum consistency 
and information processing as guiding principles for electroweak mass generation.

%% file: Section_Hadron_Structure.tex
\subsection{The Structure of Hadrons}
\label{sec:StructHN}
\noindent
As a non-perturbative composite of quarks and gluons, 
the nucleon emerges from quantum chromodynamics at low temperatures and densities
as the lowest lying baryon.
Its properties, dynamics and short-distance structure remain foci of fundamental research.
Decades of research have provided precise data about the behavior of its constituents, 
in particular high-energy scattering involving highly virtual photons and weak gauge bosons,
have cleanly established parton distribution functions of the valence quarks, and their evolution with 
resolutions scale, $Q^2$.
The current experimental program at JLab is providing precision measurements of
the valence quark distributions, 
that are essential to a complete picture of nucleon structure.
However, while such electroweak probes provide a direct interaction with the quarks, 
they only indirectly interact with the gluons, and 
the Electron-Ion Collider (EIC) is designed to 
provide clean and precise measurements of the gluon and sea-quark distributions.

Given the non-conservation of particle number through strong-interaction processes, 
and the scale dependence of the separation of the constituents into valence and sea distributions, 
one expects the nucleon wavefunction, defined in a given basis, 
to exhibit multipartite entanglement and 
quantum complexity.
First studies were performed by Kharzeev and Levin~\cite{Kharzeev:2017qzs,Kharzeev:2021yyf} in 2017
to quantify these features 
by considering the micro-states available to partons in a 1+1D model.
They made a connection between the von Neumann entropy and the gluon distribution function at small-x
(see also Refs.~\cite{Gotsman_2020,Tu_2020}), and   
suggested that this connection extends to a relation with the entanglement entropy.
Refinements away from small-x, with a kinematic function identified using 
DGLAP evolution~\cite{Gribov:1972ri,ALTARELLI1977298,Dokshitzer:1977sg} 
(for a nice discussion, see Ref.~\cite{Kovchegov_Levin_2023}),
led to an entanglement entropy of,
\begin{align}
    S(E)_{\rm hadron} & =  \log \left( x\Sigma(x, Q^2) \right)
     \rightarrow \log \left( x G(x, Q^2) \right)
    \ ,
\end{align}
where $\Sigma(x, Q^2)$ is the sea-quark distribution, and $G(x, Q^2)$ is the gluon distribution.
This relation agrees well with the available experimental hadron multiplicity data from the H1 collaboration~\cite{h1data}.

\begin{figure}[ht!]
    \centering
\includegraphics[width=0.95\linewidth]{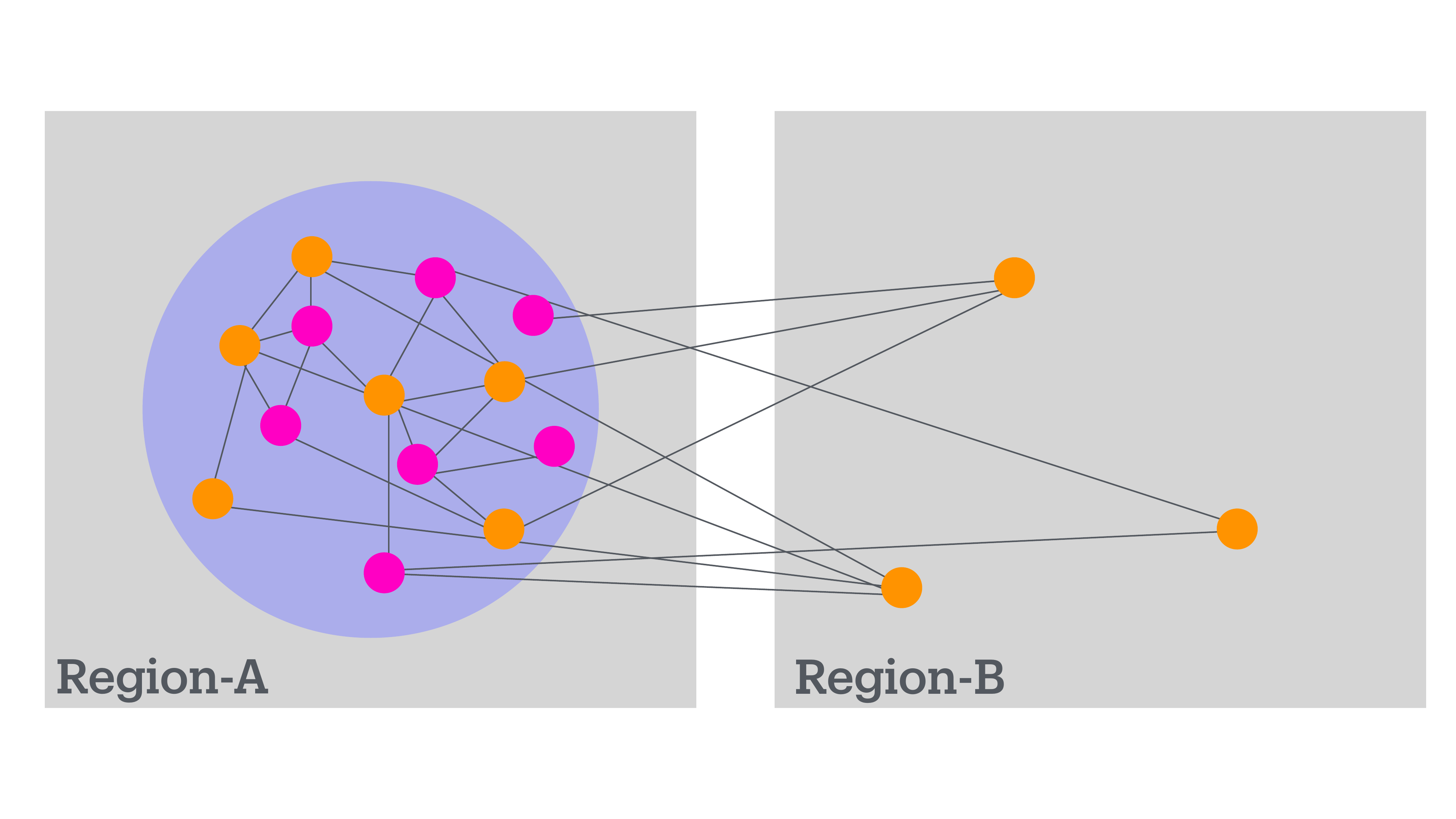}
\caption{
A schematic of the partitioning between valence and sea quarks comprising a hadron~\cite{Beane:2019loz,Ehlers_2023}.
The solid lines denote  correlations between hadrons that have been partitioned between valence and sea spaces, denoted by Regions A and B.
}
    \label{fig:Parton}
\end{figure}

Toward the chiral symmetry breaking scale,
and using light-front chiral wavefunctions,
Beane and Ehlers~\cite{Beane:2019loz,Ehlers_2023}
found that the entanglement entropy 
between partitioned valence and sea quarks
(for a schematic, see Fig.~\ref{fig:Parton})
furnishes an order parameter for chiral symmetry breaking.
Fitting a parameterized chiral model to experimental data, they found the entanglement entropy
to be near maximal.
In the simplest, two-component model,  
the spin components of the nucleon wavefunction in terms of
chiral multiplets can be mapped to two qubits,
\begin{eqnarray}
    |N,\uparrow\rangle & = & \cos\psi\  |00\rangle + \sin\psi\  |11\rangle\ ,
\end{eqnarray}
where the correspondence between the qubit states and the chiral multiplets can be found in Ref.~\cite{Beane:2019loz}.
The valence-quark contribution to the nucleon spin, ${1\over 2} \Delta\Sigma$,
the axial-current coupling constant, $g_A$, and the axial coupling to the decuplet,
${\cal C}_{\Delta N}$,
are given by
\begin{align}
\Delta\Sigma & =  \cos 2\psi
,\ 
|g_A| = {1\over 3} \left(4+\cos 2\psi\right)
,\nonumber\\
|{\cal C}_{\Delta N}| &= 2\cos\psi
\ ,
\end{align}
which,
by fitting $|{\cal C}_{\Delta N}| $ to data, gives
\begin{eqnarray}
\psi & = & 41(4)^o\ , \
\Delta\Sigma = 0.14(13)
,\ 
|g_A| = 1.38(5)
\ ,
\end{eqnarray}
at the scale of chiral symmetry breaking.
Computing the entanglement entropy~\cite{Beane:2019loz}, linear magic, non-local linear magic and antiflatness, gives
\begin{eqnarray}
S_N(\psi) & = & \sin^2\psi\  \log \sin^2\psi + \cos^2\psi\  \log \cos^2\psi\ =\ 0.98(2)
\ ,
\nonumber\\
{\cal M}_{\rm lin} & = & 
{\cal M}^{(NL)}_{\rm lin}
\ =\ 4 {\cal F}_A\ =\ 
{1\over 4} \sin^2 4\psi
\ =\ 0.024(35)
\ .
\end{eqnarray}
While the large-N$_c$ limit of QCD gives $\psi=0$, 
nature has chosen $\psi\sim \pi/4$ and hence the valence-sea 
nucleon wavefunction to be consistent with a (entangled) stabilizer state within fit uncertainties,
with (near) vanishing magic, anti-flatness and non-local magic, and maximal entanglement entropy.

An interesting comparison has been made between inclusive and exclusive cross sections in 
anti-neutrino-nucleus scattering~\cite{Iskander_2020}.  
Specifically, the differential cross sections 
for 
$\overline{\nu}_\mu + A\rightarrow \mu^+ + \pi^0 +  X$ 
and 
$\overline{\nu}_\mu + {}^{12}{\rm C} \rightarrow \mu^+ + \pi^0 + {}^{12}{\rm C}$,
where $X$ is an arbitrary strong interaction final state, and $A$ denotes a hydro-carbon target.
The idea follows from that previous discussion in this section, that the highly inelastic reaction
will be sensitive to the entanglement among the partons between regions of the nucleon, set by the 
scale of the process, while the process leaving the nucleus intact, will not.
The experimental data from the Miner$\nu$a experiment for each of these processes, 
Refs.~\cite{McGivern_2016,Le_2015} and Ref.~\cite{PhysRevD.97.032014},
exhibit the predicted features:
$\overline{\nu}_\mu + A\rightarrow \mu^+ + \pi^0 +  X$  has a clear thermal distribution on top of the hard process for $p_\pi \lesssim 1~{\rm GeV}$ for neutrino energies below $E_\nu\lesssim 10~{\rm GeV}$,  
while $\overline{\nu}_\mu + {}^{12}{\rm C} \rightarrow \mu^+ + \pi^0 + {}^{12}{\rm C}$ show no such structure.

%% file: Section_QuantumThermo.tex
\subsection{Quantum Thermodynamics}
\label{sec:Thermo}
\noindent
While the general area of quantum complexity related to thermodynamics is beyond the scope of this review,
there are interesting results  related to fundamental physics.
The evolution of non-equilibrium quantum systems is a major research area in both nuclear physics and particle physics.  
Such processes play essential roles in fundamental aspects of our universe, including
in electroweak baryogenesis, see, 
{\it e.g.}, Refs.~\cite{Cohen:1993nk,ma2024electroweakbaryogenesisdarkmatter,PhysRevD.110.043516,li2024doeselectronedmpreclude}, 
in fragmentation and hadronization in high-energy collisions, 
in the phases of matter created in heavy-ion collisions and in astrophysical events,
as well as in the evolution of black holes.  
For recent reviews of thermalization in gauge theories, see Refs.~\cite{Mueller:2024mmk,Halimeh:2025vvp}.

Early quantum simulations of matter under extreme conditions have been of 
small, low-dimensional, truncated systems, including 
studies of plaquette chains in non-Abelian lattice gauge theories.
Chains in SU(2) plaquettes with significant truncations in gauge space using the Kogut-Susskind mapping~\cite{Kogut:1974ag,Banks:1975gq} are amenable to classical simulation.
The subsystem size scaling of entanglement entropy has been computed in the 
ground states and excited states,
as has the anti-flatness and magic~\cite{Ebner:2024qtu}.
Area-law scaling is found for the ground state, while volume-law scaling is found for excited states~\cite{Ebner:2024mee} for truncations of $j_{\rm max}=1/2, 1, 3/2$.
Further, the systems exhibit {\it Page curves} with subsystem size~\cite{Page_1993}~\footnote{One of the well-established ``workhorses'' for understanding the apparent thermalization is the entanglement entropy of subsystems of pure states as a function of the size of the subsystem - giving 
rise to {\it Page curves}~\cite{Page_1993}.}.
These studies have also included searches for ``scar states'',
which are states that lie high in the spectrum, but have low entanglement entropy, and delay thermalization of quantum many-body systems - thus violating the Eigenstate Thermalization Hypothesis (ETH) and quantum ergoticity.
Scar states are found to be present in plaquette chains~\cite{Ebner:2024mee}~\footnote{Scar states are also found in  $Z_2$ plaquette chains  with periodic boundary conditions, 
with the intriguing observation that they are also stabilizer states~\cite{Hartse:2024qrv}.
This appears to be a special case, as this correspondence is absent when the size of the transverse direction of the lattice increases.}.
\begin{figure}[ht!]
    \centering
\includegraphics[width=\columnwidth]{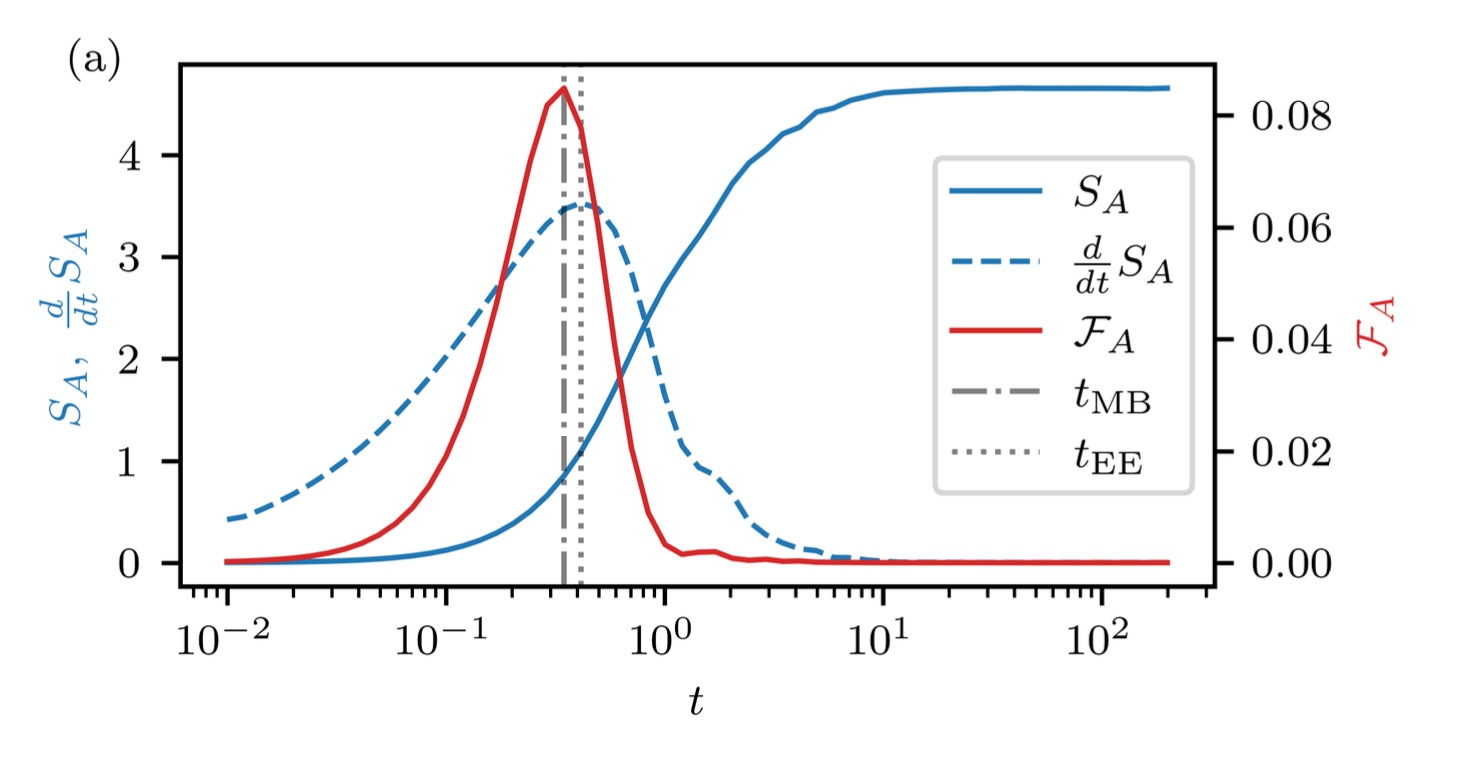}
\caption{
The real-time evolution of entanglement entropy and anti-flatness~\cite{Ebner:2025pdm}
of a two-plaquette subsystem of an asymmetric seven-plaquette system with $j_{\rm max}=1$
for a highly excited initial state chosen at random that lies 19.17~{\rm l.u.}
above the ground state.
[Figure reproduced from Ref.~\cite{Ebner:2025pdm} with permission from the authors.]
}
    \label{fig:PlaQMagic}
\end{figure}
Figure~\ref{fig:PlaQMagic} shows the entanglement entropy and anti-flatness as a function of time from a randomly selected highly-excited (electric-field computational basis) state.
The anti-flatness
(a lower bound on non-local magic), peaks 
at the same time as
the maximum rate of increase of the 
entanglement entropy.  
This behavior is further exhibited for all states within a narrow energy region high in the spectrum.
Thus ``magic barriers'', or peaks in quantum complexity,  are found
during thermalization of highly-excited states in the plaquette chains~\cite{Ebner:2025pdm}. 
Not only that, this also corresponds to a peak in the  non-local magic~\footnote{
The form of the barrier is independent of the bare lattice coupling.
}.
The authors of Ref.~\cite{Ebner:2025pdm} speculate that such barriers exist for a wide range of chaotic theories, 
suggesting that quantum simulation will be generically required for robust results 
from studies of thermalization.
An interesting perspective on the presence of complexity barriers during time evolution 
of a general out-of-equilibrium many-body system
can be found in Ref.~\cite{aaronson2014C}.

There are ongoing efforts underway to further reveal the role of entanglement in thermalization in an array of 
model systems to 
better inform expectations of the Standard Model.
Focusing on the role of entanglement, 
it has been recently realized that 
Entanglement Hamiltonians~\cite{Li_2008,Dalmonte_2022} 
provide connections between states and chaos, and 
new quantum methods, such as classical shadows~\cite{Huang:2020tih} and random sampling, 
enable quantum simulation to access such quantities.
\begin{figure}[ht!]
    \centering
\includegraphics[width=\columnwidth]{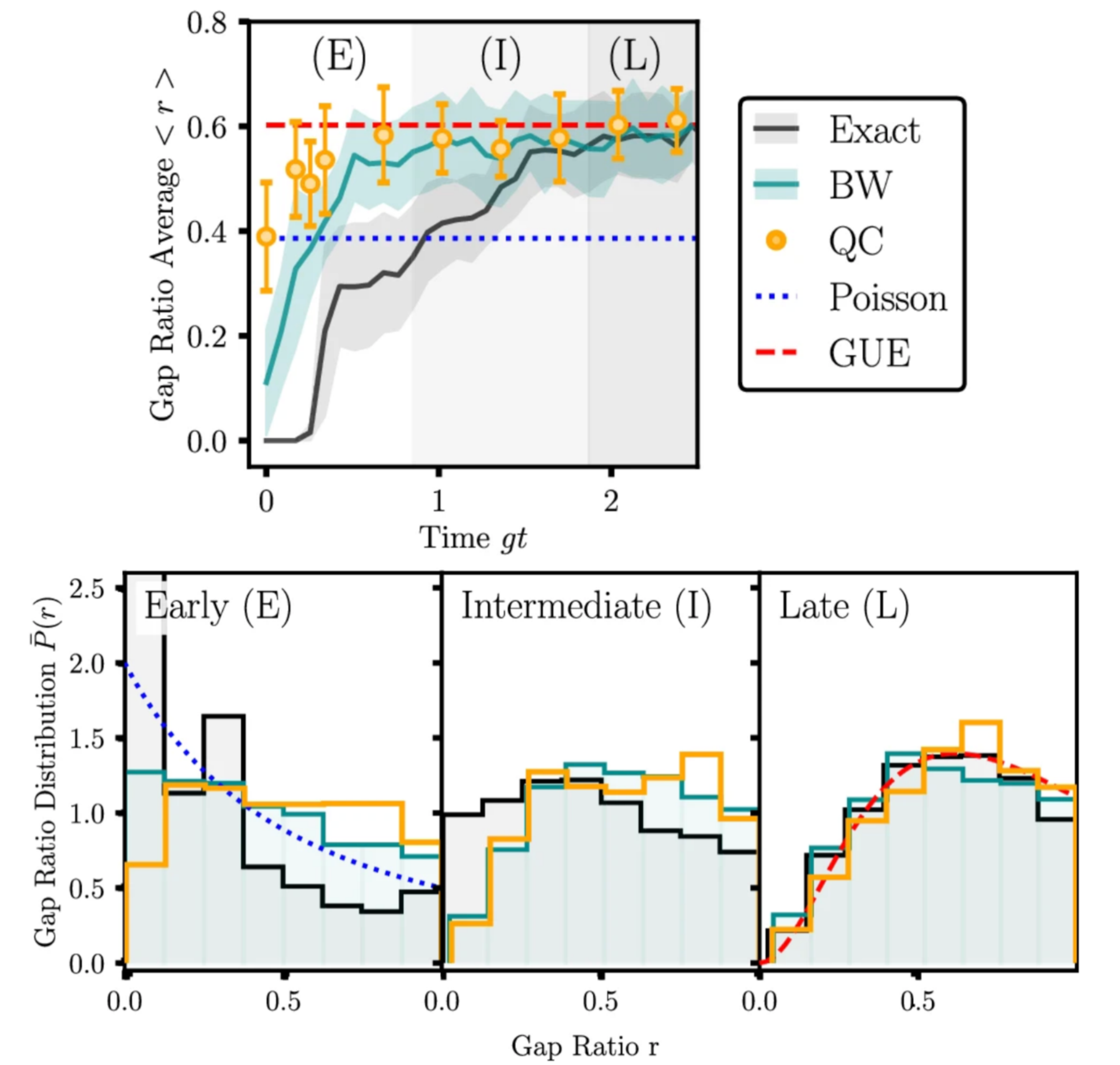}
\caption{
The distributions of gap ratios associated with bi-partitions of plaquette chains of $Z_2$ lattice gauge theory
using a trapped-ion quantum computer, along with theoretical expectations.
The upper panel shows the time evolution from early to intermediate and late times, 
along with the expectations of Poisson and GUE distributions.
The lower panel shows the corresponding distributions of gap ratios in the entanglement spectra.
For more details, see Ref.~\cite{Mueller:2024mmk}.
[Figure adapted from Ref.~\cite{Mueller:2024mmk} with permission from the authors under {\it Creative Commons Attribution 4.0 International license}~\cite{cc_by_4.0}.]
}
    \label{fig:PlaChainThermal}
\end{figure}
As an example, 
recent work was performed to simulate the time evolution of a selection of 
initial-state  gauge-invariant  plaquette chains in $Z_2$ lattice gauge theory
using a trapped-ion quantum computer~\cite{Mueller:2024mmk}.
Experimentally, they estimated 
entanglement Hamiltonians from  measured density matrices of bi-partitions for a modest number of initial states (using a Bisognano-Wichmann~\cite{Bisognano:1976za} inspired assumption about the form of the entanglement Hamiltonian), 
from which the  spacing between energy levels was extracted.
For both experiment and classical simulation,
the distributions of gap-ratios between levels, defined from adjacent energy eigenvalues, 
were found to be consistent with a Poisson distribution at early times 
(non-repulsive pair-wise levels)
and a Gaussian Unitary Ensemble (GUE) at late times (repulsive pair-wise levels),
as shown in Fig.~\ref{fig:PlaChainThermal},
indicating that the systems become chaotic at late times.

%% file: Section_StringBreaking.tex
\subsection{String-Breaking and Fragmentation}
\label{sec:SBFrag}
\noindent
String breaking is at the heart of the evolution from quarks and gluons to hadrons that takes place during 
collisions of high-energy particles and nuclei.
It is governed by the emergent phenomena of confinement and chiral symmetry breaking in QCD.
As this is an essential element in the discovery of new physics,
it has been extensively modeled and simulated
(for recent reviews, see Refs.~\cite{Halimeh:2025vvp,Kharzeev:2026jkq}).
While first-principles dynamical simulations of such processes are beyond the reach of classical computing and analytic
efforts, early progress toward quantum simulations in systems of reduced complexity have taken place~\cite{Bauer:2019qxa,Deliyannis:2022uyh}.
Aspects of string breaking can be probed using the potential between fixed background charges, 
as implemented in
lattice-QCD calculations, {\it e.g.}, Refs.~\cite{PhysRevD.59.031501,CP-PACS:1998hwq,SESAM1998209,Bali:2005fu,Pennanen:2000yk,PhysRevD.63.111501,Bulava:2019iut}. 
Recently, the nature of entanglement in such systems is beginning to be addressed with advances in 
Hamiltonian simulation~\cite{Buyens:2015tea,Berges:2017zws,Mueller:2019qqj,Gong:2021bcp,Farrell:2022wyt,Florio:2023dke,Farrell:2024mgu,Florio:2024aix,Grieninger:2025rdi,Florio:2025hoc,Janik:2025bbz,Barata:2025hgx,Artiaco:2025qqq} 
and spin chains~\cite{Verdel:2019chj,Verdel:2023mmp,Mallick:2024slg},
and now early extensions to 2+1D simulations~\cite{Cochran:2024rwe,Gonzalez-Cuadra:2024xul,Borla:2025gfs,Cataldi:2025cyo,Xu:2025abo,DiMarcantonio:2025cmf,Ciavarella:2024fzw,Ciavarella:2024cyt}. 
Further, string breaking in static and dynamic systems is being simulated using quantum computers, 
{\it e.g.}, Refs.~\cite{Crippa:2024hso,Liu:2024lut,De:2024smi,Surace:2024bht,Ciavarella:2024lsp,Alexandrou:2025vaj,Luo:2025qlg}.

The Schwinger model is the theory of electromagnetism in 1+1D
and  maps straightforwardly to spacetime lattices for numerical simulation~\cite{Schwinger:1962tp,Schwinger:1962tn,Coleman:1976uz,Coleman:1975pw,Kogut:1974ag}. 
It shares features with 3+1D QCD, including confinement and charge screening, and a 
fermion condensate, that has made it a compelling ``sandbox'' for present-day 
quantum simulations that are on a critical path toward simulating QCD.
There has been a number of fruitful studies of its entanglement structure, 
{\it e.g.}, Refs.~\cite{Buyens:2016hhu,Grieninger:2024ent,Florio:2024aix,Florio:2025gic,Florio:2025hoc,Grieninger:2026pai},
including dependence on the mass and coupling~\cite{Gould:2023onb}
and comparisons with its bosonized dual, {\it e.g.}, Ref.~\cite{Frishman:2010zz}.  
As an example, the two-fermion negativity was computed as a function of separation, 
which was found to fall as  a power-law within the confinement radius, and exponentially outside, as anticipated~\cite{Florio:2023mzk}.
Multipartite entanglement has been used  to assess the amount of Lorentz violation 
in real-time simulations of energy loss induced by the non-zero lattice spacing  in the Schwinger model~\cite{Farrell:2024mgu}, which has recently been extended to the 1+1D SU(2) theory~\cite{Li:2025sgo}.

\begin{figure}[ht!]
    \centering
\includegraphics[width=0.98\linewidth]{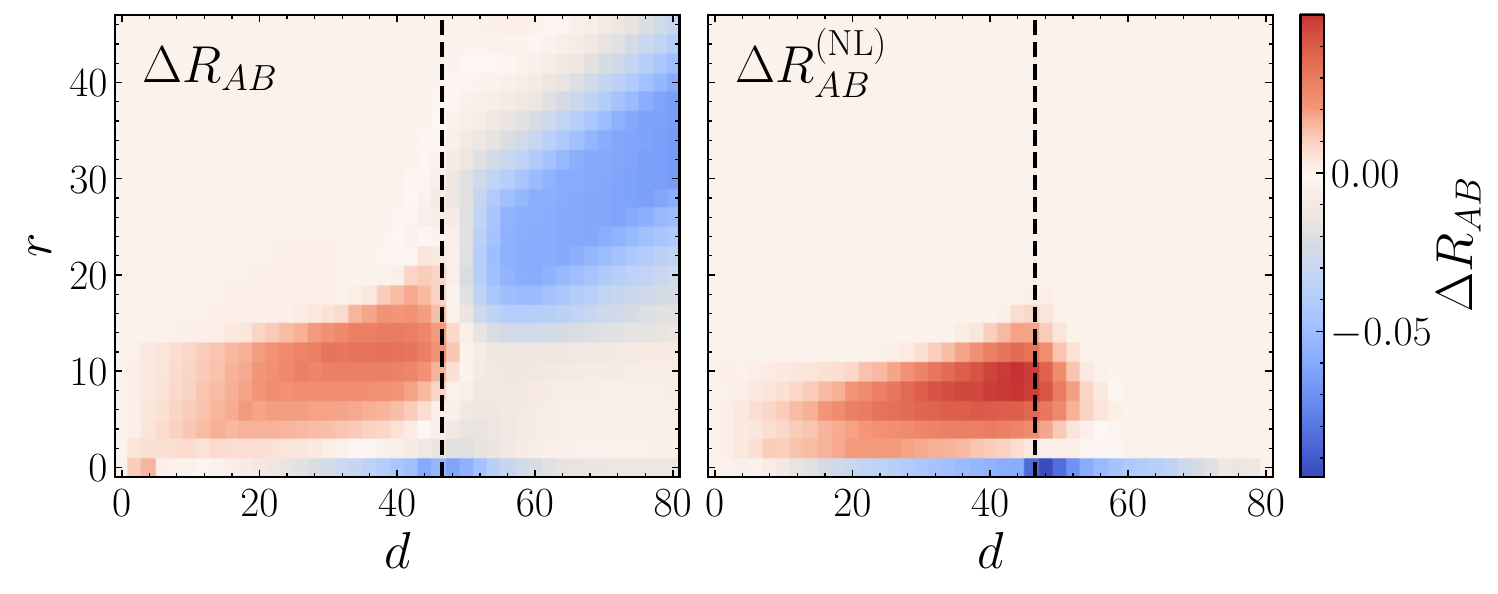}
\caption{
The 
vacuum-subtracted
total (left) and non-local (right) RoM among the $e^-$ and $e^+$ in the Schwinger model as  a 
function of separation between static background charges, $d$, 
between a spatial site at the center of the string and a site at distance $r$~\cite{Grieninger:2026bdq}. 
The parameters of the (classical) simulations were such that both lattice-spacing artifacts 
and finite-volume effects are estimated to be small.
The dashed vertical lines indicate the point of string breaking.
[Figure created by Nikita Zemlevskiy.]
}
    \label{fig:StringRoM}
\end{figure}
Motivated by the need to quantify the quantum resources required to simulate energy loss, fragmentation and hadronization in high-energy processes at scale, 
measures of magic are explored in the string-breaking process in the 
Schwinger model~\cite{Grieninger:2026bdq}. 
Specifically, 
the total and non-local Robustness of Magic (RoM), defined in Eqs.~(\ref{eq:RoMdef}) and (\ref{eq:NLMdef}),
have been computed
in the ground-state wavefunctions of the 
$e^-$ and $e^+$ in the presence of two background static charges for a range of separations, $d$.
Figure~\ref{fig:StringRoM} shows 
both the total and the non-local
RoM along the spatial dimension as a function of  $d$.
\footnote{The optimization defining the non-local RoM preserves the charge in each sector~\cite{Grieninger:2026bdq}}
As the charges separate, both the local and  non-local RoM increase (the total RoM exceeds the non-local RoM).
At the point of string breaking, 
corresponding to the ground state transitioning 
from a flux tube between the static charges
into non-interacting mesons,
the non-local RoM rapidly decreases and vanishes for larger separations.
In contrast, the local RoM persists within the individual mesons.
It is interesting to note that at the point of string breaking, 
the reduced density matrix in the region of the static charges approaches that of a thermal distribution~\cite{Grieninger:2025rdi}. 
Exploring the behavior of non-local quantum correlations in dynamical processes, 
such as the high-energy production of particle pairs, is a near-term objective.

%% file: Section_HeavyIons.tex
\subsection{Heavy-Ion Collisions}
\label{sec:HIC}
\begin{figure}[ht!]
    \centering
\includegraphics[width=0.95\linewidth]{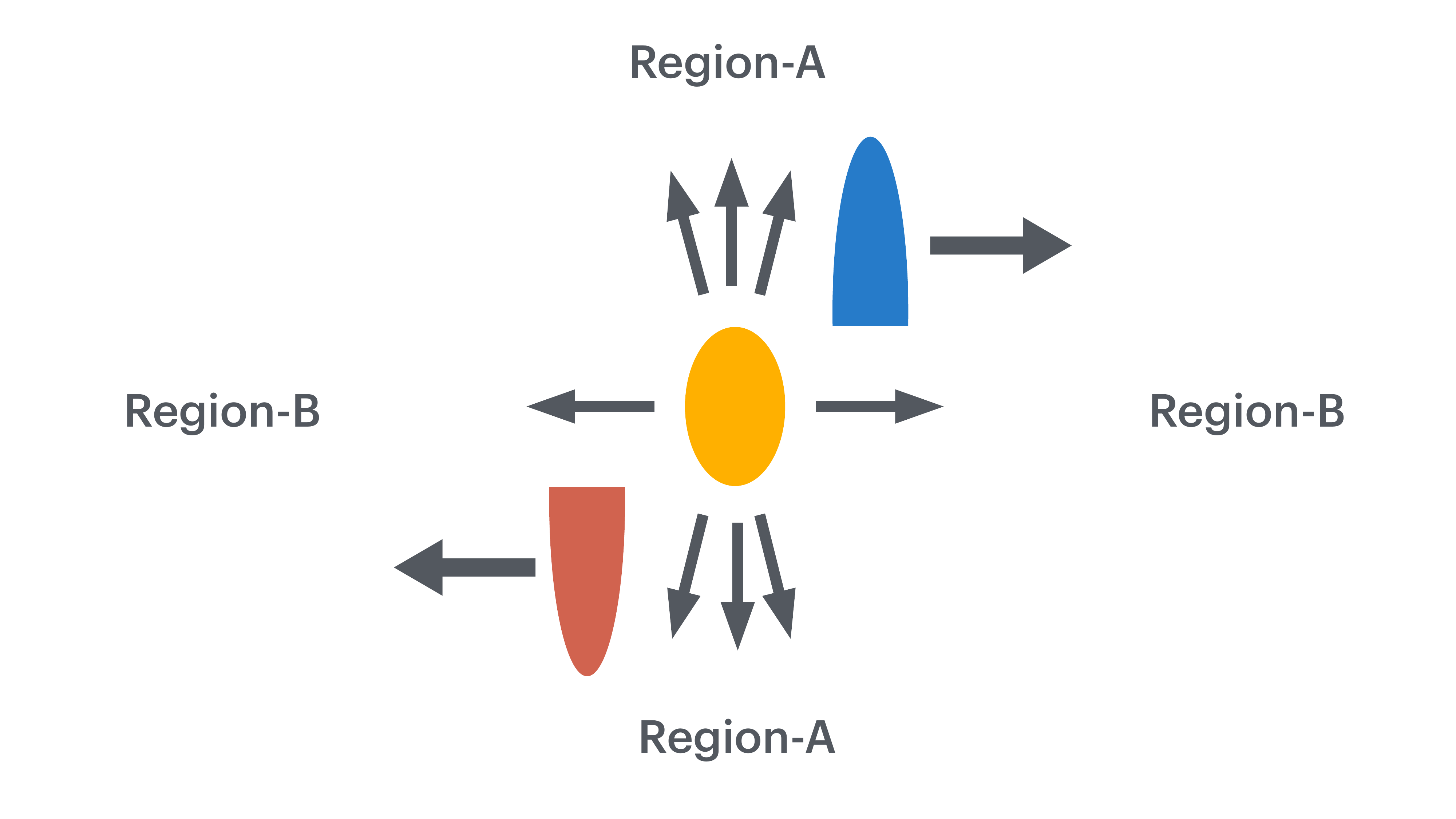}
\caption{
A schematic of a post off-axis collision of two high-energy (Lorentz contracted) 
heavy-ions (red and blue regions).
Hard and co-linear partons continue down the beam-line (partition Region-B),
while (some of) the soft central modes, transverse to the beam axis) 
enter the detection regions (partition Region-A)~\cite{Ho:2015rga}.
}
    \label{fig:HIC}
\end{figure}

\noindent
A remarkably insightful paper by Ho and Hsu~\cite{Ho:2015rga}  made the first connections among entanglement, mixed states and thermal expectation values in heavy-ion collisions (HICs).
It is observed that thermalization and an effective description of HICs by hydrodynamics become consistent with experimental data at times of order $\sim 1~{\rm fm}$ or shorter, which is quite fast on strong-interaction time scales. 
For reviews, see, {\it e.g.}, Refs.~\cite{Casalderrey-Solana:2011dxg,Berges:2020fwq}.
Recognizing that the initial state of a HIC is a pure (tensor-product) state of nuclei
constrained in energy,
and the final state is a mixed state 
(with undetected particles traveling down the beam-line, as shown schematically in Fig.~\ref{fig:HIC}),
they used the results of Ref.~\cite{Popescu_2006} 
to show that the reduced density matrix of observed particles  is,
in general, exponentially close to that of a 
maximally mixed state. 
Further, they made the connection between fast thermalization and entanglement.
These theoretical conclusions were soon investigated using experimental data 
for pp collisions at 13~TeV from the ATLAS and CMS collaborations by Baker and Kharzeev~\cite{Baker_2018}.
They found thermal components in spectra located where they were expected based on entanglement arguments, supporting a connection between rapid thermalization and entanglement.
For an updated analysis, see Ref.~\cite{varma2024a}, 
and for a discussion of modeling these spectra including entanglement, see Ref.~\cite{Trainor:2024wwe}.

It has been suggested, using simple 1+1D models, 
that quantum correlations in particular final states of HICs may provide windows into non-perturbative aspects of QCD, including string breaking and confinement.  
Specifically,  in the context of a two-flavor spin model of string evolution,
the  spin correlations of $\Lambda\Lambda$ pairs can violate (modified) CHSH inequalities~\cite{Gong:2021bcp,Barata:2023jgd}.
As decays of the $\Lambda$ are self-analyzing, 
such types of decay modes have the potential for experimental analysis.
While classical simulations of the model showed the potential of such a measurement, 
the corresponding quantum simulations were not of sufficient fidelity.
Correlations 
between heavier quarks, specifically the entanglement entropy, was examined in the situation where a background electric field propagated across the 1+1D lattice in a way to resemble that of a quark-antiquark pair rapidly separating.

There are a number of efforts highlighting the potential 
utility of the EIC as a probe of the quantum complexity of nucleons and nuclei.
One area that could be explored is the QCD scale evolution of the entanglement entropy within the nucleon,
and how it may be imprinted into the final state hadrons of an EIC deep inelastic scattering events~\cite{Hentschinski_2024}.  
A correlation was found between the evolution from QCD and experimentally determined hadron entropy.
This notion was extended to jet fragmentation, 
where a relation between jet fragmentation functions and the entropy of hadrons produced during
jet fragmentation
was verified 
by ATLAS Collaboration data~\cite{Datta_2025}.
In spin correlations in quark-antiquark pairs produced via virtual photon-gluon fusion, 
it has been calculated that longitudinally polarized photons produce maximal entanglement, 
while transversely polarized photons create significant entanglement 
near the threshold and in the ultra-high-energy regime~\cite{qi2026EIC}. 
The EIC is projected to provide a low background environment for measuring such entanglement in these processes.
It has been found that high-energy deep-inelastic scattering at the EIC also has the potential to generate final states with magic and entanglement, driven by the transversity parton distribution functions~\cite{cheng2025QIatEIC}.

%% file: Section_Neutrinos.tex
\subsection{Coherent Neutrino Oscillations}
\label{sec:neuts}
\noindent
The neutrino sector plays a key role in the evolution of core-collapse supernovae (CCSN)
(for a recent review, see Ref.~\cite{Johns_2025}).
The neutrino mass matrix and lepton-number matrix are not simultaneously diagonal, enabling a neutrino of flavor $i$ to transform into flavor $j$ during evolution in free space.
In dense matter, in addition to these vacuum oscillations, the interaction among neutrinos and matter through electroweak processes (weak charged current and neutral current interactions) 
add to these transformations, with resonant possibilities.
Impressive experimental constraints have already been established on parameters defining 
the neutrino sector, significant ongoing 
experimental programs continue to dramatically improve this knowledge.
The non-equilibrium high-density environment in the core (with potentially different phase structure),
giving way to decreasing densities as the neutrinos move outward, 
lead to highly non-trivial neutrino dynamics, 
including coherent flavor transformations over relatively large distances 
and non-forward scattering processes.
Through differential interactions with the matter, this impacts the evolution of the supernova  
in essential ways, and determines the abundance of elements that are produced.
Much of the knowledge about this transport, 
and the properties of multi-body systems of neutrinos, have been established through formalism and classical computing techniques developed over many decades, {\it e.g.}, Refs.~\cite{Qian:1994wh,Duan:2006jv,Duan:2006an,Izaguirre:2016gsx,Capozzi:2020kge,Fiorillo:2023mze,Fiorillo:2023hlk,Fiorillo:2024fnl,Shalgar:2023ooi,Johns:2023ewj}.  
However, these techniques have well-known limitations, and considerations from quantum information and quantum computing have led to new ideas and model explorations that probe the boundaries of classical techniques and reveal potentially new quantum phenomena (and potentially algorithms)~\cite{Rrapaj:2019pxz,Patwardhan:2019zta,Patwardhan:2021rej,Roggero:2021asb,Xiong:2021evk,Roggero:2021fyo,Martin:2021bri,Roggero:2022hpy,Illa:2022zgu,Bhaskar:2023sta,Martin:2023gbo,Martin:2023ljq,Neill:2024klc,Kost:2024esc,Cirigliano:2024pnm,Siwach:2024jet,Spagnoli:2025etu,Mihalikova:2025ruh,Mangin-Brinet:2025sau,kiss2025neuts,Cervia:2025pfg}.
A major question that remains to be answered 
(even if interesting quantum phenomena are identified in model systems) is:
Does the supernova environment allow for quantum effects in neutrino-flavor evolution to manifest themselves in measurable ways, or are they suppressed by other, non-forward, interactions?
And, further,
is a quantum computer required to simulate aspects of the evolution of supernova, specifically, the neutrino sector?

In present explorations, 
the flavor and momentum content of the 
low-energy effective Hamiltonian derived from the Standard Model are mapped to 
all-to-all connected spin models 
(either qubit or
qutrit, {\it e.g.}, Refs~\cite{Balantekin:2006tg,Siwach:2022xhx,Balantekin:2023qvm,Chernyshev:2024kpu,Turro:2024shh,Chernyshev:2024pqy,Spagnoli:2025etu}).
The structure and evolution of the flavor, entanglement and quantum complexity 
of a multi-neutrino system are simulated,
starting from a selected initial state.
As full 3+1D simulations of such systems are currently impractical,
systems of reduced dimensionality with ``simple'' initial states are employed for present-day quantum simulations, using superconducting qubit and qutrit systems, annealers, and trapped-ions systems~\cite{Hall:2021rbv,Yeter-Aydeniz:2021olz,Illa:2022jqb,Amitrano:2022yyn,Illa:2022zgu,Siwach:2023wzy,Chernyshev:2024kpu,Turro:2024shh,Chernyshev:2024pqy,Spagnoli:2025etu}.
In these modest-sized simulations, pure, typically tensor product, initial states are seen to evolve multipartite entanglement among the neutrinos, exceeding that of Bell-pairs alone~\cite{Illa:2022zgu,Martin:2023ljq}.
Of course, more realistic simulations require substantial improvements above present-day simulations, including working with mixed states.

\begin{figure}[!ht]
    \centering
    \includegraphics[width=0.9\columnwidth]{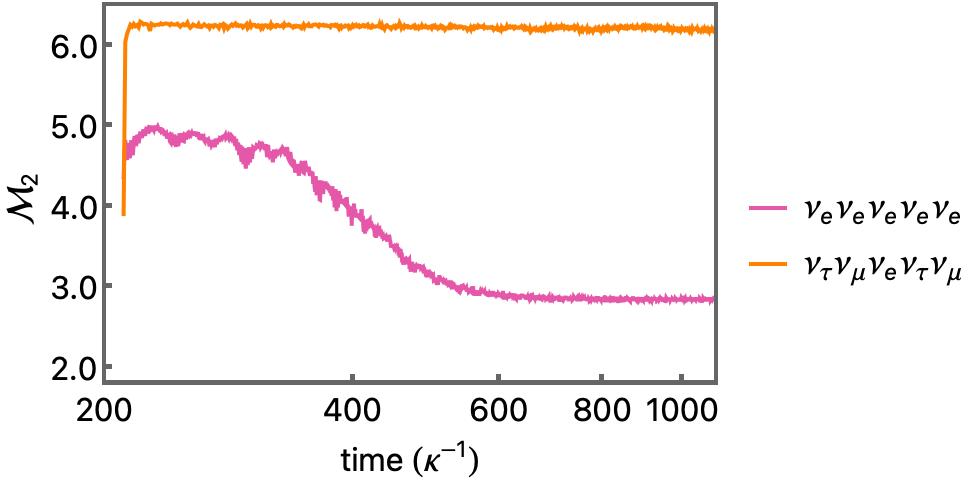} \\
 \vspace{0.5cm}
    \includegraphics[width=0.9\columnwidth]{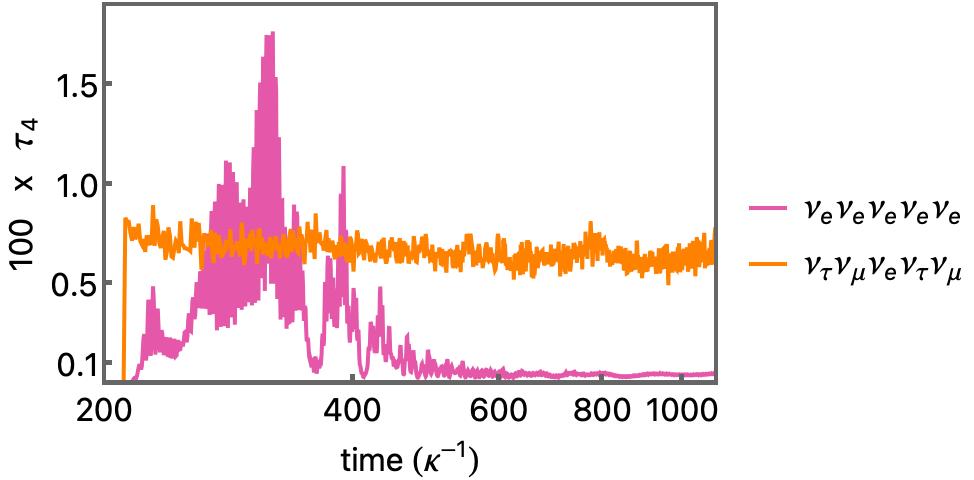}
    \caption{
    The upper panel shows the magic $\mathcal{M}_2$ in the time-evolved neutrino state starting from 
    $|\nu_e\rangle^{\otimes 5}$ and $|\nu_\tau\nu_\mu\nu_e\nu_\tau\nu_\mu\rangle$~\cite{Chernyshev:2024kpu}.
    The lower panel shows the evolution of the 4-tangle $\tau_4$.
    [Image adapted from from Ref.~\cite{Chernyshev:2024kpu} used under 
{\it Creative Commons Attribution 4.0 International license}~\cite{cc_by_4.0}.]
        }
    \label{fig:5nue}
\end{figure}

\begin{figure}[!ht]
    \centering
    \includegraphics[width=0.9\columnwidth]{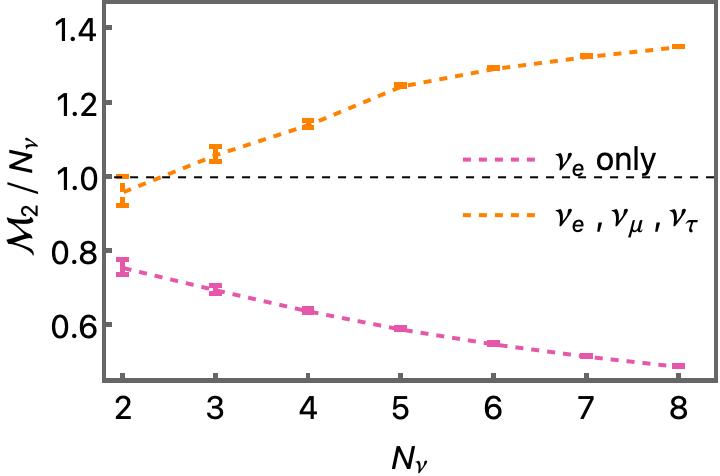}
    \caption{
    The asymptotic magic per neutrino as a function of the number of neutrinos for 
    initial states with only $\nu_e$, and for initial states with all three flavors, 
    determined from multi-qutrit simulations~\cite{Chernyshev:2024kpu}.
    The dotted horizontal lines denotes maximum magic density that can be supported by a tensor-product state.
    [Image adapted from from Ref.~\cite{Chernyshev:2024kpu} used under 
{\it Creative Commons Attribution 4.0 International license}~\cite{cc_by_4.0}.]
    }
    \label{fig:MagDens}
\end{figure}

To illustrate the potential neutrino complexity in supernovae,
we present results obtained from a commonly used model of coherent collective neutrino oscillations
for $n_f=3$ favors of neutrinos
with a model-dependent neutrino density profile (dictating the two-neutrino forward interaction)
in the single-angle limit, {\it e.g.}, Refs.~\cite{Fuller:1987gzx,savage1991neutrino,Pantaleone:1992eq,PhysRevD.46.510,Malaney:1993ah,Kostelecky:1993yt,DOlivo:1995qgv,Qian:1993hh,Fuller:2005ae,Balantekin:2006tg,Balantekin:2023qvm,Chernyshev:2024pqy}
\begin{eqnarray}
    \hat H_2(r) & = &   \mu (r) \sum_{a=1}^8 \hat T^a \otimes \hat T^a  ,    
    \nonumber\\
    \mu(r) & = &  
    \mu_0 \left( 1-\sqrt{1- (R_\nu/r)^2}\right)^2
    \ ,
    \label{eq:2bodyevol}
\end{eqnarray}
where the $\hat T^a$ are the generators of SU(3), 
with
$\mu_0=3.62 \times 10^4$ MeV, 
$\kappa R_\nu=32.2$,
and $\kappa=10^{-17}$MeV, $r(t) = r_0+t$, 
with $r_0=210.65/\kappa$ defining $t=0$.
Using a distribution of neutrino one-body energies below $E_0=10$ MeV, scaling as $E_n=E_0/n$, the 
wavefunction can be written as, 
assuming radial propagation, 
\begin{eqnarray}
|\psi (t)\rangle  & = & \hat U_2 (t,0) |\psi\rangle_0
    \ =\ T \left[ e^{-i \int_{0}^{t}\ dt^\prime\ \hat H (t^\prime)} \right] |\psi\rangle_0
    \ .
    \label{eq:NeutHt}
\end{eqnarray}
Figure~\ref{fig:5nue} shows the results of this evolution, 
for the magic $\mathcal{M}_2$ and  $4$-tangle for an initial pure state 
of $|\nu_e\rangle^{\otimes 5}$
and multi-flavored state 
$|\nu_\tau\nu_\mu\nu_e\nu_\tau\nu_\mu\rangle$
(assuming the normal mass hierarchy).
It is particularly interesting to examine the behavior of the magic per neutrino 
asymptotically in time, with increasing system size for different initial conditions. 
Figure~\ref{fig:MagDens} shows this as a function of the number of neutrinos for initial states $|\nu_e\rangle^{\otimes n}$ and states with all three flavors present.
As the system size grows, the magic per neutrino from a flavor-heterogeneous initial state exceeds 
what can be supported by tensor-product states, 
indicating increasing amounts of non-local magic as the number of neutrinos in the system grows.
Some basic attributes of 2-qutrit systems are given in App.~\ref{sec:TwoQudits}.

The flavor structure of neutrinos in supernovae 
may be surprisingly complex, 
with the potential for significant multipartite entanglement and 
quantum magic~\cite{Chernyshev:2024kpu}.
More realistic simulations are required to quantify the role of non-forward scattering and the extent to which the results obtained for pure state evolution are of importance.
Included in this is the extension of simulations to higher numbers of spatial dimensions and full particle kinetics.
First steps toward such simulations have been accomplished~\cite{Cirigliano:2024pnm}, 
where a full lattice formulation has been established, within the context of the low-energy effective field theory, and 2+1D simulations of modest-sized systems performed.  
Off-forward scattering is found to significantly impact the flavor dynamics,
with kinetic and flavor thermalization times comparable.
A more detailed exploration of quantum complexity would provide further understanding of such systems, including 
calculations of the non-local magic, which is currently under investigation in modest-sized systems~\cite{Hite2026string}.